\documentclass[11pt]{article}
\pdfoutput=1

\usepackage[utf8]{inputenc}
\usepackage{multirow}
\usepackage{booktabs}%
\usepackage{amsmath, amsfonts, amssymb}
\usepackage{comment}
\usepackage{graphicx,import}
\usepackage{psfrag}
\usepackage{amsthm}
\usepackage{mdframed}
\allowdisplaybreaks
\usepackage[dvipsnames,svgnames,table]{xcolor}
\usepackage{enumerate}
\usepackage{arydshln}
\usepackage{soul}
\usepackage{array}
\usepackage{slashed}
\usepackage{nicefrac}
\usepackage[most]{tcolorbox}
\usepackage[font=small,labelfont=bf]{caption} 
\usepackage{subcaption}
\usepackage{mathtools}
\usepackage{colortbl}
\usepackage{mathrsfs}
\usepackage{a4wide}
\usepackage{tikz}
\usepackage{tikz-cd}
\usetikzlibrary{shapes.geometric}
\usepackage{tcolorbox}
\usepackage{diagbox}

\usepackage{color}
\definecolor{dark-gray}{gray}{0.20}
\definecolor{gray}{gray}{0.30}
\definecolor{light-gray}{gray}{0.80}
\definecolor{dark-red}{rgb}{0.7,0,0}
\definecolor{dark-green}{rgb}{0.1,0.4,0}
\definecolor{dark-blue}{rgb}{0.3,0.3,0.7}
\definecolor{light-blue}{rgb}{0.8,0.8,1}
\definecolor{swamp}{RGB}{240, 199, 197}
\definecolor{landscape}{RGB}{180, 250, 199}
\definecolor{undecided}{RGB}{252, 252, 197}
\definecolor{myRED}{HTML}{FF0000}
\definecolor{myGREEN}{HTML}{00AA00}
\definecolor{myBLUE}{HTML}{0055D4}
\definecolor{nORANGE}{RGB}{245, 154, 35}
\definecolor{nBLUE}{RGB}{0, 0, 170}
\definecolor{nGREEN}{RGB}{142, 196, 79}
\definecolor{nRED}{RGB}{170, 0, 0}
\definecolor{nPURPLE}{RGB}{115, 20, 205}
\definecolor{nPINK}{RGB}{255, 126, 136}
\definecolor{nCYAN}{RGB}{127, 198, 233}
\definecolor{IIA}{RGB}{255, 204, 0}
\definecolor{hetGREEN}{RGB}{0, 212, 85}
\definecolor{iBLUE}{RGB}{0, 102, 255}
\definecolor{mthRED}{RGB}{255, 0, 50}
\definecolor{iibBLUE}{RGB}{0, 204, 255}

\usepackage{pifont}
\usepackage{newunicodechar} 

\usepackage{setspace}

\newcommand{\beq}{\begin{equation}}  \newcommand{\eeq}{\end{equation}}
\newcommand{\bal}{\begin{aligned}}   \newcommand{\eal}{\end{aligned}}
\newcommand{\be}{\begin{equation}}
\newcommand{\ee}{\end{equation}}

\def\be{\begin{equation}}
\def\ee{\end{equation}}
\def\bea{\begin{eqnarray}}
\def\eea{\end{eqnarray}}

\newcommand{\ctm}{{\vec{\zeta}}}
\newcommand{\EE}{{E_8 \times E_8}}

\newcommand{\I}{{\text{I}}}
\newcommand{\Ip}{{\text{I}'}}

\newcommand{\II}{{\text{II}}}

\newcommand{\IIB}{{\text{IIB}}}
\newcommand{\Mpl}[1]{ M_{\text{Pl},#1} }
\newcommand{\KK}{{\rm KK}}
\newcommand{\unw}{{\rm unw}}

\newcommand{\h}{{\text{h}}}
\newcommand{\str}{{\text{s}}}
\newcommand{\osc}{{\rm osc}}
\newcommand{\win}{{\rm win}}
\newcommand{\half}{\frac{1}{2}}
\newcommand{\inv}[1]{\frac{1}{#1}}
\newcommand{\Vol}{{\rm Vol}}
\newcommand{\sfrac}[2]{\sqrt{\frac{#1}{#2}}}

\newcommand{\ra}{\rightarrow}

\newcommand{\rasp}{\phantom{a} \Longrightarrow \phantom{a}} 

\renewcommand{\a}{\alpha}

\newcommand{\R}{\mathbb{R}}

\newcommand{\Z}{\mathbb{Z}}
\newcommand{\T}{\mathbb{T}}
\renewcommand{\S}{\mathbb{S}}
\renewcommand{\P}{\mathbb{P}}
\newcommand{\E}{\mathbb{E}}

\newcommand{\abs}[1]{\lvert #1 \rvert}

\newcommand{\dd}{\mathrm{d}}

\def\simleq{\; \raise0.3ex\hbox{$<$\kern-0.75em
      \raise-1.1ex\hbox{$\sim$}}\; }
   \def\simgeq{\; \raise0.3ex\hbox{$>$\kern-0.75em
      \raise-1.1ex\hbox{$\sim$}}\; }

\numberwithin{equation}{section}

\usepackage{jheppub}
\usepackage{hyperref}

\hypersetup{
	colorlinks=true,
	linkcolor=dark-blue,
	citecolor=dark-red,
	urlcolor=dark-green,
	linktoc=page
}

\theoremstyle{remark}

\title{\centering Reconstructing string dualities through\\ symmetries of moduli spaces}
\author{Ignacio Ruiz$^1$,}\affiliation{$^1$CERN, Theoretical Physics Department, 1211 Meyrin, Switzerland}
\author{Germ\'an Varona$^{2,3}$}
\affiliation{$^2$Department of Physics, Universidad de Oviedo, C/ Leopoldo Calvo Sotelo 18, 33007 Oviedo, Spain \\
$^3$Instituto Universitario de Ciencias y Tecnologías Espaciales de Asturias (ICTEA), 
C/ de la Independencia 13, 33004 Oviedo, Spain}
\emailAdd{ignacio.ruiz.garcia@cern.ch}
\emailAdd{varonaluis@uniovi.es}

\preprint{CERN-TH-2026-269}

\abstract{
We perform a bottom-up analysis of the possible duality descriptions characterizing the perturbative regimes associated to the asymptotic regions of quantum gravity moduli spaces. Our approach relies on the symmetries of such moduli spaces as scalar manifolds, as well as the Emergent String Conjecture, which constrains the possible towers of states becoming light, and thus the interpretation of the dual theories emerging at infinite distance. For supergravity theories with 32 and 16 supercharges, our reconstruction uniquely matches those duality frames associated to top-down toroidal compactifications of M-theory/Type II string theory and heterotic string theory, respectively. 
For theories with less supersymmetry, moduli spaces exhibit fewer symmetries, and quantum corrections can render certain directions of the classical moduli space inaccessible; nonetheless, we are still able to constrain the possible arrangements of dualities.
While these are mostly realized by known top-down constructions, there are some intriguing additional possibilities which might lay in unexplored corners of the string landscape. 
}

\begin{document}

\hypersetup{pageanchor=false}
\makeatletter
\let\old@fpheader\@fpheader

\makeatother

\maketitle

\hypersetup{
    pdftitle={Reconstructing string dualities through symmetries of moduli spaces},
    pdfauthor={Ignacio Ruiz, German Varona},
    pdfsubject={Reconstructing string dualities through symmetries of moduli spaces}
}

\newcommand{\remove}[1]{\textcolor{red}{\sout{#1}}}

\section{Introduction}
\label{sec.Intro}

One of the main lessons of working in string theory (this being a general expectation of Quantum Gravity, see \cite{Harlow:2018jwu,Harlow:2018tng,Harlow:2018fse,Banks:2010zn,Cordova:2019jnf,McNamara:2020uza}) is the absence of continuous \textit{free parameters}, such as gauge couplings or particle masses. Instead, these are found to be functions of the vacuum expectation value (v.e.v.) of scalar fields $\{\varphi^a\}_a$ within the low-energy EFT description. At the two-derivative level, their dynamics are governed by the effective $d$-dimensional action\footnote{Throughout this work we take the $\{\varphi^a\}_a$ to have mass dimension 0.}
\begin{equation}\label{eq. eff act}
    S^{(d)}\supset\int\dd^dx\sqrt{-g_d}\left\{\frac{1}{2\kappa^2_d}\left[\mathcal{R}_g-\mathsf{G}_{ab}(\vec\varphi)\partial_\mu\varphi^a\partial^\mu\varphi^b\right]-V(\vec\varphi)\right\}\,.
\end{equation}
From a top-down perspective, their origin can be very varied, associated with higher $(D=d+n)$-dimensional couplings such as the string dilaton $\Phi_D$, reduced components of the metric setting the volume or shape of the internal space $X_n$, reduction of higher-dimensional $p$-forms over $p$-cycles of $X_n$, the position and orientation of the branes along the internal space, etc. These fields can be understood as functions $\varphi^a:\mathbb{R}^{1,d-1}\to\mathcal{M}$, where $\mathcal{M}$ is a scalar manifold known as \textit{parameter} or \textit{moduli space} parameterized by $\{\varphi^a\}_a$, whose geometry is set by the kinetic matrix $\mathsf{G}_{ab}$, defining a Riemannian metric over $\mathcal{M}$.\footnote{Additionally, a scalar potential $V(\vec\varphi):\mathcal{M}\to\mathbb{R}$ can be defined over the moduli space making the scalars $\{\varphi^a\}_a$ massive and possibly stabilizing them or preventing some trajectories over $\mathcal{M}$ from being taken. In the coming sections and along this work, we will consider settings with enough supersymmetry that can assure the vanishing $V(\vec\varphi)\equiv 0$, in such a way that the scalars remain massless, making them true \textit{moduli}.}\\
Using the geodesic distance the metric $\mathsf{G}_{ab}$ defines on $\mathcal{M}$, we can classify the scalars as either \textit{compact} or \textit{non-compact}. The latter admit trajectories for which the geodesic distance $\Delta\varphi$ from a given initial point in $\mathcal{M}$ can be made arbitrarily large. We say that such trajectories probe \textit{asymptotic regions}, with the value of the associated EFT parameters being very large or small. This allows for a \textit{perturbative description} of (at least some sector of) the EFT, with a power expansion in terms of the scalar being sent to infinite distance, thus attaining some level of parametric control. Interestingly, many observed physical quantities in our universe take surprisingly small values in Planck units, which raise questions about their origin and stability. Examples include the cosmological constant $\Lambda_{\rm cc} \sim 10^{-120} M_{\rm Pl,4}^4$, the Higgs vacuum expectation value $v_H \sim 10^{-17} M_{\rm Pl,4}$, and neutrino masses $m_\nu \lesssim 10^{-30} M_{\rm Pl,4}$ \cite{ParticleDataGroup:2026mpi}. Thus, understanding the behavior of asymptotic regimes of moduli spaces obtained from string theory or quantum gravity could show us important lessons about our universe.

It is precisely in asymptotic regions of moduli space that the Swampland program \cite{Vafa:2005ui,Brennan:2017rbf,Palti:2019pca,vanBeest:2021lhn,Grana:2021zvf,Harlow:2022ich,Agmon:2022thq} becomes much more restricting if one expects the EFT in question to admit a UV completion into Quantum Gravity. It is in this regimes where QG effects can become important much below the Planck scale, thus affecting the properties of the low-energy effective description. Thus, as per the \textbf{Swampland Distance Conjecture} (SDC) \cite{Ooguri:2006in}, working in these perturbative limits of moduli space comes with a drawback: as one moves along an asymptotic trajectory in $\mathcal{M}$ towards an infinite distance limit, an \textit{infinite tower} of weakly coupled states, with mass scale $m_*$, becomes exponentially light with the traveled geodesic distance,
\begin{equation}\label{eq. distance conj}
    \frac{m_*}{M_{{\rm Pl},d}}\sim e^{-\alpha\,\Delta\varphi}\to 0\quad\text{as}\quad \Delta\varphi\to\infty\,,
\end{equation}
where $\alpha$ is some $\mathcal{O}(1)$ positive number. This conjecture has been tested and sharpened in many settings \cite{Baume:2016psm,Klaewer:2016kiy, Blumenhagen:2017cxt, Grimm:2018ohb,Heidenreich:2018kpg, Blumenhagen:2018nts, Grimm:2018cpv, Buratti:2018xjt, Corvilain:2018lgw, Joshi:2019nzi,  Erkinger:2019umg, Marchesano:2019ifh, Font:2019cxq,Andriot:2020lea,Gendler:2020dfp, Lanza:2020qmt,Bedroya:2020rmd,Klaewer:2020lfg, Lee:2021qkx,Lee:2021usk,Rudelius:2022gbz,Alvarez-Garcia:2023gdd,Alvarez-Garcia:2023qqj,Rudelius:2023mjy,Aoufia:2024awo,Hassfeld:2025uoy,Monnee:2025ynn,Monnee:2025msf,Raucci:2026fzp,Monnee:2026bsj}, and additionally, universal patterns have been observed \cite{Lee:2019wij,Etheredge:2022opl,Castellano:2023stg,Castellano:2023jjt,Etheredge:2024tok}: not only does such a light infinite tower always appear, but its microscopic nature and the exponential rate $\alpha$ at which it becomes light are highly restricted. In particular, the tower can always be interpreted either as a \textit{Kaluza–Klein} tower (associated to the decompactification of extra dimensions) or as a tower of \textit{oscillator modes} of a critical fundamental string, in what is known as the \textbf{Emergent String Conjecture} (ESC) \cite{Lee:2019wij}, also extensively tested, see e.g.~\cite{Lee:2018urn, Lee:2019xtm,Baume:2019sry, Xu:2020nlh, Baume:2020dqd, Perlmutter:2020buo, Lanza:2021udy, Baume:2023msm, Rudelius:2023odg, Calderon-Infante:2024oed, Grieco:2025bjy,Kaufmann:2026rcn}. 

Going back to perturbative regimes, not all the asymptotic limits of a given moduli space are associated to the same emergent theory (in the sense of a higher-dimensional description or a tensionless string). This is perhaps pretty obvious, since in principle, one could imagine accessing regimes at strong coupling or small volume, which also correspond to asymptotic regions of moduli space and, as such, a (possibly different) perturbative description would be expected. Precisely in this regard, the second superstring revolution, occurring during the mid-1990s, was driven by the discovery of dualities, revealing deep connections between the five previously distinct superstring theories \cite{Font:1990gx,Sen:1994fa,Sen:1994wr,Hull:1994ys,Witten:1995ex,Townsend:1995kk,Horava:1995qa,Horava:1996ma,Sen:1998kd}, together with the realization that these theories are actually different limits of a more fundamental, eleven-dimensional theory known as M-theory.

In this sense, the SDC is also referred to as a \textit{Duality Conjecture} (see, e.g., \cite{Montero:2022prj}), as the predicted light degrees of freedom are associated to the \textit{dual} weakly coupled theory that is \textit{emerging} as we move towards infinite distance. For decompactification limits, the KK tower corresponds to the momentum modes along the growing direction(s), that eventually become massless and free as we decompactify. If the light tower is given by the string oscillator modes, then the \textit{dual} theory has a tensionless string, coming with an infinite set of massless higher spin modes which prevent any local EFT description.

Thus, any perturbative regime associated with an asymptotic region of moduli space featuring one or more KK towers signals that the EFT can be obtained through the compactification of some higher-dimensional theory where the internal scale is large in terms of the fundamental length upstairs. Conversely, an emergent string limit admits a weakly coupled string theory description, while its lack indicates that the UV theory is not weakly coupled (think about 11d M-theory or F-theory outside regimes such as the Sen limit).\\

In \cite{Etheredge:2024tok}, the authors used the ESC to propose a set of \textit{taxonomy rules} which, under some assumptions, govern the arrangement of different perturbative descriptions/duality frames along the asymptotic regimes of moduli space. Such reconstruction is, in a way, of a local character, since it only requires the existence of the asymptotic limit. In this work, we will go one step further, and use the \textit{symmetries} of the moduli space $\mathcal{M}$ to further sharpen the possible dual descriptions of the asymptotic regimes of the moduli space. As we will argue, both the arrangement of duality frames and the light towers must respect such symmetries, which in turn severely restricts the possible descriptions. We will exploit this to argue that only a very reduced set of possible dualities are allowed once the symmetry group $\Gamma$ of $\mathcal{M}$ is known. These reproduce exactly the perturbative descriptions in top-down constructions with maximal and half-maximal supersymmetry (32 and 16 supercharges), obtained from toroidal compactification of M-theory/Type II string theory or heterotic string theory, respectively. For theories with less supercharges, quantum corrections become important and not all classically allowed limits are approachable, but still important lessons can be obtained. We extend the approach of \cite{Etheredge:2025ahf} (see also \cite{Etheredge:2026jun,Etheredge:2026cha}) by allowing ``warped'' decompactifications to running solutions, as explained in \cite{Etheredge:2023odp,Raucci:2026fzp}.

Our paper is framed within similar works which aim to justify the existence of string dualities from a bottom-up approach, such as \cite{Bedroya:2023xue,Delgado:2024skw}. Indeed, at least for enough supersymmetry, it seems to point out that the possible perturbative arrangements are precisely those realized by top-down string constructions. This is in line with similar findings in \cite{Grieco:2026yip}, and seem to point that there is some notion of ``string universality'' in the allowed perturbative descriptions.\\

The paper is structured as follows: In Section \ref{sec.Pert reg} we discuss the possible universal description of perturbative regimes from the point of view of the ESC, and discuss their range of validity. In Section \ref{sec.sym} we discuss how symmetries of moduli space can constrain the arrangement of duality frames, explicitly testing it in the moduli spaces of supergravity theories with 32 and 16 supercharges in Sections \ref{sec. Udual} and \ref{sec. half max}, respectively. In Section \ref{sec. low} we explain the possible issues that arise in theories with low or no SUSY, and finally we conclude in Section \ref{sec.conc} with a summary and outlook. In Appendix \ref{app. Details Het T2} we perform some of the details of heterotic string theory on $\mathbb{T}^2$, and finally in Appendix \ref{ap.extra} we discuss additional (likely spurious) arrangements once some assumptions are relaxed.

\section{Perturbative regimes and dimensional reduction\label{sec.Pert reg}}

In very simple cases, such as 10d Type IIA string theory, the parameter space $\mathcal{M}$ is given by the real line $\R$, parameterized by the 10d dilaton $\Phi_{10}$ (which is related to the string coupling by $g_\str = e^{\Phi_{10}}$), and whose infinite distance points are just $\Phi_{10}\to \pm \infty$. General moduli spaces, however, are multidimensional, and their asymptotic structure is considerably more involved. 
Following \cite{Klingenberg+1995} (see also \cite{Castellano:2023jjt,Baines:2025upi}), we can understand the set of infinite distance points $\infty(\mathcal{M})$ as the set of equivalence classes consisting of geodesic rays $c,c':[0,\infty)\to\mathcal{M}$ for which there exists some positive $a_{cc'}>0$ such that the geodesic distance between $c(\lambda)$ and $c'(\lambda)$ is bounded by $a_{cc'}$ for all $\lambda\in [0,\infty)$. Generically we have $\dim\infty(\mathcal{M})<\dim\mathcal{M}$, see \cite{Ooguri:2006in,Raman:2024fcv} for more properties of the asymptotic regions of moduli space.

Similarly, for more than one modulus, the decay rate $\alpha$ in \eqref{eq. distance conj} depends on the asymptotic trajectory taken. This motivates the introduction of \textit{scaling vectors} \cite{Calderon-Infante:2020dhm} (also known as \textit{scalar charge-to-mass ratio vectors}, see \cite{Etheredge:2022opl, Etheredge:2023odp, Calderon-Infante:2023ler,Etheredge:2023usk,Etheredge:2025ahf} for more results),
\begin{equation}\label{eq. zeta vector}
    \vec\zeta_\star=-\vec{\nabla}_{\vec\varphi}\log\frac{m_\star(\vec\varphi)}{M_{\rm {Pl},d}}\,,
\end{equation}
where the gradient is taken with respect to the moduli coordinates of $\mathcal{M}$. Considering $\hat{T}$ the normalized tangent vector for a given asymptotic trajectory, then the scaling $\alpha$ from \eqref{eq. distance conj} is straightforwardly obtained by projecting
\begin{equation}\label{eq. projection}
    \alpha(\hat{T})=\hat{T}\cdot \vec\zeta_\star\,.
\end{equation}
Since asymptotic trajectories defining the same infinite distance point are expected to feature the same tangent vector $\hat{T}$ in tangent space, and possible changes or ``sliding'' in $\vec\zeta_\star$ along parallel asymptotic trajectories are expected to occur only in directions perpendicular to $\hat{T}$ \cite{Etheredge:2023odp,Castellano:2023jjt,Raucci:2026fzp,Grieco:2026yip}, it follows that $\alpha(\hat{T})$ is a well-defined quantity on $\infty(\mathcal{M})$. 

From the point of view of the effective action \eqref{eq. eff act} describing the low-energy gravitational theory, one generally takes a higher-derivative expansion of quantum-gravitational origins\footnote{As noted in \cite{Calderon-Infante:2025ldq}, more generally one expects an expansion of the form 
\begin{equation}
     S_d\supseteq\frac{1}{2\kappa_d^2}\int\dd^d x\sqrt{-g}\left\{\mathcal{R}_g-\mathsf{G}_{ab}\partial_\mu\varphi^a\partial^\mu\varphi^b-2\kappa_d^2V(\vec\varphi)+\sum_{n>2}\frac{\mathcal{O}_n(\mathcal{R})}{\Lambda_{\rm QG}(\varphi)^{n-2}}\right\}+\sum_i\int\dd^d x\sqrt{-g}\sum_{n>2}\frac{\widetilde{\mathcal{O}}_n(\mathcal{R})}{M_i^{n-d}}\;,
\end{equation} where $\{M_i\}_i$, with $M_i\lesssim\Lambda_{\rm QG}$, are a set of scales of \emph{field theoretic} (rather than \emph{quantum gravitational}) origin, purely coming from integrating out massive states present in our theory, such as KK modes along compact dimensions. When the scale $M_i$ is reached, the original EFT breaks down, but this can be cured by integrating in massive states to a (higher-dimensional) EFT. This is opposed to the $\Lambda_{\rm QG}$, where purely quantum gravity effects \emph{kick in} and the possibility of a local EFT description is lost.} of the form 
\begin{equation}
    S_d\supseteq\frac{1}{2\kappa_d^2}\int\dd^d x\sqrt{-g}\left\{\mathcal{R}_g-\mathsf{G}_{ab}\partial_\mu\varphi^a\partial^\mu\varphi^b-2\kappa_d^2V(\vec\varphi)+\sum_{n> 2}\frac{\mathcal{O}_n(\mathcal{R})}{\Lambda_{\rm QG}(\varphi)^{n-2}}\right\}\;,
\end{equation}
where $\mathcal{O}_n(\mathcal{R})$ is a dimension $n$ higher-curvature operator and $\Lambda_{\rm QG}(\varphi)$ is the energy scale suppressing the operators $\mathcal{O}_n(\mathcal{R})$. At energy scales of order $\Lambda_{\rm QG}(\varphi)$, an infinite set of higher-dimensional operators becomes important, rendering the low-energy, local effective field theory description of our gravitational theory invalid.  
A more fundamental quantum gravity description is then required, precluding any semiclassical description in terms of Einstein gravity. This sets $\Lambda_{\rm QG}(\varphi)$ as the highest possible EFT cut-off, which in general will vary over moduli space. 
Naively, one would expect \cite{Donoghue:1994dn} that this QG scale is given by the $d$-dimensional Planck mass, $\Lambda_{\rm QG}\overset{?}{=}M_{{\rm Pl},d}=\kappa_{d}^{-2/(d-2)}$. However, as noted in \cite{Arkani-Hamed:2005zuc, Distler:2005hi, Dimopoulos:2005ac, Dvali:2007hz, Dvali:2007wp,Dvali:2010vm}, 
this expectation breaks down in the presence of a large number $N$ of light states---precisely the situation encountered in asymptotic regions of moduli space, where an infinite tower of states becomes light. In such regimes, $\Lambda_{\rm QG}$ is dramatically lowered to the so-called \emph{species scale}, parametrically below $M_{{\rm Pl},d}$:
$\Lambda_{\rm QG} \simeq M_{{\rm Pl},d}\, N^{-\frac{1}{d-2}} \overset{N\to\infty}{\longrightarrow} 0$. 
From the perspective of the ESC, $\Lambda_{\rm QG}$ can be set either by the Planck scale of the higher-dimensional theory we are decompactifying to, $M_{{\rm Pl},d+n}$, or by the string scale, $M_{\rm s}$. In the former case, the polynomial degeneracy of the KK spectrum leads to a parametric separation between the quantum gravity scale and the KK scale, such that $\Lambda_{\rm QG}\simeq M_{{\rm Pl},d+n}\gg m_{{\rm KK},n}$. In the latter case, the exponential degeneracy of the string spectrum implies that $\Lambda_{\rm QG}\simeq M_{\rm s}\sim m_{\rm osc}$, where $m_{\rm osc}$ denotes the characteristic mass scale of the string oscillator states; see, e.g., \cite{Castellano:2022bvr}.

Within some asymptotic perturbative regime, the characteristic scales involved in the construction of the EFT---such as the internal lengths and curvatures of the internal manifold 
$X_n$---must be parametrically separated from $\Lambda_{\rm QG}$. Following the nomenclature introduced in \cite{Etheredge:2024tok}, the ESC allows for the following two possible asymptotic regimes:
\begin{itemize}
    \item \textbf{Stringy regime}: In the presence of an emergent string limit, the fundamental scale is set by the string scale, $\Lambda_{\rm QG}\simeq M_{\rm s}=\sqrt{\mathcal{T}_{\rm s}}=(2\pi\alpha')^{-1/2}$, or equivalently the string length $\ell_{\rm s}=2\pi\sqrt{\alpha'}$. Moreover, taking the string coupling $g_{\rm s}=e^{\Phi_{10}}$ (where $\Phi_{10}$ is the 10d dilaton), the perturbative regime corresponds to
    \begin{equation}
       \label{eq. cond str} \boxed{g_{\rm s}\ll1\,,\quad \Vol(C_k)\gg \ell_{\rm s}^k\quad\text{for any }k\text{-cycle }C_k\subseteq X_n\,.}
    \end{equation}
    \item \textbf{Planckian regimes}: In this setting there is no emergent string limit, and all infinite distance limits correspond to (possibly partial) decompactifications of the internal space $X_n$. 
    From the relation
    \begin{equation}
        M_{{\rm Pl},d}\sim M_{{\rm Pl},d+n}\big(\Vol(X_n)\ell_{d+n}^{-n}\big)^\frac{1}{d-2}\,,\quad\text{where}\quad \ell_d\sim M_{{\rm Pl},d}^{-1}\,,
    \end{equation}
    we can see that at large volume $M_{{\rm Pl},d+n}\ll M_{{\rm Pl},d}$ parametrically, thus setting the fundamental scale $\Lambda_{\rm QG}=M_{{\rm Pl},d+n}$. The perturbative regime is then given by
    \begin{equation}
        \label{eq. cond planck} \boxed{\Vol(C_k)\gg \ell_{d+n}^k\quad\text{for any }k\text{-cycle }C_k\subseteq X_n\,.}
    \end{equation}
    Note that, in the presence of different internal cycles of $X_n$, these can grow at different rates, and thus result in parametrically separated higher-dimensional Planck masses lighter than $M_{{\rm Pl},d}$. It is easy to see that, within a perturbative regime, the highest-dimensional Planck scale, i.e., $M_{{\rm Pl},d+n}$, will be the lightest of them all.
\end{itemize}

It is clear from \eqref{eq. cond str} and \eqref{eq. cond planck} that a given perturbative regime can only describe a subsector of the asymptotic regions of our moduli space. We can study the span of each type of regime for a generic description. For this, we will assume that our moduli space has two asymptotic directions,\footnote{For $\mathcal{M}$ with larger number of non-compact directions, we can consider our two-dimensional case as a \textit{slice} of the whole moduli space. One can then reconstruct the span of the perturbative regime by taking successive 2d slices and gluing them together.} and compute the angle on $T_{\vec\varphi}\mathcal{M}$ between the two directions limiting the given perturbative regime.

\subsubsection*{Reduction for the stringy regime}
We will consider first the case in which there is an emergent string limit and a decompactification limit associated to $n$ growing internal dimensions. This in turn implies that the fundamental string uplifts to the ($D=d+n$)-dimensional theory, so that we can write the universal string frame action in the higher-dimensional theory:
\begin{equation}
    S^{(D)}\supset\frac{1}{2\kappa^2}\int\dd^Dx\sqrt{-g_{D}^{\rm (s)}}e^{-2\Phi_{D}}\left\{\mathcal{R}_D^{\rm (s)}+4(\partial\Phi_D)^2\right\}\,,
\end{equation}
where $\kappa^2=\ell_{\rm s}^{D-2}$. We now write the following generic ansatz for the string frame line element:
\begin{equation}
    \dd s^2_{D,{\rm (s)}}=g^{(\rm s)}_{MN}\dd x^M\dd x^N=g_{\mu\nu}^{(\rm s)}\dd x^\mu\dd x^\nu+e^{2\varrho_n(x)}\dd \vec y^2\,,
\end{equation}
where $g^{(\rm s)}_{MN}$ and $g_{\mu\nu}^{(\rm s)}$ are respectively the $D$- and $d$-dimensional string frame metrics, and $\varrho_n$ measures the characteristic length of the compact space in string units ($\dd^n y$ has unit volume), i.e.,
\begin{equation} \label{eq. varrho}
    \frac{\Vol (X_n)}{\ell_{\rm s}^n}=\int_{X_n}\dd^ny\sqrt{g^{\rm (s)}_n}=e^{n\varrho_n}\,.
\end{equation}
We then can expand
\begin{equation}
    \sqrt{-g_{D}^{\rm (s)}}\mathcal{R}_D^{\rm (s)}=\sqrt{-g_{d}^{\rm (s)}}e^{n\varrho_n}\left\{\mathcal{R}_d^{\rm (s)}+e^{-2\varrho_n}\mathcal{R}_n^{\rm (s)}-2n\square\varrho_n-n(n+1)\partial_\mu\varrho_n\partial^\mu\varrho_n\right\}\,.
\end{equation}
The integral of the internal curvature $e^{-2\varrho_n}\mathcal{R}_n^{\rm (s)}$ would contribute to a scalar potential in the $d$-dimensional EFT. For the cases we will consider along this work, there is no lower-dimensional potential associated to the internal curvature, and thus we can ignore such term in what follows.\\
Denoting the lower-dimensional dilaton as
\begin{equation} \label{eq. lower dim dilaton}
    \Phi_d=\Phi_D-\frac{n}{2}\varrho_n=\Phi_D-\frac{1}{2}\log\frac{\Vol(X_n)}{\ell_{\rm s}^n}\,,
\end{equation}
the lower-dimensional string action is given, after integration by parts, by
\begin{equation}\label{eq. string frame}
    S^{(d)}\supset\frac{1}{2\kappa^2}\int\dd^dx\sqrt{-g_{d}^{\rm (s)}}e^{-2\Phi_{d}}\left\{\mathcal{R}_d^{\rm (s)}+4(\partial\Phi_d)^2-n(\partial\varrho_n)^2\right\}\,.
\end{equation}
Through a Weyl rescaling $g^{\rm (s)}_{\mu\nu}=e^{\frac{4}{d-2}\Phi_d}g_{\mu\nu}$ to Einstein frame we finally obtain
\begin{align}
    S^{(d)}&\supset\frac{1}{2\kappa_d^2}\int\dd^dx\sqrt{-g_{d}}\left\{\mathcal{R}_d-\frac{4}{d-2}(\partial\Phi_d)^2-n(\partial\varrho_n)^2\right\}\notag\\&=\frac{1}{2\kappa_d^2}\int\dd^dx\sqrt{-g_{d}}\left\{\mathcal{R}_d-(\partial\hat\Phi_d)^2-(\partial\hat\varrho_n)^2\right\}\,,
\end{align}
with $\kappa_d^2=\kappa^2 e^{2\Phi_d}$, or equivalently $M_{{\rm Pl,}d}=\ell_{\rm s}^{-1}e^{-\frac{2}{d-2}\Phi_d}$. The canonically normalized moduli are then defined as
\begin{equation}\label{eq. can norm}
    \hat{\Phi}_d=\frac{2}{\sqrt{d-2}}\Phi_d\,,\quad \hat{\varrho}_n=\sqrt{n}\varrho_n=\frac{1}{\sqrt{n}}\log\frac{\Vol (X_n)}{\ell_{\rm s}^n}\,.
\end{equation}
Within this flat slice of moduli space, the perturbative conditions \eqref{eq. cond str} translate into 
\begin{equation}
    \boxed{\hat{\Phi}_d<-\sqrt{\frac{n}{d-2}}\hat\varrho_n\,,\quad \hat\varrho_n>0}
\end{equation}

\subsubsection*{Reduction for the Planckian regime}
We can now consider a similar approach, though in this case the absence of a higher-dimensional dilaton means that we can simply take the higher dimensional action in Einstein frame,
\begin{equation}
    S^{(D)}=\frac{1}{2\kappa_D^2}\int\dd^Dx\sqrt{-g_{D}}\mathcal{R}_D\,,
\end{equation}
and compactify it on an $n$-manifold that can be well-approximated (either because it is a genuine Cartesian product or because we are in some asymptotic regime where the fibration structure of the manifold is well described by a product, see e.g., \cite{Oguiso1993ONAF,Grimm:2019bey,Liu:1998tha}) as $X_n=X_{n_1}\times\dots\times X_{n_k}$, with $\sum_{i=1}^k n_i=n$, and perform a usual KK reduction on the metric
\begin{equation}
    \dd s_D^2=G_{MN}\dd x^M\dd x^N=e^{-\frac{2}{d-2}\sum_{i=1}^kn_i\sigma^i}g_{\mu\nu}\dd x^\mu\dd x^\nu+\sum_{i=1}^ke^{2\sigma^i}h^{(i)}_{ab}\dd y^a\dd y^b
\end{equation}
where $g_{\mu\nu}$ is the lower-dimensional Einstein frame metric and $h^{(i)}_{ab}$ are internal metrics normalized to have unit volume in $M_{{\rm Pl},D}$ units. Similarly, $\{\sigma_i\}_{i=1}^k$ are a set of radions, depending only on the external $x^\mu$ coordinates, measuring the volume of the internal spaces in higher-dimensional Planck units,
\begin{equation} \label{eq. sigma}
   \sigma_i=\frac{1}{n_i}\log\frac{{\rm Vol}(X_{n_i})}{\ell_D^{n_i}}\,.
\end{equation}
Using now that, modulo total derivatives, the higher-dimensional Ricci scalar can be expanded as
\begin{equation}
    \mathcal{R}_D=e^{\frac{2}{d-2}\sum_{i=1}^kn_i\sigma^i}\left\{\mathcal{R}_d-\sum_{i,j=1}^k\Big(n_i\delta_{ij}+\frac{n_in_j}{d-2}\Big)\partial_\mu\sigma^i\partial^\mu\sigma^j\right\}+\sum_{i=1}^ke^{-2\sigma^i}\mathcal{R}_{h^{(i)}}\,,
\end{equation}
we obtain (ignoring again the internal curvatures, which would give rise to scalar potentials acting on the radions)
\begin{equation}\label{eq. red to d}
    S_d\supseteq\frac{1}{2\kappa_d^2}\int\dd^dx\sqrt{-g_d}\left\{\mathcal{R}_d-\sum_{i,j=1}^k\Big(n_i\delta_{ij}+\frac{n_in_j}{d-2}\Big)\partial_\mu\sigma^i\partial^\mu\sigma^j\right\}\,.
\end{equation}
Note that the scalars above are not canonically normalized (as a matter of fact the metric is not even diagonal). In order to do this, we define, similar to \cite[Appendix A.1]{Etheredge:2024tok}
\begin{equation}
    \hat\rho^i=\sqrt{n_i}\left[\sigma^i+\frac{1}{n}\left(\sqrt{\tfrac{d+n-2}{d-2}}-1\right)\sum_{j=1}^k n_j\sigma^j\right]\;\Leftrightarrow\; \sigma^i=\frac{\hat\rho^i}{\sqrt{n_i}}-\frac{1}{n}\left(1-\sqrt{\tfrac{d-2}{d+n-2}}\right)\sum_{j=1}^k \sqrt{n_j}\hat\rho^j\,,
\end{equation}
where now $\{\hat\rho^i\}_{i=1}^k$ are canonically normalized. Then, the condition \eqref{eq. cond planck} on the internal volumes, i.e., $\sigma ^i>0$, translates into
\begin{equation}
    \boxed{\hat{\rho}^i>\frac{\sqrt{n_i}}{n}\left(1-\sqrt{\tfrac{d-2}{d+n-2}}\right)\sum_{j=1}^k\sqrt{n_j}\hat\rho^j\quad\forall\,i=1,\dots,k\,,}
\end{equation}
which for $k=2$ simplifies to
\begin{equation}
    \hat{\rho}^1>\frac{\sqrt{n_1 n_2}}{\left(1-\sqrt{\frac{d-2}{d+n-2}}\right)^{-1}n-n_1}\hat{\rho}^2\,,\quad \hat{\rho}^2>\frac{\sqrt{n_1 n_2}}{\left(1-\sqrt{\frac{d-2}{d+n-2}}\right)^{-1}n-n_2}\hat{\rho}^1\,.
\end{equation}
\vspace{0.5cm}

With the above expressions it is now straightforward to compute the angular span within the asymptotic trajectories that a perturbative regime spans:
\begin{subequations}\label{eq. angle reg}
\begin{align}
    \textbf{Stringy regime}:\;\theta_{n}&=\arccos\sqrt{\frac{n}{d+n-2}}\,,\\ \textbf{Planckian regime}:\;\theta_{n_1,n_2}&=\arccos\sqrt{\frac{n_1n_2}{(d+n_1-2)(d+n_2-2)}}\,,
\end{align}
\end{subequations}
precisely recovering the results from \cite{Etheredge:2024tok} in the context of \textit{taxonomy rules} for the scaling vectors  \eqref{eq. zeta vector} of light towers. 
Similarly, the exponential rates $\alpha$ in \eqref{eq. distance conj}---which govern how rapidly the light states predicted by the ESC become light---can be obtained from the relations derived from the universal coupling of the $d$-dimensional dilaton in the string frame action, \eqref{eq. string frame}, as well as from Weyl's law: 
\begin{subequations} \label{eq. osc and KK towers}
\begin{align}
 {M_{\rm s}}&=m_{\rm osc}\sim {M_{{\rm Pl},d}}\,e^{\frac{2}{d-2}\Phi_d}\,,\\
 m_{{\rm KK},n}&\sim\Vol(X_n)^{-\frac{1}{n}}\sim \left(\tfrac{\Vol(X_n)}{\ell_{D}^n}\right)^{-\frac 1n}{M_{{\rm Pl},D}}\sim \left(\tfrac{\Vol(X_n)}{\ell_{D}^n}\right)^{-\frac{d+n-2}{n(d-2)}}{M_{{\rm Pl},d}}\,,
 \end{align}
\end{subequations}
resulting in the universal scalings \cite{Etheredge:2022opl,Etheredge:2024tok}
\begin{equation}\label{eq. univ scalings}
    \alpha_{\rm osc}=|\vec\zeta_{\rm osc}|=\frac{1}{\sqrt{d-2}}\,,\qquad \alpha_{{\rm KK},n}=|\vec\zeta_{{\rm KK},n}|=\sqrt{\frac{d+n-2}{n(d-2)}}\,,
\end{equation}
where the exponential rate $\alpha_i$ is given along the pure emergent string/decompactification limit, where $\hat{T}\propto\vec\zeta_i$, see \eqref{eq. projection}. For the sake of completeness (this will prove useful in the coming sections), we also consider the span of the sector of the perturbative Planckian regime associated to $n_1+n_2$ compact dimensions where the $n_1$ dimensions grow much faster than the other $n_2$ directions:
\begin{equation}
        \theta_{n_1\gg n_2}=\arccos\sqrt{\frac{n_1(d+n_1+n_2-2)}{(d+n_1-2)(n_1+n_2)}}\,,\quad \text{with}\; \theta_{n_1\gg n_2}+\theta_{n_2\gg n_1}=\theta_{n_1n_2}\,.
    \end{equation}

This expression will prove valuable in the context where $n_1=n_2$ and the two volumes can be identified by some discrete symmetry, or the growth of some of the $n_2$-volume faster than that of the $n_1$-dimensional part is obstructed (as it can be the case in non-trivial fibrations, see e.g., \cite{Grieco:2025bjy}).\\

One important caveat must be considered. In the derivation of the above universal relations it is implicitly assumed that the internal space decompactifying does so in a homogeneous way, such that the possible internal warped profiles dilute fast enough and the resulting higher-dimensional theory corresponds to Minkowski space. However, as found out in \cite{Etheredge:2023odp} and studied in more detail in \cite{Raucci:2026fzp}, this might not be the case for the decompactification of single internal directions, where the backreaction of codimension-1 defects can be strong enough such that the resulting $D=d+1$ theory does not correspond to Minkowski space but rather a spacetime with a running scalar.\footnote{\label{fn.warping}Following \cite[Section 3]{Raucci:2026fzp}, the higher-dimensional spacetime can be parameterized as
\begin{equation}
    \dd s^2=e^{2\rho(y)}\eta_{\mu\nu}\dd x^\mu\dd x^\nu+ e^{2\sigma(y)}\dd y^2,\quad\text{with }y\in[0,1]\,.
\end{equation}
In \cite{Raucci:2026fzp} the general backgrounds $\dd s_D^2=G_{MN}\dd x^M\dd x^N=e^{2\rho(y)}\eta_{\mu\nu}\dd x^\mu \dd x^\nu+e^{2\sigma(y)}\mathsf{M}_{ij}\dd y^i \dd y^j$ are studied, with $\mathsf{M}_{ij}$ of unit volume. Thus only overall decompactification limits are considered.
}

Such warping can be parameterized through a potential $\hat{V}(\hat\varphi)$ acting on a (canonically normalized) higher-dimensional scalar (which we can interpret as a higher dimensional dilaton or radion), whose backreaction warps the internal space, see Figure \ref{fig. geometries} for an illustrative example. For potentials of the form $\hat{V}(\hat\varphi)\sim V_0 \exp(-\lambda\hat{\varphi})$, with $\lambda>0$,\footnote{As discussed in \cite{Raucci:2026fzp} consistency with Swampland predictions such as the Sharpened Distance Conjecture \cite{Etheredge:2022opl} or the de Sitter conjecture \cite{Obied:2018sgi,Agrawal:2018rcg,Rudelius:2021azq} requires $\lambda\geq\frac{2}{\sqrt{D-2}}=\frac{2}{\sqrt{d-1}}$.} the scaling of the internal space is modified to
\begin{equation}\label{eq. warp scaling}
    \alpha_{\rm KK,1}^{(\lambda)}=\sqrt{\frac{d-1}{d-2}}\left(1+\frac{4(d-2)}{(d-1)\lambda^2}\right)^{-1/2}\,,
\end{equation}
and similarly the range of validity of the perturbative regime:
\begin{subequations}\label{eq.warping}
\begin{align}
    \theta_{\rm 1}^{(\lambda)}&=\arctan\left( \frac{\sqrt{d-2}\left(\lambda \sqrt{d-1}-2\right)}{2(d-2) + \lambda \sqrt{d-1}} \right)\,,\\
       \theta^{(\lambda)}_{\rm 1\leftrightarrow1} &=\arctan\left(\frac{2}{\lambda}\sqrt{\frac{d-2}{d-1}}\right)\,,\\
       \theta^{(\lambda)}_{n,1}&=\arctan\left[\sqrt{\frac{(d+n-1)(d-2)}{n}}\left(1+\frac{4(d+n-2)}{(d+n-1)\lambda^2}\right)^{-1/2}\right]\,.
\end{align}
    \end{subequations}
Here, $\theta_{\rm 1}^{(\lambda)}$ corresponds to a stringy regime featuring both an emergent string limit and a warped decompactification limit, $\theta^{(\lambda)}_{\rm 1\leftrightarrow1}$ describes a Planckian regime where both warped and unwarped decompactifications of the same 1-cycle are present, and $\theta^{(\lambda)}_{n,1}$ corresponds to a Planckian regime with an unwarped decompactification of $n$ directions to a $(d+n)$-theory with an additional warped decompactification limit. Note that for $\lambda\to\infty$ (i.e., vanishing warping potential) we recover the unwarped results \eqref{eq. angle reg}, with $\lim_{\lambda\to\infty }\theta^{(\lambda)}_{\rm 1\leftrightarrow1}=0$, since now both limits are the same unwarped decompactification.  Note that for finite $\lambda$ the range of validity of the perturbative description is lower than in the unwarped case. As argued in \cite{Raucci:2026fzp}, higher codimension defects result in much weaker warpings that dilute away as we decompactify. 
Thus, we expect that the range of validity of the perturbative regime can only be affected for decompactification of a single direction.

\begin{figure}[htb ]
    \centering
    \resizebox{\textwidth}{!}{%
    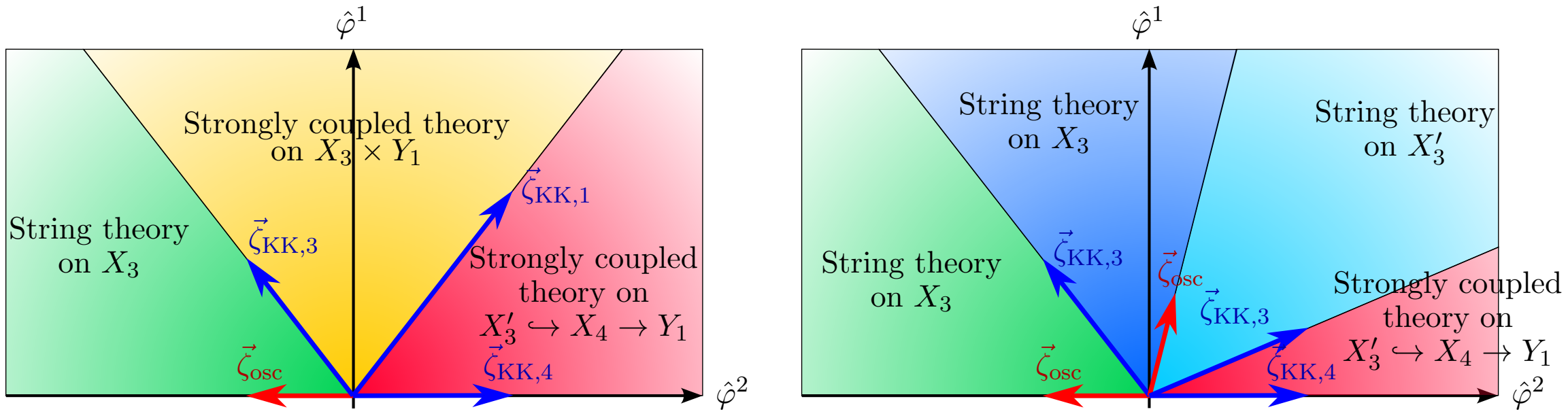
    }
    \caption{Illustrations of different perturbative regimes in $d=7$ filling the asymptotic regimes of a moduli space with two (canonically normalized) moduli $\{\hat\varphi^1,\hat\varphi^2\}$. Starting from the left, the different span of each arrangement is $(\theta_3,\theta_{3,1},\theta_{3\gg 1})$ and $(\theta_3,\theta_3,\theta_3,\theta_{3\gg 1})$, in both cases adding up to a total of $\pi$. We also depict the scaling $\zeta$-vectors associated to the different light towers (in red for oscillator and blue for KK towers), see Section \ref{sec.sym} for their interpretation between common duality frames.}
    \label{fig.example}
\end{figure}

\section{Dualities and symmetries on moduli space\label{sec.sym}} 

\subsection{Dualities and the marked moduli space}

In the previous section, we discussed how the Emergent String Conjecture allowed for two generic classes of perturbative descriptions within the asymptotic regions of moduli space, see \eqref{eq. cond str} and \eqref{eq. cond planck}, and we were able to obtain their generic ranges of validity in terms of the asymptotic trajectories or directions within moduli space, \eqref{eq. angle reg}. However, it is clear that generically a single perturbative regime is not able to describe all the possible asymptotic directions of a given moduli space. As discussed in the introduction of this paper, different perturbative descriptions might be related by dualities mapping different asymptotic regimes of moduli space between each other, in such a way that the small volume/strong coupling regime of a perturbative description is identified with the large volume/weak coupling of another. \\
A paradigmatic example of such dualities is \textbf{T-duality}, which exchanges the KK and winding modes of the fundamental string along a circle of radius $R$, thus relating small and large volume descriptions, with\footnote{See, e.g., \cite{Bergshoeff:1994dg,Alvarez:1995np}, for a more detailed description of the transformation of the rest of the metric components, as well as other fields such as the Kalb-Ramond 2-form, under T-duality.}
\begin{equation}\label{eq. Tdual}
    \mathsf{T}:\; R\to\frac{\alpha'}{R}\,,\quad \Phi\to\Phi+\log\frac{\sqrt{\alpha'}}{R}\,,
\end{equation}
where the dilaton transformation assures that the lower-dimensional dilaton remains invariant under the transformation. Similarly, \textbf{S-duality} relates strong and weak coupling descriptions, with
\begin{equation}\label{eq. Sdual}
    \mathsf{S}:\; \Phi\to-\Phi\,,\quad \ell_s\to e^{-\frac{1}{2}\Phi}\ell_s\,,
\end{equation}
where in the right-hand side we have the \textit{transformed} dilaton. The transformation of the string length is such that the Einstein frame metric is left unaffected by the transformation. Generalization of the above symmetries, such as \textit{Mirror Symmetry} (see \cite{Hori:2003ic} for a review) between Type IIA and Type IIB string theory compactified on mirror CY 3-folds, or Type IIB $SL(2,\mathbb{Z})$ self-duality, extend the above transformations to more involved geometries or field configurations. Similarly, maximal supergravity (obtained by torus compactification of \mbox{M-theory} or Type II string theory) also enjoys a rich \textbf{U-duality} structure, which will be the focus of Section \ref{sec. Udual}.

\begin{figure}[htb ]
    \centering
    \resizebox{0.66\textwidth}{!}{%
    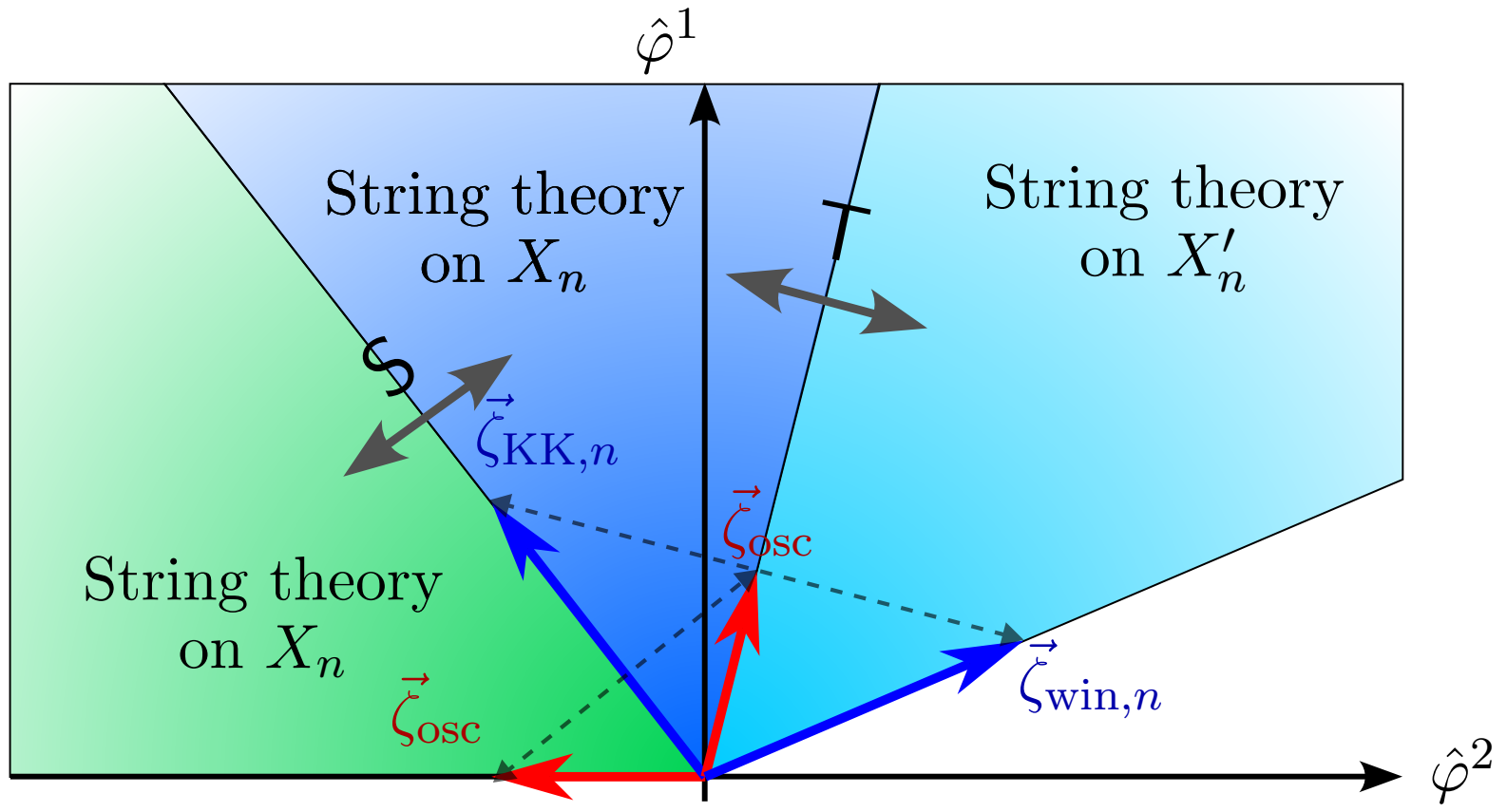
    }
    \caption{Illustrations of the $\mathbb{Z}_2$ transformations corresponding to T- and S-duality, mapping the regime of validity of different perturbative descriptions to each other, including the mapping of the $\zeta$-vectors (in red for oscillator and blue for KK towers) under such dualities. For illustration purposes, we take a 2-dimensional moduli space parameterized by canonically normalized $(\hat\varphi^1,\hat\varphi^2)$, with $d=7$ and $n=3$.}
    \label{fig.TSdual}
\end{figure}

Regarding the above transformations, it is interesting to notice that duality transformations always leave some sector of the spectrum invariant. For example, while T-duality interchanges KK and winding modes, the mass scale of the string oscillator modes is left invariant. Similarly, while S-duality does not leave invariant the string scale, the fact that the Einstein frame metric is not affected means that the KK scale remains invariant under such transformation. Within the more geometrical context of Section \ref{sec.Pert reg}, under the duality transformations discussed above, one has that the scaling $\zeta$-vectors \eqref{eq. zeta vector} transform as\footnote{In order to derive the transformation of $\vec\zeta_{\rm osc}$ under S-duality, we notice that
\begin{equation}
    \frac{m_{\rm osc}}{M_{{\rm Pl},d}}=\frac{m_{\rm osc}}{M_{{\rm Pl},D}}\frac{M_{{\rm Pl},D}}{M_{{\rm Pl},d}}=\frac{m_{\rm osc}}{M_{{\rm Pl},D}}\left(\frac{\Vol(X_n)}{\ell_D^n}\right)^{-\frac{1}{d-2}}=\frac{m_{\rm osc}}{M_{{\rm Pl},D}}\left(\frac{m_{{\rm KK},n}}{M_{{\rm Pl},d}}\right)^{\frac{n}{d+n-2}}\,,
\end{equation}
so that $\vec\zeta_{\rm osc}=-\vec\nabla_{\vec\varphi}\log\frac{m_{\rm osc}}{M_{{\rm Pl},D}}+\frac{n}{d+n-2}\vec\zeta_{{\rm KK},n}$. Since $\frac{m_{\rm osc}}{M_{{\rm Pl},D}}$ depends only on the higher dimensional dilaton, which under S-duality changes sign while leaving invariant the KK vector, we arrive to the desired result.
}
\begin{equation}
    \mathsf{T}:\;\vec{\zeta}_{\rm KK,1}\leftrightarrow\vec\zeta_{\rm win,1}\,,\;\vec{\zeta}_{\rm osc}\to \vec{\zeta}_{\rm osc}\,,\qquad\mathsf{S}:\;\vec{\zeta}_{{\rm KK},n} \to \vec{\zeta}_{{\rm KK},n}\,,\;\vec\zeta_{\rm osc}\to-\vec\zeta_{\rm osc}+\tfrac{2n}{d+n-2}\vec\zeta_{{\rm KK},n}\,.
\end{equation}
Several T-dualities can be performed if our space corresponds to the product of several circles (this is, $\mathbb{T}^n$). In that case, the bound states of the $n$ KK and winding states are mapped to one another, $\mathsf{T}:\vec{\zeta}_{{\rm KK},n}\leftrightarrow\vec{\zeta}_{{\rm win},n}$. The above transformations are depicted in Figure \ref{fig.TSdual}.

The above $\mathsf{T}$ and $\mathsf{S}$ transformations individually generate a $\mathbb{Z}_2$ duality group acting on non-compact scalars (radion and dilaton, each parameterizing $\mathbb{R}$ after canonically normalizing them). In order to generalize this to more involved moduli spaces, following \cite{Raman:2024fcv} we take the notion of \textit{marked moduli space}, $\hat{\mathcal{M}}$, corresponding to the space parameterizing the vacua of a given theory together with a basis of observables. Acting over it we find the duality group $\Gamma$, whose elements leave the EFT invariant while acting non-trivially on its spectrum/observables. The moduli space is then defined by the left-quotient
\begin{equation}
    \mathcal{M}=\Gamma\backslash\hat{\mathcal{M}}\,,
\end{equation}
after modding out the equivalent sectors of the theory. See \cite{Delgado:2024skw} for a bottom-up argument  on the need of including enough duality transformations in $\Gamma$ based on a physical \textit{compactifiability} criterion of $\mathcal{M}=\Gamma\backslash\hat{\mathcal{M}}$. As an example, we can consider 10d Type IIB string theory, with a single complex modulus consisting in the axio-dilaton $\tau=C_0+ie^{-\Phi}$, taking values on the upper-half plane $\mathbb{H}=\{z\in\mathbb{C}\,:\,{\rm Im}(z)>0\}$. The Type IIB action is invariant under the $SL(2,\mathbb{Z})$ action
\begin{equation}\label{eq. SL2Z}
    \tau\to \frac{a\tau+b}{c\tau+d}\,,\quad\begin{pmatrix}
        C_2\\B_2
    \end{pmatrix}\to M\begin{pmatrix}
        C_2\\B_2
    \end{pmatrix}\,,\quad C_4\to C_4\,,\quad g_{\mu\nu}\to g_{\mu\nu}\,
\end{equation}
for $M=\begin{psmallmatrix}
        a&b\\c&d
    \end{psmallmatrix}\in SL(2,\mathbb{Z})$, which on the other hand acts non-trivially on the charge lattice of $(p,q)$-7-branes by $(p,q)^\intercal\to M(p,q)^\intercal$. Thus, the Type IIB moduli space is given by $\mathcal{M}=SL(2,\mathbb{Z})\backslash\mathbb{H}$, with one infinite distance limit at $\tau\to +i\infty$. We will now generalize how this works for more general moduli spaces.

\subsection{The geometry and symmetries of the asymptotic moduli space}

As described above with the prototypical Type IIB example, we build the moduli space $\mathcal{M}$ as the quotient of the marked space $\hat{\mathcal{M}}$ by the duality group $\Gamma$, following \cite{Raman:2024fcv,Delgado:2024skw}. 

Consider that $\hat{\mathcal{M}}$ is equipped with a moduli space metric $\mathsf{g}_{ij}$, which defines a geodesic distance
\begin{equation}\label{eq. geo dist}
    \begin{array}{rccl}
         \Delta:&\hat{\mathcal{M}}\times \hat{\mathcal{M}}&\longrightarrow&[0,\infty)  \\
         & (p,q)&\longmapsto& \inf_{\gamma\in\Gamma_{p,q}}\left\{\int_0^1\sqrt{\mathsf{g}_{ij}(\gamma^i)'(\gamma^j)'}\dd\tau\,\right\}
    \end{array}
\end{equation}
where $\Gamma_{p,q}=\{\gamma:[0,1]\to\hat{\mathcal{M}}\;\text{geodesic with}\;\gamma(0)=p,\,\gamma(1)=q\}$. With this definition in mind, in what follows geodesic paths will have unit tangent vectors. This allows us to define the asymptotic boundary
\begin{equation}
    \infty(\hat{\mathcal{M}})=\{\text{unit speed geodesic ray }\gamma:[0,\infty)\to\hat{\mathcal{M}}\}/\sim
\end{equation}
with
\begin{equation}
    \gamma_1\sim\gamma_2\;\Longleftrightarrow\;\exists\;\epsilon>0\::\:\Delta\big(\gamma_1(t),\gamma_2(t)\big)<\epsilon\:\text{ as }\;t\to\infty.
\end{equation}
This allows to define the compactification of $\hat{\mathcal{M}}$ as
\begin{equation}
    {\rm cp}(\hat{\mathcal{M}})=\hat{\mathcal{M}}\cup\infty(\hat{\mathcal{M}})\,,
\end{equation}
which is compact with respect to the topology induced by the geodesic distance \eqref{eq. geo dist}. The above construction gets simplified through the conjectures of \cite{Raman:2024fcv} on the geometry of physical marked moduli spaces, proposing that \textbf{(1)} $\hat{\mathcal{M}}$ is contractible, and \textbf{(2')} for any $p,\,q\in \infty(\hat{\mathcal{M}})$, there is at most one geodesic (class) between them.\footnote{As explained in \cite{Raman:2024fcv}, a stronger property that implies the above \textbf{(1)} and \textbf{(2')} is that \textbf{(i)} there is at most one geodesic between any pair of points in $\hat{\mathcal{M}}$, and \textbf{(ii)} there is a unique shortest path between pairs of points in $\hat{\mathcal{M}}$, with such path being piecewise geodesic.} This for example removes the possibility of 1-cycles of positive minimal length.

For a given effective theory of quantum gravity with marked moduli space $\hat{\mathcal{M}}$, we denote $\{G^{(p)}\}_{p\geq 0}$ the $p$-form gauge symmetries,  see \cite{Delgado:2024skw}. Then, the duality group is defined as the group of connected components of $G^{(0)}$
\begin{equation}
    \Gamma=\pi_0( G^{(0)})\,,
\end{equation}
which prevents us from considering gauge redundancies on 1-form fields within our duality group, which due to charge quantization and instanton effects is expected to be discrete, and in general non-Abelian. Within the moduli space $\mathcal{M}=\Gamma\backslash\hat{\mathcal{M}}$, the distance is 
\begin{equation}
    \Delta ([p],[q])=\inf_{\alpha\in\Gamma}\{\Delta(p,\alpha\cdot q)\}\,,
\end{equation}
which in general affects the structure of the set of infinite distance points. As an example, in Type IIB, $\infty(\mathbb H)=\mathbb{R}\cup\{\infty\}$, while $\infty(SL(2,\mathbb{Z})\backslash\mathbb{H})=\{\infty\}$. We take now a slice $\Sigma\subseteq\hat{\mathcal{M}}$. Given $\bar{\Sigma}^{ {\rm cp}(\hat{\mathcal{M}})}$ its closure within $ {\rm cp}(\hat{\mathcal{M}})$, the associated infinite distance points
\begin{equation}
    \infty^{\hat{\mathcal{M}}}(\Sigma)=\bar{\Sigma}^{ {\rm cp}(\hat{\mathcal{M}})}\cap \infty (\hat{\mathcal{M}})
\end{equation}
correspond to the classes of asymptotic trajectories lying within $\Sigma$. We can also identify the class of transformations that act on $\Sigma$,\footnote{The use of $\mathcal{W}(\Sigma)$ for the reduced group of dualities acting on $\Sigma$ is done in analogy with the Weyl group that we will shortly describe for coset marked moduli spaces.}
\begin{equation}\label{eq. weyl group}
    \mathcal{W}_\Gamma(\Sigma)=\frac{N_\Gamma(\Sigma)}{Z_\Gamma(\Sigma)}=\frac{\{\alpha\in\Gamma:\alpha\cdot\Sigma=\Sigma\}}{\{\alpha\in\Gamma:\alpha\cdot s=s\;\forall s\in\Sigma\}}\,,
\end{equation}
which in general will correspond to a non-Abelian discrete group. Note that choices of $\Sigma$ for which $\alpha\cdot\Sigma\neq\Sigma$ for any $\alpha \in\Gamma-\{e\}$ results in $\mathcal{W}(\Sigma)$ being the trivial group. One can then take a fundamental domain
\begin{equation}
    \mathcal{W}(\Sigma)\backslash\Sigma
\end{equation}
and divide its asymptotic regions $\infty\big( \mathcal{W}(\Sigma)\backslash\Sigma\big)$ by a set of perturbative duality frames as those described in Section \ref{sec.Pert reg}. 

It is crucial to notice that the fixed loci of $\Sigma$ under $\mathcal{W}(\Sigma)$ are precisely the (finite distance) boundaries of $\mathcal{W}(\Sigma)\backslash\Sigma$. Functions defined over $\Sigma\subset\hat{\mathcal{M}}$ are also left invariant under the elements of $\mathcal{W}(\Sigma)$, in particular the mass of the towers becoming light towards asymptotic trajectories on $\Sigma$. This means that when moving in an asymptotic trajectory located along a fixed duality locus, the scaling vector $\vec\zeta$ \eqref{eq. zeta vector} of the leading light tower points along such locus, since orthogonal components are killed. Similarly, the boundaries of the fundamental domain $ \mathcal{W}(\Sigma)\backslash\Sigma$ will correspond with boundaries of the regime of validity of perturbative descriptions. Since the span of such duality frames takes value in a finite set of possibilities, the asymptotic regions $\infty\big( \mathcal{W}(\Sigma)\backslash\Sigma\big)$ of a given fundamental domain can only be divided through a finite set of possible perturbative descriptions.

\subsection{Symmetric spaces and maximal Euclidean flat slices\label{ssec.symmetric}}

In general, for an arbitrary marked moduli space $\hat{\mathcal{M}}$ there is not a unique maximal choice for $\Sigma$ that accounts for all non-compact directions of moduli space, since the properties (such as dimensionality) of the particular asymptotic region need not be universal and depend on the infinite distance degeneration being proben.\footnote{For example, for 4d $\mathcal{N}=1$ or 2 theories, the K\"ahler potential $K$, from where the hermitian metric $\mathsf{G}_{a\bar b}=\partial_a\partial_{\bar b} K$ is defined, takes the asymptotic form $K\sim -\log P(s)$, where $P(s)$ is a homogeneous polynomial on the saxions. Depending on the particular asymptotic regime, the number of saxions entering the leading term in $P(s)$ can change (the remaining saxions become subleading and are arbitrarily suppressed at infinite distance), thus affecting the effective dimensionality of the asymptotic region of moduli space, see \cite{Grimm:2018cpv,Corvilain:2018lgw,Grieco:2025bjy}.} However, in simpler cases the marked moduli space is globally given by a symmetric space (see \cite{Baines:2025upi,Baines:2026aug} for more properties of such spaces)
\begin{equation}
\hat{\mathcal{M}}=G/K\,,
\end{equation}
where $G=\mathbb{R}^p\times G_1\times\dots\times G_s$ is a \textit{reductive} Lie group with non-compact simple factors $\{G_a\}_{a=1}^s$ and $K=K_1\times\dots\times K_s$ is the maximally compact subgroup.\footnote{This structure for $\hat{\mathcal{M}}=G/K$ uniquely 
determines the moduli space metric to be
\begin{equation}
    \mathsf{G}_{ij}\dd\varphi^i\otimes\dd\varphi^j=\dd s^2_0+\sum_{a=1}^s\kappa_a\dd \tilde s_a^2\,,\quad\text{for some positive }\kappa_a>0\,,
\end{equation}
where $\dd s^2_0$ is the flat metric on $\mathbb{R}^p$ and $\dd \tilde s_a^2$ are Einstein metrics on $G_a/K_a$ with $\tilde{\mathcal{R}}_{ij}^{(a)}=-\frac12\tilde{\mathsf{g}}_{ij}^{(a)}$. 
} Then the Cartan decomposition of the Lie algebra of $G$ with respect to the maximally compact subgroup $K$ is
\begin{equation}
    \mathfrak{g}= \mathbb{R}^p\oplus\mathfrak{g}_1\oplus\dots\oplus\mathfrak{g}_s=\mathbb{R}^p\oplus(\mathfrak{k}_1\oplus\mathfrak{p}_1)\oplus \dots\oplus (\mathfrak{k}_s\oplus\mathfrak{p}_s) \, .
\end{equation}
Considering the maximal Abelian subalgebras $\mathfrak{a}_i\subset \mathfrak{p}_i$, and the basepoints $o_i=e_iK_i\in G_i/K_i$, we can define the \textit{standard maximal Euclidean flat} slice $F\subset G/K$ via the exponentiation
\begin{equation}
    F=\mathbb{R}^p\times\big(\exp(\mathfrak{a}_1)\cdot o_1\big)\times\dots\times\big(\exp(\mathfrak{a}_s)\cdot o_s\big)\simeq \mathbb{R}^{p+r_1+\dots+ r_s}\,,
\end{equation}
where $r_i={\rm rank}_{\mathbb{R}}(G_i/K_i)$. In this case one obtains
\begin{equation}
    \infty^{\hat{\mathcal{M}}}(F)\simeq \mathbb{S}^{p+r_1+\dots+ r_s-1}\,,
\end{equation}
with all infinite distance limits being unobstructed due to the symmetric character of $\hat{\mathcal{M}}$. If $G$ is a Chevalley group, see \cite{Steinberg2016LecturesOC}, as it is the case in many string realizations, we take $\Gamma=G(\mathbb{Z})$, then the inherited duality transformations \eqref{eq. weyl group} on the maximal Euclidean flat slice $F\subset G/K$ is given by \cite{gerasimov2022normalizers}
\begin{equation}\label{eq.weyl abs}
    \mathcal{W}_{G(\mathbb{Z})}(F)=W(G_1,K_1)\times\dots\times W(G_s,K_s)\,,
\end{equation}
where $W(G,K)=N_K(A)/Z_K(A)$ is the restricted Weyl group, with $A=\exp(\mathfrak{a})$. The residual group of symmetries $\mathcal{W}_{G(\mathbb{Z})}(F)$ is discrete and finite, and generally non-Abelian.
Note that the Abelian $\mathbb{R}^p$ part does not enter in the above expression, and thus is left invariant under any of the elements of $\mathcal{W}_{G(\mathbb{Z})}(F)$. This will prove quite useful when considering the half-maximally supersymmetric theories from Section \ref{sec. half max}.

With this in mind, we can compute the fundamental domain
\begin{equation}
   \mathcal{F}= \infty\big(\mathcal{W}_{G(\mathbb{Z})}(F)\backslash F\big)=\mathbb{S}^{p+r_1+\dots+r_s-1}/\big(W(G_1,K_1)\times\dots\times W(G_s,K_s)\big)\,,
\end{equation}
spanned by a simplicial cone, whose generating vectors correspond to fixed points of subgroups of $\mathcal{W}_{G(\mathbb{Z})}(F)$. We can then obtain the possible ways of filling these spherical sectors with the perturbative regimes described in Section \ref{sec.Pert reg}. Note that in order to do so, the following \textit{reconstruction} or ``\textit{bootstrapping}'' rules must be followed:
\begin{itemize}
    \item In the interface between two duality frames within $ \mathcal{F}$, the $\vec\zeta$ scaling vector of the tower of light states common to both perturbative descriptions is invariant and points perpendicular to the duality direction. This results in a common property between the two duality frames, such as the presence of $n_i$ large compact dimensions or a common light string. 
    \item The moduli space symmetries establish that some $\zeta$-vectors of leading light towers must point in the direction given by the generators of the simplicial cone spanning $\mathcal{F}$. 
    \item Similar as in the reconstruction procedures of \cite{Etheredge:2024tok,Grieco:2026yip}, the maximal resulting spacetime dimension after decompactification can be $D=10$ (in the presence of light strings, following the critical dimension of the superstring) or $D=11$ (in the absence of light strings, since in 11d SUGRA/M-theory there are no moduli that can describe an additional asymptotic direction).
    \item The span of perturbative description within the different 2-dimensional slices of $\mathcal{F}$ is quantized except for warped codimension-one decompactifications, see \eqref{eq.warping}, where the span explicitly depends on the coupling $\lambda$ of the warping scalar with respect the higher-dimensional potential $\hat{V}(\hat\varphi)\sim V_0 \exp(-\lambda\hat{\varphi})$. Thus, we will find ourselves in the event of requiring the use of perturbative description featuring warped codimension one decompactification, whose coupling $\lambda$ will be quantized by the angular span of $\mathcal{F}$, itself determined by the duality group.
\end{itemize}
In \cite{Baines:2025upi,Baines:2026aug}, an analysis of the asymptotic regions of symmetric moduli spaces and their light towers was carried out, taking a representation theoretical approach to systematically describe the allowed light towers for a given duality group acting on a symmetric space. However, there the focus was in ``\textit{perturbative}'' descriptions, working at fixed values of the $\mathbb{R}^p$ Abelian factors which are generally not affected by the $\Gamma$. As we will see in Section \ref{sec. half max}, such factor can correspond to the lower-dimensional dilaton, and thus, not moving in such direction allows one to remain in the weakly coupled string theory description. As it was explicitly found in \cite{Etheredge:2023odp}, in the strong coupling regions some of the general assumptions one would expect from the weak coupling regime, such as unwarped decompactifications, see \eqref{eq. angle reg} and \eqref{eq. univ scalings}, no longer hold, and thus the general analysis becomes more complicated. However, as we will see in the following sections, since for systems with enough supercharges the symmetric space description and its associated symmetries remain valid even in the strong coupling regimes, one can still obtain information about the arrangement of towers and duality frames.\\

Along the following sections, we will consider different (marked) moduli spaces and duality groups acting on them. We will see that the symmetries acting on non-compact scalars are enough to recover the different allowed duality frames that match those obtained from a full UV embedding of the theories.

\section{Maximal Supergravity: M-theory on \texorpdfstring{$\mathbb{T}^n$}{Tn}\label{sec. Udual}}
We start our examples with maximal supergravity, i.e., M-theory compactified on $\mathbb{T}^n$, or equivalently torus compactifications of Type II string theory, all preserving 32 supercharges. The symmetries of these spaces and their relation with the different duality frames and taxonomy rules were already discussed in \cite{Castellano:2023jjt,Etheredge:2024tok}, and thus in this section we will summarize their results and recast them in the language of this paper.

The moduli space for these theories is given by \cite{Cremmer:1978ds,Cremmer:1979up,Julia:1980gr,Obers:1998fb}
\begin{equation}\label{eq. mod space udual}
    \mathcal{M}_d=E_{n(n)}(\mathbb{Z})\big\backslash E_{n(n)}/K_n\,,\quad\text{with }\;d=11-n\,,
\end{equation}
where $E_{n(n)}$ is the split real form of $E_n$, with maximal compact subgroup $K_n$. The U-duality group corresponds to the integral lattice $\Gamma=E_{n(n)}(\mathbb{Z})$. Since ${\rm rank}_{\mathbb{R}}(E_{n(n)}/K_n)=n$, the standard maximal Euclidean flat slice is given by $F=\mathbb{R}^n$, parameterized by the individual radii of the $n$-torus. Since $E_{n(n)}$ is a Chevalley group, then $\mathcal{W}_{E_{n(n)}(\mathbb{Z})}(F)=W(E_{n(n)},K_n)$, dividing the $\infty^{E_{n(n)}/K_n}(F)\simeq\mathbb{S}^{n-1}$ sphere into $|W(E_{n(n)},K_n)|$ equivalent subsectors. The explicit form of the different groups involved is given in Table \ref{tab. U duality}. Following \cite{Obers:1998fb}, the U-duality groups can be expressed as
\begin{equation}\label{eq. knit product}
    E_{n(n)}(\mathbb{Z})=SL(n,\mathbb{Z})\bowtie SO(n-1,n-1,\mathbb{Z})\,,
\end{equation}
where $\bowtie$ is a generalization of the semidirect product denoting the group generated by two non-commuting subgroups. The first factor can be interpreted as the modular group of the $\mathbb{T}^n$ on which we compactify M-theory, while the second is associated with the T-dualities of Type IIA string theory upon double compactification on $\mathbb{T}^{n-1}$. Note that  the cardinality of $W(E_{n(n)},K_n)$ is always no greater than that of product of the individual Weyl groups of $SL(n,\mathbb{Z})$ and $SO(n-1,n-1,\mathbb{Z})$, respectively $|S_n|=n!$ and $|\mathbb{Z}_2^{n-2}\rtimes S_{n-1}|=2^{n-2}(n-1)!$, due to the non-trivial relations involved in the \eqref{eq. knit product} product.\\
\begin{table}[ht]
    \centering
    \begin{tabular}{|c|c|c|c|c|}\hline
       $d=11-n$  & $E_{n(n)}$ & $K_n$  & $W(E_{n(n)},K_n)$ & $|W(E_{n(n)},K_n)|$ \\\hline
       10  & $SO(1,1)\simeq\mathbb{R}^+$ & $\{1\}$  & $\{1\}$  & $1$ \\
        9 & $SL(2,\mathbb{R})\times\mathbb{R}^+$ & $SO(2)\simeq U(1)$  & $\mathbb{Z}_2$ & 2 \\
        8 & $SL(3,\mathbb{R})\times SL(2,\mathbb{R})$  &  $SO(3)\times SO(2)$& $S_3\times \mathbb{Z}_2$  & 12\\
        7 &$SL(5,\mathbb{R})$  & $SO(5)$ & $S_5$ &120 \\
        6 &$SO(5,5,\mathbb{R})$ & $SO(5)\times SO(5)$ & $\mathbb{Z}_2^4\rtimes S_5$ &1\,920 \\
        5 &$E_{6(6)}$  &  $Sp(4)/\mathbb{Z}_2$& $O^-(6,\mathbb{F}_2)$ & 51\,840\\
        4 &$E_{7(7)}$  & $SU(8)/\mathbb{Z}_2$ & $\mathbb{Z}_2\times Sp(6,\mathbb{F}_2)$ &290\,3040 \\
        3 &$E_{8(8)}$& $Spin(16)/\mathbb{Z}_2$ & $2\cdot O^+(8,\mathbb{F}_2)$ &696\,729\,600 \\\hline
    \end{tabular}
    \caption{Real split forms of the exceptional groups $E_{n(n)}$, together with their maximal compact subgroups and associated Weyl groups, as well as the cardinality of the latter. For $E_{6(6)}$, $E_{7(7)}$ and $E_{8(8)}$, $K_n$ is given by the quotient of the R-symmetry group by the appropriate central $\mathbb{Z}_2$ subgroup, since the $-I$ element acts trivially on the scalars (though not on the spinors). Their Weyl groups are described in \cite{conway1985atlas}, with $\mathbb{F}_2$ the field of 2 elements and $2\cdot O^+(8,\mathbb{F}_2)$ corresponding to the double cover of $O^+(8,\mathbb{F}_2)$.}
    \label{tab. U duality}
\end{table}

\begin{figure}[ht]
\begin{center}
\begin{subfigure}[b]{0.5\textwidth}
\centering
\captionsetup{width=.95\linewidth}
     \resizebox{0.95\textwidth}{!}{%
    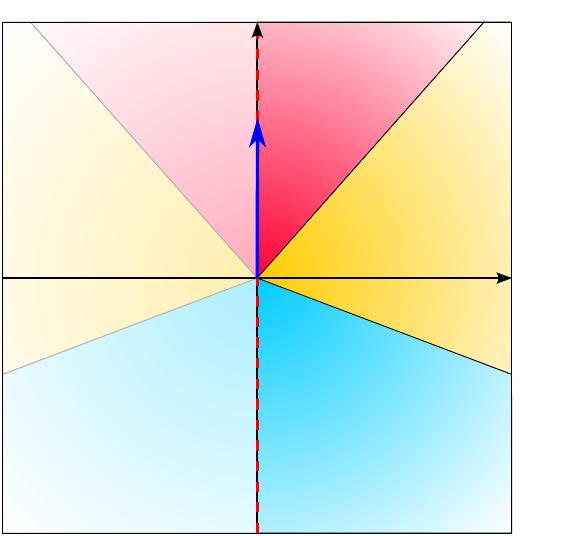
    }
\caption{Arrangement of perturbative descriptions for  maximal SUGRA in $d=9$ (M-theory on $\mathbb{T}^2$). We use the volume $V_2$ and $\tau_2$ of the torus as moduli. The self-dual line is depicted in red.} \label{ff.Mth2}
\end{subfigure}
\begin{subfigure}[b]{0.49\textwidth}
\captionsetup{width=.95\linewidth}
\centering
     \resizebox{\textwidth}{!}{%
    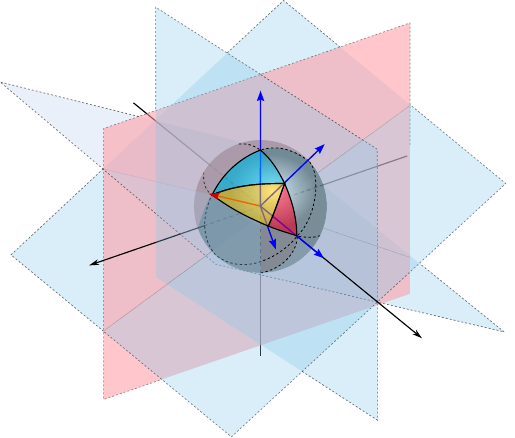
    }
\caption{Arrangement of perturbative descriptions for the fundamental domain of maximal SUGRA in $d=8$ (M-theory on $\mathbb{T}^3$). We use the total volume $V_3$, together with two orthogonal shape factors $\sigma^1$ and $\sigma^2$ as moduli. We depict the different fixed planes of the group $\mathbb{Z}_2\times S_3=\langle x\,|x^2\rangle\times\langle y,a\,|\,y^2,a^3, (ya)^2\rangle$.} \label{ff.Mth3}
\end{subfigure}
\caption{Arrangement of duality frames/perturbative descriptions for the maximal Euclidean flat slice of maximal SUGRA in $d=9$ and 8 (i.e., M-theory on $\mathbb{T}^2$ and $\mathbb{T}^3$). We depict the fundamental domain and the scaling $\zeta$-vectors of the associated light towers (in red for oscillator and blue for KK modes), as well as their image under the different discrete dualities acting on the non-compact moduli, respectively $\mathbb{Z}_2$ and $S_3\times \mathbb{Z}_2$. All quantities are depicted in terms of canonically normalized moduli.
\label{fig.Mth}}
\end{center}
\end{figure}

Now, as explained in more detail in \cite{Castellano:2023jjt,Etheredge:2024tok}, once the sector of the $\infty^{E_{n(n)}/K_n}(F)\simeq\mathbb{S}^{n-1}$ sphere corresponding to the fundamental domain of $W(E_{n(n)},K_n)$ has been computed, there is a \textit{unique} way of filling it with a series of perturbative regimes by following the rules described in Section \ref{ssec.symmetric}. We can illustrate this for the first case, where the low dimensionality of the maximal Euclidean flat slice allows for a pictorial representation:
\begin{itemize}
    \item $\mathbf{n=1}$: Here $F=\mathbb{R}$ and $\infty^{E_{1(1)}/K_1}(F)\simeq\{\pm\infty\}$, with $W(E_{1(1)}/K_1)$ the trivial group. We can thus not identify the two infinite distance limits and their asymptotic description. It is then natural to have them correspond to a decompactification limit to 11d and an emergent string limit, precisely as it is the case of M-theory on $\mathbb{S}^1\,\equiv$ 10d Type IIA string theory. By contrast, if we had considered Type IIB string theory, with
    \begin{equation}
        \mathcal{M}_{\rm IIB}=SL(2,\mathbb{Z})\big\backslash \mathbb{H}= SL(2,\mathbb{Z})\big\backslash SL(2,\mathbb{R})/SO(2)\,,
    \end{equation}
    the Weyl group is precisely $\mathbb{Z}_2$, identifying the two limits at strong and weak coupling of the dilaton, associated both with emergent string limits of Type IIB F1 and D1-strings, interchanged under the S-duality $\Phi\to-\Phi$.
    \item $\mathbf{n=2}$: Now $F=\mathbb{R}^2$ and $W(E_{2(2)}/K_2)\simeq\mathbb{Z}_2$, so that the fundamental domain is the half 2-plane.
    When trying to fill such asymptotic region with perturbative descriptions, the only possible arrangement consists in two T-dual string theory descriptions with a single compact dimension, together with an additional Planckian regime with a compact space $X_2=Y_1\times Y_1'$ with a $\mathbb{Z}_2$ symmetry that exchanges both 1-cycles. As depicted in Figure \ref{ff.Mth2}, this has the straightforward interpretation of the different frames of Type IIB on $\mathbb{S}^1$ or M-theory on $\mathbb{T}^2=\mathbb{S}^1\times\mathbb{S}^1$ (equivalently Type IIA on $\mathbb{S}^1$), where the $\mathbb{Z}_2$ symmetry is understood as the S-duality transformation in the Type IIB frame, or the radii interchange in the M-theory description (this, sending the complex structure $\tau=\tau_1+i\tau_2 \to -\frac{1}{\tau}$). 
    \item $\mathbf{n=3}$: Similarly, in this case $F=\mathbb{R}^3$ and $W(E_{3(3)}/K_3)\simeq S_3\times\mathbb{Z}_2$, which divides the asymptotic $\infty^{E_{3(3)}/K_3}(F)\simeq\mathbb{S}^{2}$ in $3!\cdot2=12$ equivalent fundamental domains. As seen in Figure \ref{ff.Mth3}, such region is filled in a unique way by two equivalent stringy regimes compactified on $X_2=Y_1\times Y_1'$ (again with a $\mathbb{Z}_2$ symmetry interchanging the two cycles), which are related by T-duality, and a third description corresponding to a Planckian regime with compact space $X_3=Y_1\times Y_1'\times Y_1''$, with the $S_3$-symmetry interchanging the 1-cycles. Of course, the UV interpretation of said frames are Type IIA and IIB string theories on $\mathbb{T}^2$, as well as M-theory on $\mathbb{T}^3$.
\end{itemize}

Similar constructions can be continued as we lower the dimensionality of the EFT, with symmetries having the interpretation of S-dualities/T-dualities and interchange of internal radii, following the interpretation of Type IIA/IIB or M-theory toroidal compactifications.

It is interesting to notice that at no point warped codimension-one backgrounds are needed in order to fill the asymptotic regions with perturbative descriptions. With the hindsight that these theories with 32 supercharges are obtained by toroidal compactifications of M-theory, this should not be particularly surprising, since there are no light scalars in 11d supergravity that could produce a space-varying vacuum energy resulting in the warping of the compact dimension.\\

What is more surprising is the fact that, as shown in \cite[Appendix A.3]{Etheredge:2024tok}, simply considering the flat slice $F=\mathbb{R}^n$ (with $n=11-d$), if one tries to fill the asymptotic regions $\infty(\mathbb{R}^n)\simeq\mathbb{S}^{n-1}$ with the perturbative regimes from Section \ref{sec.Pert reg} (and for the sake of simplicity, not considering warped decompactifications), there is a \emph{unique} way of filling the sphere.\footnote{For $n=1$, i.e., $d=10$, since $\mathbb{S}^0\simeq\{\pm\infty\}$ has two disconnected components, there are actually \textit{three} possible non-equivalent arrangements, corresponding to combinations of single decompactification directions and emergent string limits. Two of them correspond to Type IIA and IIB string theory, while the third one (featuring two decompactification limits) is not featured in the known string landscape.} Even more, the symmetry group of said arrangement precisely corresponds with $W(E_{n(n)},K_n)$. A possible explanation for this might be in the fact that having a $F=\mathbb{R}^{11-d}$ flat slice (for example parameterized by the $d$-dimensional dilaton and the different radions) without warped compactifications is only possible in theories with 32 supercharges. Since the different discrete U-duality groups are uniquely defined from charge quantization and the classical symmetries of the supergravity action, see \cite{Hull:1994ys,Ferrara:1997ci}, it is then not surprising that the inherited discrete symmetries acting on non-compact scalars are the only allowed ones.

\section{Half-maximally supersymmetric theories: Heterotic string theory on \texorpdfstring{$\mathbb{T}^n$}{Tn}\label{sec. half max}}

We now turn our attention to moduli spaces parameterizing the vacua of supersymmetric theories with 16 supercharges, particularly arising from toroidal compactifications of heterotic string theory.\footnote{The same theory, and moduli space, is recovered from compactifications of Type I string theory (a Type IIB 10d orientifold) on tori. However, we will stick to the heterotic description, since the interpretation of the symmetries and their fixed loci is more evident there.} Compactifying on an $n$-torus results in the ($d=10-n$)-dimensional theory having the following moduli space:
\begin{equation}\label{eq.moduli space het}
\mathcal{M}_n = O(n,16+n;\Z) \, \big\backslash \,
\frac{ O(n,16+n;\R) }{ O(n;\R) \times O(16+n;\R) }
\times \R \, ,\quad \text{for}\;n=10-d\in\{1,\dots, 5\}\,,
\end{equation}
where the $\mathbb{R}$ factor parameterizes the $d$-dimensional dilaton $\Phi_d$, see \eqref{eq. lower dim dilaton}. For lower dimensionalities ($d=4$ and $3$) the above structure is no longer valid, due to the additional scalars resulting from the dualization of the Kalb-Ramond 2-form and the gauge vectors. We will describe this in Section \ref{sec. het d4 and d3}. An immediate difference with respect to the maximal supergravity moduli space \eqref{eq. mod space udual} is precisely this additional Abelian factor $\mathbb{R}$, which is left invariant under the duality group $O(n,16+n;\mathbb{Z})$, a fact that will be important on the study of dualities at strong coupling. The dimension of the moduli space is 
\begin{equation}\label{eq. dim mod het}
    \dim\mathcal{M}_n=\overbrace{\underbrace{1}_{\Phi_d}+\underbrace{n}_{\text{radii}}}^{\text{non-compact}}+\overbrace{\underbrace{\frac{n(n-1)}{2}}_{\text{torus angles}}+\underbrace{\frac{n(n-1)}{2}}_{B_{ij} \text{ on 2-cycles}}+\underbrace{16n}_{\text{Wilson lines}}}^{\text{compact}}=n(n+16)+1\,,
\end{equation}
with the maximal Euclidean flat slice given by $F=\mathbb{R}^{n+1}=\mathbb{R}\times\mathbb{R}^{n}$, parameterized by the $d$-dimensional heterotic dilaton and the radii of $\mathbb{T}^n$ in string units. Through the construction outlined around \eqref{eq.weyl abs}, using $K_n=O(n)\times O(16+n)$, one obtains
\begin{equation}
    \mathcal{W}_{O(n,16+n;\Z)}(F)=\mathbb{Z}_2^n\rtimes S_n\,,\quad \text{with}\quad|\mathcal{W}_{O(n,16+n;\Z)}(F)|=2^n\cdot n!\,,
\end{equation}
corresponding to the T-dualities along the different radii and their interchange within the \mbox{$n$-torus}. The $\mathbb{R}$ factor parameterizing the $d$-dilaton $\Phi_d$ is left invariant by any of the $O(n,16+n;\Z)$ transformations (and particularly those of $\mathcal{W}_{O(n,16+n;\Z)}(F)=\mathbb{Z}_2^n\rtimes S_n$), and thus is usually dubbed as \textbf{self-dual line}, corresponding to the intersection of all the hyper planes fixed by the individual $\mathbb{Z}_2^n\rtimes S_n$ transformations. It is associated with a gauge enhancement along it,\footnote{If the rest of Wilson line and $B_{ij}$ components are set to 0, the enhancement is $U(1)^n\to SU(2)^n$, but this can be generalized to simply laced Lie groups for particular choices of the additional compact moduli, see \cite{Narain:1986am,PhysRevD.35.648,Aspinwall:1995zi,Font:2020rsk}.} which in the weak coupling heterotic description $\Phi_d\to-\infty$ corresponds to the $n$ internal radii being string sized: 
$R_i=\sqrt{\alpha'}$. 
As described in Section \ref{sec.sym}, the scaling vector of two light towers must point along the two directions of such self-dual line, $\Phi_d\to-\infty$ and $+\infty$. The former corresponds to the oscillator modes of the heterotic string $\vec\zeta_{\rm osc}$, while the latter will depend on $n$. Since the compactification of both $SO(32)$ and $E_8\times E_8$ heterotic string theories results in the same moduli space (as one can change the Wilson lines to interpolate from one theory to another), the tower in the strong coupling limit should be the same, since the self-dual line is left invariant under such ``rotations''. 
Therefore, the self-dual line is common for both theories, while the intermediate perturbative descriptions between the weakly coupled heterotic string theory on $\mathbb{T}^n$ and the final description at $\Phi_d\to+\infty$ might differ.\\

Here is where we are first able to go beyond the previous results from \cite{Baines:2025upi,Baines:2026aug}, which focused in the dualities that stayed within the perturbative heterotic description. From our bottom-up approach we will be able to obtain information about the consistent completions of moduli space at strong heterotic coupling. To our surprise, only two possibilities are allowed, respectively corresponding to the different duality frames of $SO(32)$ and $E_8\times E_8$ heterotic string theories on $\mathbb{T}^n$. Our analysis will stay within the maximal Euclidean flat slice $F=\mathbb{R}^{n+1}$, where the different compact moduli are set to zero, and thus will not consider the effect of axion or Wilson line variations. While this is a significant caveat that must be acknowledged, our results will still strongly constrain the possible duality descriptions.

\subsection{A reminder: the heterotic string on \texorpdfstring{$\S^1$}{S1}\label{sec.het S1}}

We start considering the simplest case, consisting in heterotic string theory on a circle, for which
\begin{equation}
    F=\mathbb{R}^2\,,\quad \mathcal{W}_{O(1,17;\mathbb{Z)}}(F)=\mathbb{Z}_2\,,
\end{equation}
parameterized by the 9d dilaton $\hat{\Phi}_9$ and the radion $\hat\varrho$ associated to the radius of the circle in string units, respectively describing the directions parallel and perpendicular to the self-dual line. The angle subtended by $\mathbb{R}/\mathbb{Z}_2$ is $\pi$, of which $\theta_1=\arccos\frac{1}{\sqrt{8}}$ correspond to the weakly coupled heterotic string on $\mathbb{S}^1$. Since at most we can decompactify to 10 or 11 dimensions, the possible angles associated to the other perturbative descriptions can be
\begin{equation}\begin{array}{cc}
     \theta_1=\arccos\frac{1}{\sqrt{8}}\,,\quad&\theta_{1,1}=2\,\theta_{1\gg 1}=\arccos\frac{1}{8}\,,\\\theta_1^{(\lambda)}=\arctan\frac{\sqrt{14} \lambda -\sqrt{7}}{\sqrt{2} \lambda +7}\,,\quad&
      \theta_{1 \leftrightarrow 1}^{(\lambda)}=\arctan\frac{2\sqrt{7}}{\lambda\sqrt{8}} \,.
\end{array}
\end{equation}
A straightforward check shows that it is not possible to add the above angles up to $\pi$ while doing a consistent gluing of the duality frames if only unwarped regimes are used.\footnote{Note that for the 9d maximal supergravity, where the symmetry group over the flat $\mathbb{R}^2$ slice was also $\mathbb{Z}_2$, we were able to fill the fundamental domain with $\theta_{1\gg1}+\theta_1+\theta_1$, with an emergent string limit between the two weakly coupled string descriptions, each with span $\theta_1$. However, this is not possible in the case we study now, since the emergent string limit is located along the self-dual line.} However, if one allows for warped backgrounds, there are two unique possible arrangements:
\begin{subequations}
    \begin{align}
        \textbf{Arrangement O}&:\theta_1+\theta_1+\theta_1^{(\frac{5}{\sqrt{2}})}=\arccos\frac{1}{\sqrt{8}}+\arccos\frac{1}{\sqrt{8}}+\arccos{\frac{3}{4}}=\pi\\
        \textbf{Arrangement E}&:\theta_1+\theta_{1,1}+\theta_{1\leftrightarrow1}^{(\frac{5}{\sqrt{2}})}=\arccos\frac{1}{\sqrt{8}}+\arccos\frac{1}{8}+\arctan\frac{\sqrt{7}}{5}=\pi
    \end{align}
\end{subequations}

\begin{figure}[ht]
\begin{center}
\begin{subfigure}[b]{0.5\textwidth}
\centering
     \resizebox{0.95\textwidth}{!}{%
    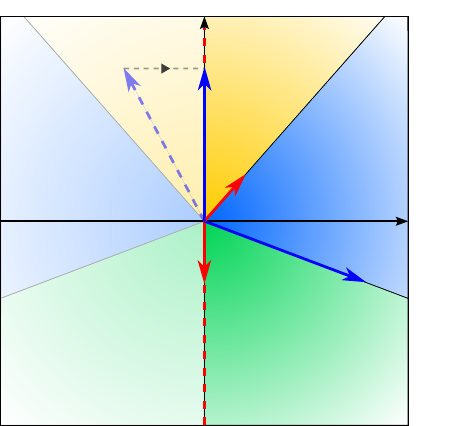
    }
\caption{Arrangement \textbf{O} in $d=9$.} \label{ff.HO1}
\end{subfigure}
\begin{subfigure}[b]{0.49\textwidth}
\centering
     \resizebox{\textwidth}{!}{%
    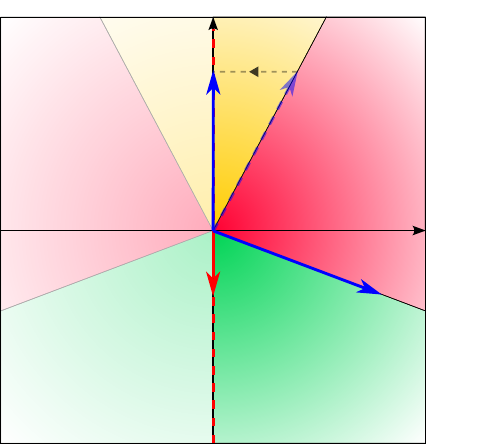
    }
\caption{Arrangement \textbf{E} in $d=9$.} \label{ff.HE1}
\end{subfigure}
\caption{Only two possible arrangements for a two-dimensional Euclidean flat slice in $d=9$ with a $\mathbb{Z}_2$ symmetry and weakly coupled description with a compact dimension. We depict the matching with the two heterotic $\mathbb{S}^1$ arrangements of duality frames. We depict the self-dual line $\hat\varrho=0$ in red, together with the scaling $\zeta$-vectors (in red for oscillator and blue for KK towers), including the sliding of $\vec\zeta_{\rm KK,I'}$ according to \eqref{eq.sliding}. All quantities are depicted in terms of canonically normalized moduli.
\label{fig.Het1}}
\end{center}
\end{figure}

Arrangement \textbf{O} corresponds to a \textbf{(O.1)} the original weakly coupled string theory on a $X_1$ manifold, \textbf{(O.2)} a S-dual string theory description in the same $X_1$ manifold, and \textbf{(O.3)} a weakly coupled string theory description on a warped $X_1'$ manifold. On the other hand, for arrangement \textbf{E}, after the original \textbf{(E.1)} weakly coupled string theory on $Y_1$, we have a strongly coupled description with compact space $Y_1\times Y_1'$ \textbf{(E.2)}, followed by a strongly coupled description \textbf{(E.3)} with a warped compact space $Y_1'$. This is pictured in Figure \ref{fig.Het1}. Importantly, notice that the numerical coupling in the higher-dimensional potential $\hat{V}(\hat\varphi)\sim V_0 \exp(\lambda\hat{\varphi})$ causing the warping of the compact space is the same for both cases, $\lambda=\frac{5}{\sqrt{2}}$. This is self-consistent with the fact that the self-dual line $\R$ is unique along moduli space, while there can be different maximal Euclidean flat slices differing in the constant values of the compact scalars, having the same inherited elements from $O(1,17;\mathbb{Z})$, but different duality descriptions. In any case, our results make compatible the same description along the strong coupling direction of the self-dual line. We can state, in a clearer way, that our findings imply
\begin{mdframed}\centering
     Given a (slice of a) 9d moduli space with form $F=\mathbb{R}\times\mathbb{Z}_2\backslash \mathbb{R}$, if an asymptotic direction in the first $\mathbb{R}$ factor is an emergent string limit, then $F$ admits two non-analogous arrangements of duality frames. In both cases the strong coupling limit in the $\mathbb{R}$ factor involves a warped decompactification to a 10d running solution with potential $\hat{V}(\hat\varphi)\sim V_0 \exp(\frac{5}{\sqrt{2}}\hat{\varphi})$, with $\hat{\varphi}$ a 10d scalar.
\end{mdframed}

The top-down embedding of the above perturbative descriptions is well known, as the arrangement of dualities in the 9d dilaton-radion moduli space has been studied in detail in the literature \cite{Polchinski:1995df,Aharony:2007du,Etheredge:2023odp}. These are different depending on the gauge group of the heterotic theory considered:
\begin{itemize}
    \item{$SO(32)$}: Starting from the weakly coupled HO string theory on a large $\mathbb{S}^1$, the strong-coupling regime is better described via S-duality by Type I string theory (a Type IIB orientifold) on the same circle. The small radius limit of Type I is then better studied by T-dualizing to Type I$'$, corresponding to a Type IIA orientifold on a $\mathbb{S}^1/\mathbb{Z}_2$ interval with an O8$^-$-plane on each endpoint, together with a total of 32 D8-branes required to cancel the RR tadpole.\footnote{
In fact, it is sufficient to specify the positions of only half of the D8-branes, since the locations of the remaining ones are fixed by the orientifold action.
} The $SO(32)$ gauge group is realized by 16 D8-branes placed on top of one of the O8$^-$-planes, see \cite{Polchinski:1995df}. This corresponds precisely with the arrangement \textbf{O} described above.
    \item{$E_8\times E_8$}: Similarly, the strong-coupling limit of HE string theory on $\mathbb{S}^1$ is given by M-theory on $\mathbb{S}^1\times \mathbb{S}^1/\mathbb{Z}_2$, where the second factor corresponds to the Ho\v{r}ava-Witten interval \cite{Horava:1995qa,Horava:1996ma}. The limit in which the original circle becomes small is then better described by Type IIA on $\mathbb{S}^1/\mathbb{Z}_2$, precisely again a Type I$'$ construction. The two  $E_8$ gauge factors are non-perturbatively realized by placing seven 
D8-branes on top of the O8$^-$-plane at each boundary, with the remaining two D8-branes symmetrically located in the precise points of the interval bulk so that the 10d dilaton diverges precisely at the orientifold planes, see e.g. \cite{Aharony:2007du}.\footnote{
In general, each D8-brane contributes a $U(1)$ gauge factor, while a stack of $n$ D8-branes coincident with an O8$^-$-plane enhances the gauge symmetry to $SO(2n)$. If the dilaton diverges at the orientifold plane, the symmetry is further enhanced to the corresponding exceptional group $E_{n+1}$ due to the presence of massless D0-branes. 
} This is precisely the previous arrangement \textbf{E}.
\end{itemize}
    In both cases, the presence of D8-branes along the interval of the Type IIA orientifold/Type I$'$ description results in a non-zero Romans mass $m_0$ over the compact direction. In the 10d Einstein frame, this results in a $m_0\,e^{-\frac{5}{\sqrt{2}}\hat{\Phi}_{10}}$ term in the action, with the numerical coupling to the (canonically normalized) 10d dilaton $\hat{\Phi}_{10}$ \eqref{eq. can norm} being $\lambda=\frac{5}{\sqrt{2}}$, precisely the exponent predicted from the point of view of perturbative descriptions consistent with the symmetries of the moduli space. In both cases sending the 9d heterotic dilaton to strong coupling while keeping the internal space fixed in string units results in a decompactification to 10d massive Type IIA string theory with non-vanishing Romans mass, totally in line with the predictions from the symmetry group.\\

    Before finishing this section, let us discuss how one of the tenets outlined in Section \ref{sec.sym} holds non-trivially in this setting, namely the expectations that scaling $\zeta$-vectors point along the fixed loci of the duality group spanning the fundamental domain. Parameterizing $F\simeq\mathbb{R}^2$ by the canonically normalized $(\hat{\Phi}_9,\hat{\varrho})$, said directions are $\hat{\Phi}_9\to\pm\infty$. Common to the $SO(32)$ and $E_8\times E_8$ heterotic theories, along said slice of moduli space one finds the oscillator modes of the BPS heterotic string, as well as the heterotic KK and winding modes (also BPS):\footnote{In writing the coordinates of the $\zeta$-vectors we identify the flat $F\simeq\mathbb{R}^2$ with its tangent space. See \cite{Etheredge:2023odp} for more details on the compactification.}
    \begin{equation}
        \vec\zeta_{\rm osc,h}=\left(-\tfrac{1}{\sqrt{7}},0\right)\,,\quad \vec\zeta_{\rm KK,h}=\left(-\tfrac{1}{\sqrt{7}},1\right)\,,\quad \vec\zeta_{\rm w,h}=\left(-\tfrac{1}{\sqrt{7}},-1\right)\,.
    \end{equation}
    It is clear that under T-duality: 
    $\hat{\varrho}\to-\hat\varrho$, so 
    $\vec\zeta_{\rm osc,h}\to \vec\zeta_{\rm osc,h}$ and $\vec\zeta_{\rm KK,h}\leftrightarrow\vec\zeta_{\rm w,h}$. For the perturbative descriptions charting the regions of moduli space at strong heterotic coupling, the additional light towers (the Type I oscillator modes in $SO(32)$ and the M-theory interval KK modes in $E_8\times E_8$) are non-BPS and non-invariant under the $\mathbb{Z}_2$ symmetry. For the final Type I$'$/Type IIA orientifold description the expression of the KK modes associated to the growing warped interval is quite involved (see \cite{Etheredge:2023odp} for a detailed derivation), with
    \begin{equation}\label{eq.sliding}
    \vec\zeta_{{\rm KK,I}'}^{\, {\rm (HO)}}=\Big(\tfrac{5}{2\sqrt{7}},-\tfrac{3}{2}\Big[1+\tfrac{2}{\sqrt{1-e^{-4\hat{\varrho}}}}\Big]^{-1}\Big)\,,\qquad\vec\zeta_{{\rm KK,I}'}^{\, {\rm (HE)}}=\Big(\tfrac{5}{2\sqrt{7}},\tfrac{1}{2}-\tfrac{2}{1+3e^{2\hat{\varrho}}}\Big)\,.
    \end{equation}
   Note that in both cases the expression of the scaling vector changes depending on the value of the moduli. However, for the fixed locus $\hat{\varrho}=0$ line, the $\hat{\varrho}$ component vanishes and $\vec\zeta_{{\rm KK,I}'}^{\,{\rm (HO)}}=\vec\zeta_{{\rm KK,I}'}^{\,{\rm (HE)}}=\big(\frac{5}{2\sqrt{7}},0\big)$ precisely pointing along the self-dual line, and with the magnitude \eqref{eq. warp scaling} for $d=9$ and $\lambda=\frac{5}{\sqrt{2}}$. The explicit expressions \eqref{eq.sliding} cannot be predicted directly from the duality symmetries, but as we have seen, their value along the fixed locus is set.

\subsection{An illustrative example: heterotic string on \texorpdfstring{$\T^2$}{T2}\label{sec. het t2}}

Before systematically studying particular slices of the moduli spaces \eqref{eq.moduli space het} in lower dimensions, we will study in more detail the $d=8$ case (this is, the heterotic string compactified on $\mathbb{T}^2 \simeq \S^1_x \times \S^1_y$), since it is the last example where the dimensionality of a piece of paper allows for a clear visualization of the full set of asymptotic directions within the maximal Euclidean flat slice. As discussed above, we have
$F = \R^3$, parametrized by the 8d dilaton $\hat\Phi_8$ and two radions $\hat\varrho_x$ and $\hat\varrho_y$. The flat slice inherits the discrete symmetry group
\begin{equation}
    \mathcal{W}_{O(2,18;\Z)}(F) = \mathbb{Z}_2^2\rtimes S_2 \cong D_4 \, \,,
\end{equation}
precisely the group of symmetries of the square. This gives rise to $\abs{D_4} = 8$ equivalent fundamental domains, related by T-dualities and exchange of radii in the heterotic on $\mathbb{T}^2$ frame, see Figure \ref{fig. slice 3d T2}. 

Starting from the heterotic frames, we will now cover the remaining asymptotic directions as we move towards strong $\Phi_8$ coupling. We will later compute the two-dimensional slices fixed under $\mathcal{W}_{O(2,18;\Z)}(F)\simeq D_4$, and compare them with those obtained from our reconstruction rules. Here we only collect the relevant duality frames and the geometry of each regime, while the explicit computation of the towers of states is carried out in detail in Appendix~\ref{app. Details Het T2}.

\begin{figure}[ht]
    \centering
    \resizebox{\textwidth}{!}{%
    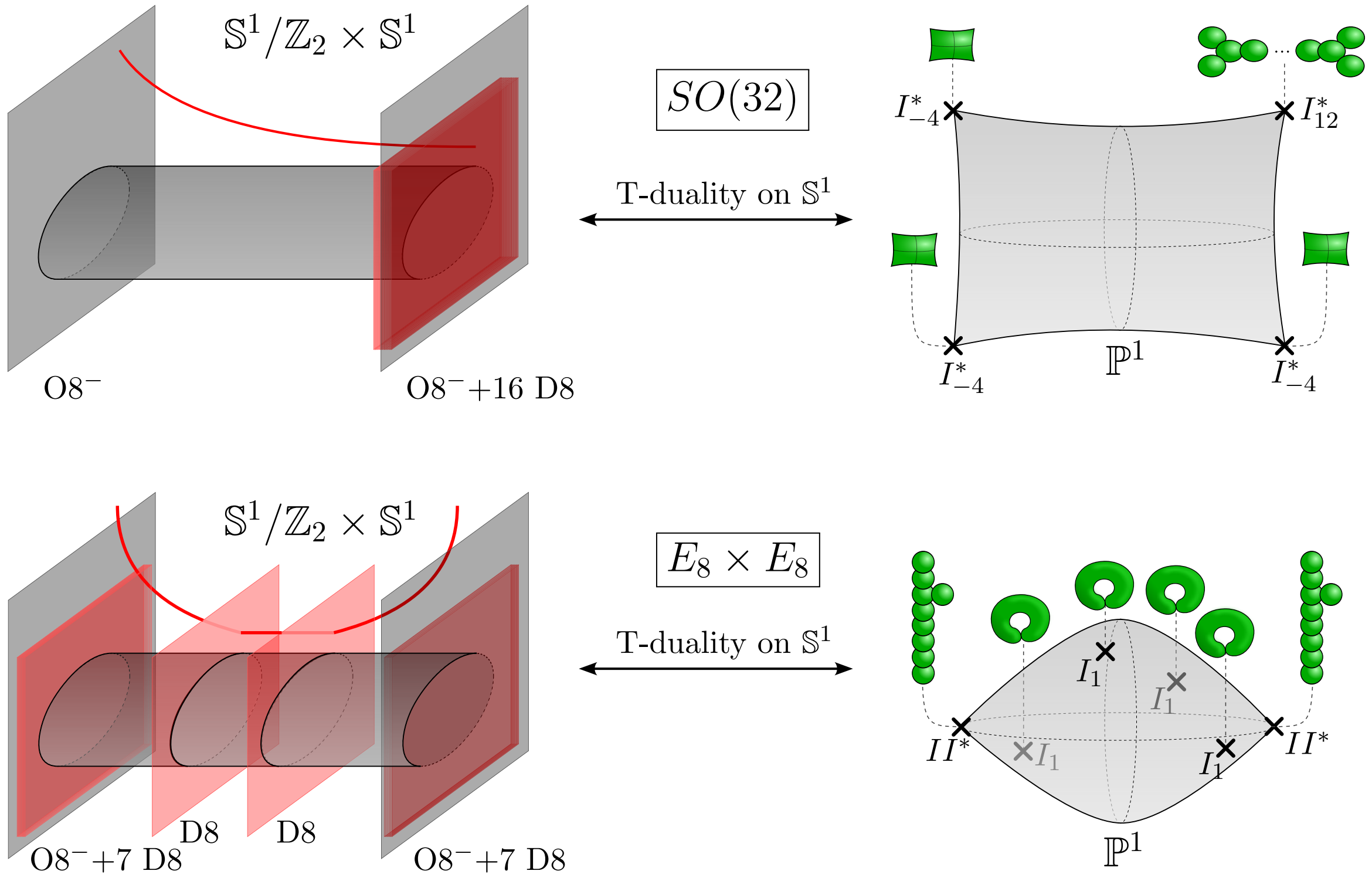
    }
    \caption{
    Geometrical arrangements in the Type IIA (left) and F-theory (right) regions. In the former we depict the D8 brane arrangement, together with the 10d dilaton $\Phi_{10}$ profile (in red), while for the latter we sketch the arrangement of singularities of the elliptic fiber, see e.g, \cite{Weigand:2018rez} and Appendix \ref{app. F-theory}.    
    \label{fig. geometries}}
\end{figure}

\begin{figure}[ht]
    \centering
     \resizebox{\textwidth}{!}{%
    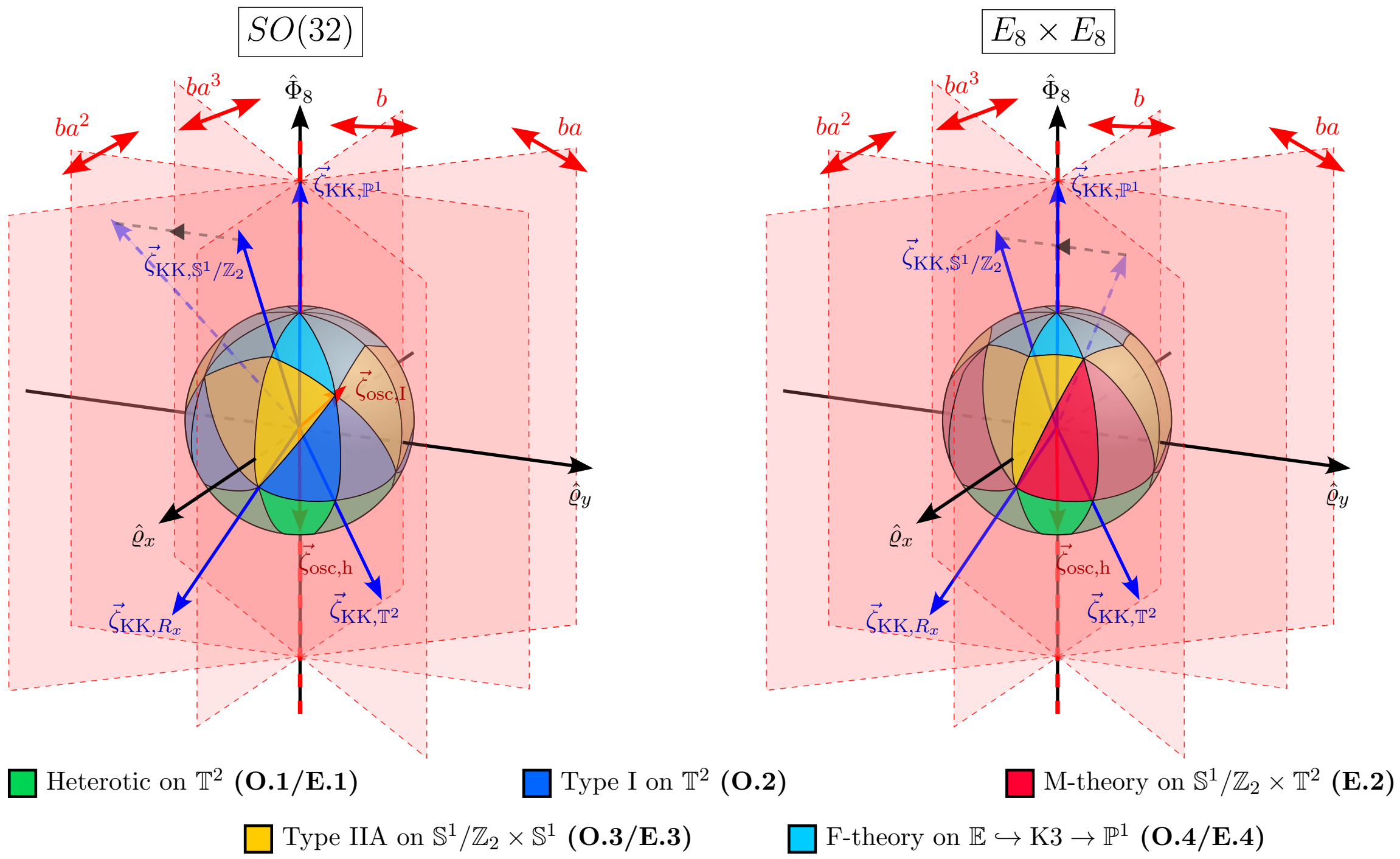
    }
    \caption{Arrangement of duality frames in the flat slice of moduli space of heterotic string theory on $\mathbb{T}^2$ at zero axions and Wilson lines, parameterized by the 8d dilaton $\Phi_8$ and the radions $\{\varrho_x,\varrho_y\}$ in string units, which we depict canonically normalized. In red we depict the fixed planes by different elements of the symmetry group $D_4=\langle a,b\,|\,a^4,b^2,(ab)^2\rangle$. The self-dual line, parameterized by $\Phi_8$, is depicted in red dashed lines. Along the fundamental domain we depict the scaling $\zeta$-vectors of the different light towers (in red for oscillator and blue for KK modes), as well as the sliding of the $\vec\zeta_{{\rm KK},\mathbb{S}^1/\mathbb{Z}_2}$. Note that in the plane fixed by $b$, $\vec\zeta_{{\rm KK},\mathbb{S}^1/\mathbb{Z}_2}$ does not lie along the interface between the Type IIA and F-theory descriptions. 
    }
    \label{fig. slice 3d T2}%
\end{figure}

\begin{itemize}
    \item $SO(32)$: Starting with the HO theory on $\mathbb{T}^2$ \textbf{(O.1)}, its strong coupling regime is given by the S-dual Type I string theory on the same $\mathbb{T}^2$ \textbf{(O.2)}. T-dualizing along one of the $\mathbb{T}^2$ circles results in Type I$'$ string theory on the remaining $\mathbb{S}^1$, this is, (massive) Type IIA string theory on $\mathbb{S}^1/\mathbb{Z}_2\times\mathbb{S}^1$ \textbf{(O.3)}. The brane and O-plane arrangement is inherited from the 9d case described in Section \ref{sec.het S1}, corresponding to two O8$^-$-planes wrapped along the two $\mathbb{S}^1$ boundaries, with a stack of 16 wrapped D8-branes placed on top of one of the wrapped O8$^-$-planes. Up to this point, the duality chain is simply the one for HO on $\mathbb{S}^1$, followed by an additional compactification on a second circle. The new feature in the two-dimensional case is that a second T-duality can now be performed along the remaining circle, mapping the theory to Type IIB compactified on $\T^2/\Z_2$, the so-called \textit{pillowcase manifold} \cite{Thurston1979TheGA, eskin2005pillowcasesquasimodularforms}, see Figure \ref{fig. geometries}. 
    This space is homeomorphic to $\P^1\simeq \mathbb{S}^2$ and has four fixed points, each corresponding to an O7$^-$-plane, with the RR tadpole canceled by 16 D7-branes (along with their images), which are again taken to coincide with one of the orientifold planes in order to preserve the $SO(32)$ gauge group. From the F-theory perspective, this corresponds to a compactification on an elliptically fibered $\mathbb{E}\hookrightarrow{\rm K3}\to \mathbb{P}^1$ \textbf{(O.4)}. As we discuss in more detail in Appendix \ref{app. F-theory}, the Type IIB orientifold compactification corresponds to the perturbative Sen limit of F-theory on K3 \cite{Sen:1997gv,Vafa:1996xn}, thus allowing for an emergent string limit. 
    \item $\EE$: In this case, starting with HE theory on $\mathbb{T}^2$ \textbf{(E.1)}, the strong-coupling limit is described by M-theory compactified on $\S^1/\Z_2 \times \T^2$ \textbf{(E.2)}. Moreover, when one of the radii of the $\T^2$ factor becomes small, the regime is better described by Type IIA on $\S^1/\Z_2 \times \S^1$ \textbf{(E.3)}, with the same D8 and O8$^-$-brane arrangement as in the HE on $\mathbb{S}^1$ case from Section \ref{sec.het S1}, now simply wrapped on the additional circle. Performing a T-duality along this remaining circle leads again to Type IIB on the pillowcase, better described by F-theory on an elliptically fibered K3 \textbf{(E.4)}. Unlike in the $SO(32)$ case, this background does not admit a weakly coupled orientifold description, as the two needed $II^*$ singularities cannot be built by simply stacking D7-branes and O7$^-$-planes, see Appendix \ref{app. F-theory} for more details. Thus, in this case there is no emergent string limit in this slice of moduli space. 
\end{itemize}

The two arrangements of duality frames/perturbative regimes along the maximal Euclidean flat slice, together with the scaling $\zeta$-vectors associated to the light towers, are depicted in Figure \ref{fig. slice 3d T2}. The different planes fixed by the symmetry group $D_4=\langle a,b\,|\,a^4,b^2,(ab)^2\rangle$ are equivalent for the two theories and are presented in Figures \ref{fig.Het2} and \ref{fig.slice1t2}. We give more detail on the derivation of the different perturbative regimes and light towers in Appendix \ref{app. Details Het T2}.

The above duality chains are already known in the literature \cite{Witten:1995ex,Aspinwall:1996mn,Horava:1995qa}, and indeed fill the fundamental region of the maximal Euclidean flat slice. However, as we will discuss now, we can see how said two arrangements are the \emph{unique} ones allowed in $d=8$ consistent with the symmetries and the weakly coupled description at $\hat\Phi_8\to-\infty$. In order to show this, we will study the two-dimensional slices that are left invariant by elements of $D_4$, and apply on them the reconstruction rules and arguments of Sections \ref{sec.Pert reg} and \ref{sec.sym}.

\subsubsection*{Dilaton--Overall volume slice of Heterotic on \texorpdfstring{$\mathbb{T}^2$}{T2}}

We will first focus on the slice parametrized by the lower-dimensional dilaton $\hat{\Phi}_d$ and the overall volume of the torus in string units $\hat \varrho_2$, moduli that are particularly convenient, as they control the overall compactification scale and the effective string coupling. Such slice(s),\footnote{Note that there are two equivalent $(\hat{\Phi}_d,\hat \varrho_2)$ slices, see Figure \ref{fig. slice 3d T2}.} where the two radii have the same size, is left invariant under the $\mathbb{Z}_2$ exchange of the circles. The inherited symmetry of this slice is $\mathbb{Z}_2$, corresponding to a double T-duality on the heterotic on $\mathbb{T}^2$ description, and leaving invariant the self-dual line parameterized by $\hat\Phi_8$.
In order to cover the $\pi$ angle subtended by the fundamental domain, and considering that the weakly coupled heterotic string on $\T^2$ subtends an angle $\theta_2 = \arccos \frac{1}{2}$, only two possibilities are allowed:
\begin{subequations}
    \begin{align}
        \textbf{Arrangement O} & :
        \theta_2+\theta_2+\theta_2  = 3 \arccos \half 
        = \pi \\
        \textbf{Arrangement E} & :
        \theta_2 + \theta_{2,1} + \theta_{1 \gg 1}
         = \arccos\half + \arccos\frac{1}{2\sqrt{7}} + \half \arccos \frac{1}{7}  = \pi
    \end{align}
\end{subequations}
There are no possible arrangements involving sensible warped decompactifications.\footnote{\label{fn. lower}As explained in Appendix \ref{ap.extra}, since the angular span of warped codimension-one backgrounds is always smaller than the unwarped case, it is always possible to obtain a $\lambda$ value that results in same angular span of perturbative regimes associated with more compact dimensions. For example, $\theta^{(\frac{2 \left(7 \sqrt{2}+8\right)}{\sqrt{7}})}_1=\theta_2$ and $\theta^{(4\sqrt{\frac27})}_{1\leftrightarrow1}=\theta_{1\gg 1}$, but the $\lambda=\frac{2 \left(7 \sqrt{2}+8\right)}{\sqrt{7}}$ and $4\sqrt{\frac27}$ couplings to a 9d potential $\hat{V}(\hat\varphi)\sim V_0 \exp(-\lambda\hat{\varphi})$ are not known in the string landscape. We refer the reader, however, to \cite[Section 4]{Bergshoeff:2003sy}, where the authors propose a double compactification of Type IIB on $\mathbb{S}^1$ on a further interval with D7/S-dual charges and orientifold sources, while leaving the construction of detailed boundary conditions and string-theory consistency open.}

\begin{figure}[ht]
\begin{center}
\begin{subfigure}[b]{0.5\textwidth}
\centering
     \resizebox{0.95\textwidth}{!}{%
    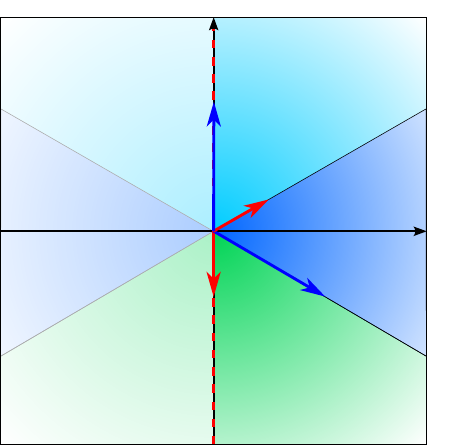
    }
\caption{Arrangement \textbf{O} in $d=8$.} \label{ff.HO2}
\end{subfigure}
\begin{subfigure}[b]{0.49\textwidth}
\centering
     \resizebox{0.97\textwidth}{!}{%
    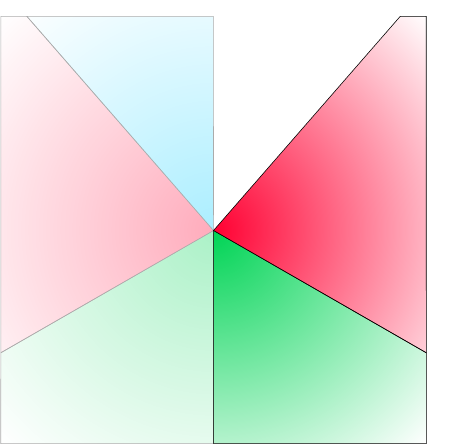
    }
\caption{Arrangement \textbf{E} in $d=8$.} \label{ff.HE2}
\end{subfigure}
\caption{Only two possible arrangements for a two-dimensional Euclidean flat slice in $d=8$ with a $\mathbb{Z}_2$ symmetry and weakly coupled description with two compact dimensions. We depict the matching with the two heterotic on $\mathbb{T}^2$ arrangements of duality frames. We depict the self-dual line $\hat\varrho_2=0$ in red, together with the scaling $\zeta$-vectors (in red for oscillator and blue for KK towers). All quantities are depicted in terms of canonically normalized moduli.
\label{fig.Het2}}
\end{center}
\end{figure}

Arrangement \textbf{O} corresponds to \textbf{(O.1)} the original weakly coupled string theory on a manifold $X_2$, \textbf{(O.2)} an S-dual string theory description on the same manifold $X_2$, and \textbf{(O.4)} another weakly coupled string theory description on a different manifold $X_2'$. On the other hand, for arrangement \textbf{E}, after the original \textbf{(E.1)} weakly coupled string theory on $Y_2$, we have \textbf{(E.2)} a strongly coupled description with compact space $Y_2 \times X_1$, followed by \textbf{(E.4)} a strongly coupled description with a compact space that is well-approximated by a product $X'_1 \times X''_1$, where $\Vol (X_1')$ grows much faster than $\Vol (X_1'')$ (the other limit $\Vol (X_1') \ll \Vol (X_1'')$ is either obstructed if the manifold is actually a fibration $X''_1 \hookrightarrow Y_2'\to  X'_1$, or symmetric under a $\mathbb{Z}_2$ exchange $X'_1 \leftrightarrow X''_1$ around the self-dual line). These two nonequivalent moduli spaces are depicted in Figure \ref{fig.Het2}, being immediate that they recover the duality arrangement of $SO(32)$ and $E_8\times E_8$ heterotic string theory on $\mathbb{T}^2$, respectively. 

The different KK and string oscillator towers for both theories are also depicted in Figure \ref{fig.Het2}. Since the arrangement of perturbative descriptions is different, and the scaling $\zeta$-vectors of the towers are generically located along the interface between different duality descriptions, it is understandable that both \textbf{O} and \textbf{E} feature different towers and vectors. One satisfactory check is that the \textbf{O} and \textbf{E} tower distribution precisely correspond with the polytopes $\Pi_{(8,\I)}$ and $\Pi_{(8,\II)}$ presented in Table~1 of \cite{Etheredge:2024tok}, the only two $d=8$ consistent tower distribution on a flat $\mathbb{R}^2$ slice, which incidentally feature $\mathbb{Z}_2$ symmetries, whose fixed locus is interpreted as the self-dual line in our case.

\subsubsection*{Dilaton--single radius slice of Heterotic on \texorpdfstring{$\mathbb{T}^2$}{T2}}

In the above analysis we have restricted to the two-dimensional slice parameterized by the 8d dilaton and the overall volume in string units with equal radii $\Vol({\S^1_x}) = \Vol({\S^1_y})$ (i.e., $\hat\varrho_x = \hat\varrho_y$). While this is a fixed locus by some of the elements of $\mathcal{W}_{O(2,18;\Z)}(F) = \mathbb{Z}_2^2\rtimes S_2 \cong D_4$ (in particular those exchanging the radii), it does not fully grasp all the different duality frames, as for example it does not cover the Type IIA on $\mathbb{S}^1/\mathbb{Z}_2\times\mathbb{S}^1$ and the warped decompactification limits with sliding towers. In order to study those, we can take the remaining fixed locus, parameterized by the 8d dilaton and one of the radions, say $\hat\varrho_x$, with the other fixed at $\hat\varrho_y = 0$, as this corresponds to the slice that is invariant under T-duality along ${\S^1_y}$: $\Vol({\S^1_y}) = \ell_{\rm s}$. 
We thus expect that when moving along this slice the scaling $\zeta$-vectors remain inside it. 

\begin{figure}[ht]
\begin{center}
     \resizebox{0.45\textwidth}{!}{%
    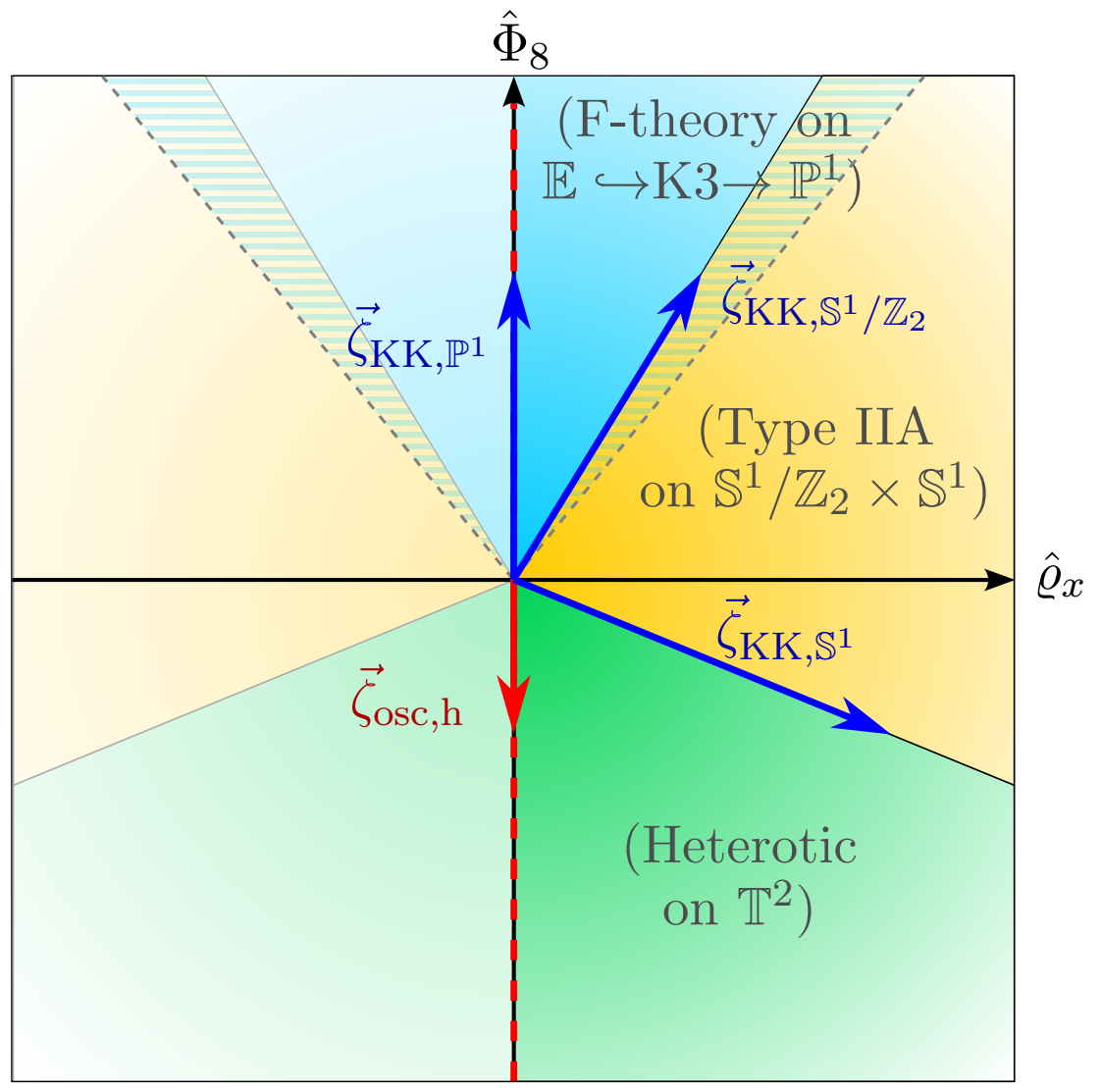
    }
\caption{Common arrangement of dual descriptions in the two-dimensional Euclidean flat slice in $d=8$ parameterized by the 8d dilaton $\Phi_8$ and the radius of one circle, with the additional radius fixed in string units, canonically normalized. Note that both $SO(32)$ and $E_8\times E_8$ heterotic theories feature the same qualitative arrangement, with geometric differences being the D8-brane arrangement in the Type IIA picture and the K3 fiber singularities in the F-theory description. We depict the self-dual line in red, together with the different scaling $\zeta$-vectors (in red for oscillator and blue for KK towers). In hatched regions we depict the apparent mismatch between the F-theory and Type IIA ranges of validity, see text for more details. 
\label{fig.slice1t2}}
\end{center}
\end{figure}

Such arrangement, qualitatively the same for both heterotic theories, is depicted in Figure \ref{fig.slice1t2}. As we expect, the three dual descriptions are heterotic on $\mathbb{T}^2$, (massive) Type IIA on $\mathbb{S}^1/\mathbb{Z}_2\times\mathbb{S}^1$ (where the interval metric is warped) and F-theory on an elliptically fibered $\mathbb{E}\hookrightarrow{\rm K3}\to\mathbb{P}^1$. Since the self-dual line is the same as in the slice depicted in Figure \ref{fig.Het2}, $\Phi_8\to-\infty$ corresponds to an emergent string limit, while $\Phi_8\to+\infty$ decompactifies the F-theory K3 base to 10d. It is clear then that the heterotic and Type IIA frames have a span given by $\theta_1=\arccos\frac{1}{\sqrt{7}}$ and $\theta_{1,1}^{(\frac{5}{\sqrt{2}})}=\arctan\left(5 \sqrt{\frac{3}{2}}\right)$, respectively. This leaves the following angle subtended by the F-theory frame:
\begin{equation}
    \theta_{\rm F-th}=\pi-\arccos\tfrac{1}{\sqrt{7}}-\arctan\left(5 \sqrt{\tfrac{3}{2}}\right)=\arccos\left(2\sqrt{\tfrac{2}{11}}\right).
\end{equation}
It is interesting to notice that the F-theory frame does not adjust to any of the possible warping regimes of \eqref{eq.warping}. In fact, understanding $\mathbb{P}^1\simeq\mathbb{S}^2$ as a fibration $\mathbb{S}^1\hookrightarrow\mathbb{S}^2\to\mathbb{S}^1/\mathbb{Z}_2$, the warped decompactification corresponds to making the base large at fixed fiber radius. Such geometry does not correspond to the backreaction of a scalar over the base interval, and thus it is out of the scope of the type of warped geometries studied in \cite{Raucci:2026fzp}, see Footnote \ref{fn.warping}. 
This is specially evident when looking at the $\zeta$-vector associated to the growth of the interval in the Type IIA string theory on $\mathbb{S}^1/\mathbb{Z}_2 \times \S^1$ or the anisotropic decompactification of the $\mathbb{S}^1\hookrightarrow\mathbb{P}^1\to\mathbb{S}^1/\mathbb{Z}_2$ base in F-theory, which is not located along the naive boundary between the two descriptions, as obtained by taking the $\hat\varrho_y = 0$ slice of Figure \ref{fig. slice 3d T2}. 

However, using the symmetries of moduli space, we have been able not only to obtain the range of validity of such involved compactification, but also the associated light towers, which are fixed by dualities. This is depicted in Figure \ref{fig.slice1t2}, with the ``mismatch'' between descriptions in blue hatched regions. While the classical naive expectation places such region in the F-theory part, with the boundary between the two regions at $\sqrt{2}\hat{\Phi}_8=\sqrt{3}\hat\varrho_x$, the placement of the scaling vector $\vec\zeta_{\rm KK,\mathbb{S}^1/\mathbb{Z}_2}$ suggests that such limits are actually more accurately described by the Type IIA on a cylinder description, with $\sqrt{3}\hat{\Phi}_8=2\sqrt{2}\hat\varrho_x$.\\

We refer to Appendix \ref{app. Details Het T2} for more details on the range of validity of the different duality frames and towers in the complete $F\simeq\mathbb{R}^3$ Euclidean flat slice of heterotic string theory on $\mathbb{T}^2$.

\subsection{Two-dimensional slices of heterotic moduli spaces}

\begin{figure}
    \centering
    \begin{subfigure}[b]{0.78\textwidth}
\centering
     \resizebox{\textwidth}{!}{%
    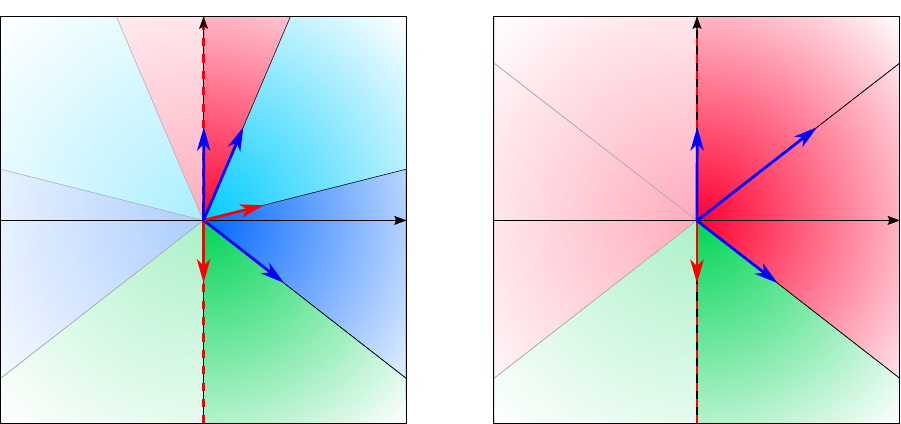
    }
\caption{Arrangements \textbf{O} and \textbf{E} in $d=7$.} \label{ff.Het7}
\end{subfigure}
\begin{subfigure}[b]{0.78\textwidth}
\centering
     \resizebox{\textwidth}{!}{%
    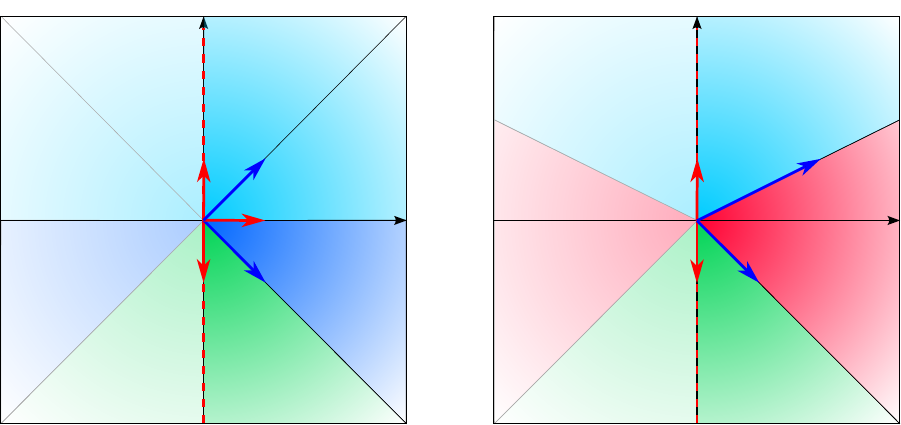
    }
\caption{Arrangements \textbf{O} and \textbf{E} in $d=6$.} \label{ff.Het6}
\end{subfigure}
\begin{subfigure}[b]{0.78 \textwidth}
\centering
     \resizebox{\textwidth}{!}{%
    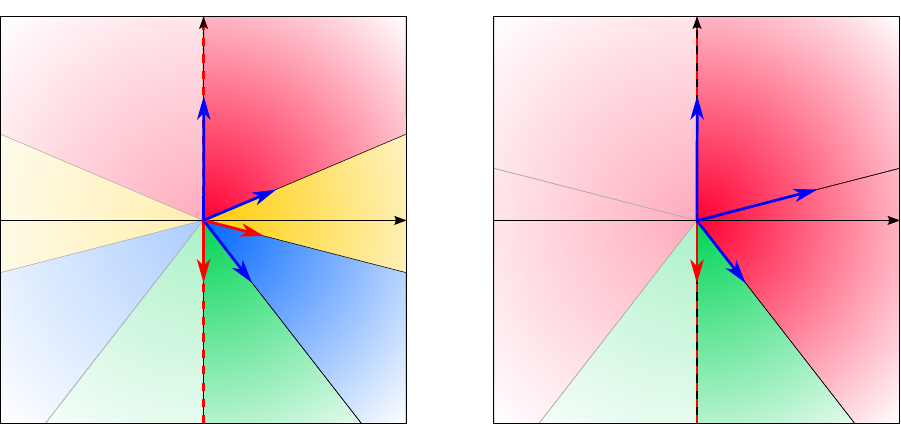
    }
\caption{Arrangements \textbf{O} and \textbf{E} in $d=5$.} \label{ff.Het5}
\end{subfigure}
    \caption{Only two possible arrangements of duality frames for the two-dimensional Euclidean flat slice for half-maximally supersymmetric theories in $d=7$ to 5, with a weakly coupled string limit and a $\mathbb{Z}_2$ symmetry. We depict the matching with the duality frames for both heterotic string theories on $\mathbb{T}^{10-d}$, with the self-dual line (in red) and the scaling $\zeta$-vectors of the light towers (in red for oscillator and blue for KK modes). All quantities are depicted in terms of canonically normalized moduli.
    }
    \label{fig. HetTn}
\end{figure}

The reconstruction of duality frames along flat Euclidean slices can also be carried out for heterotic string theory on $\mathbb{T}^n$ as $n\geq 3$. As explained before, the maximal Euclidean flat slice is $F=\mathbb{R}^{n+1}$ and the inherited dualities acting on it are $\mathcal{W}_{O(n,16+n;\Z)}(F)=\mathbb{Z}_2^n\rtimes S_n$. For simplicity of representation, we will focus on the smaller two-dimensional slice parameterized by the $(d=10-n)$-dimensional dilaton $\Phi_d$ and the volume $\hat\varrho_n$ of the \textit{square} $n$-torus (this is, with the $n$ radii having the same length) in string units. Acting on such slice there is a $\mathbb{Z}_2$ action, corresponding to performing T-duality along the $n$ 1-cycles, that leaves invariant the self-dual line parameterized by $\Phi_d$. As done before, starting from the assumption that in the asymptotic region $\Phi_d\to-\infty$ there is a perturbative description given by (heterotic) string theory on an $n$-manifold (this is, $\mathbb{T}^n$), we can fill the rest of the $\mathbb{R}^2/\mathbb{Z}_2$ half plane with the perturbative descriptions from Section \ref{sec.Pert reg}. As it was the case for $d=9$ and 8, there are \textit{only} two possible consistent arrangements, which as in the $d=8$ case do not require warped decompactifications. Similarly as in the previous cases, while both arrangements show different perturbative descriptions, the $\Phi_d\to+\infty$ limits are equivalent. Such arrangements are depicted in Figure \ref{fig. HetTn}.

As we review now, these duality frame arrangements are precisely in one-to-one correspondence with the dual descriptions of toroidal compactifications of string theory at strong coupling, which have long been known in the literature \cite{Witten:1995ex,Aspinwall:1996mn,Horava:1995qa}:
\begin{itemize}
    \item $\mathbf{d=7}$: We consider first the HO case. We start with $SO(32)$ heterotic string theory on $\mathbb{T}^3$ \textbf{(O.1)}, which in the strong coupling regime is S-dual to Type I on $\mathbb{T}^3$ \textbf{(O.2)}. Upon performing a triple T-duality on the 3-torus, the new perturbative description \textbf{(O.3)} corresponds to Type IIA string theory on $\mathbb{T}^3/\mathbb{Z}_2$ (with an O6$^-$-plane on each of the 8 fixed points, and 16 D6-branes on top of one of such points, in order to cancel the RR tadpole and realize the gauge group). In the strong coupling limit, \textbf{(O.4)} the theory is described by M-theory on K3, where each of the D6-branes uplifts to a Taub-NUT space (this is, a KK monopole), and the O6$^-$-planes come from Atiyah-Hitchin manifolds. As for the HE part, we start with heterotic $E_8\times E_8$ on $\mathbb{T}^3$ \textbf{(E.1)}, which in the strong coupling regime is dual to M-theory on $\mathbb{S}^1/\mathbb{Z}_2\times\mathbb{T}^3$ \textbf{(E.2)}. Upon shrinking the 3-torus, the new perturbative description, \textbf{(E.3)}, is given by M-theory on K3.\footnote{One can understand this by seeing that an elliptically fibered $\mathbb{T}^2\hookrightarrow\text{K3}\to(\mathbb{P}^1\simeq\mathbb{S}^2)$ can be understood as a $\mathbb{T}^3$ fibration over the interval, since $\mathbb{S}^1\hookrightarrow\mathbb{S}^2\to \mathbb{S}^1/\mathbb{Z}_2$.  In the initial frame the theory is locally given along the interval bulk by M-theory on $\mathbb{T}^3$, whose U-duality group maps a shrinking $\mathbb{T}^3$ to a growing one \cite{Obers:1998fb,Calderon-Infante:2023ler}. However, in the endpoints of the interval the negation of the base coordinate makes the $\mathbb{T}^3$ fiber degenerate, resulting in the non-trivial fibration for K3.} 
    Thus, for both theories the $\Phi_7\to\infty$ limit corresponds to a decompactification to 11d M-theory after the K3 manifold grows.
    \item $\mathbf{d=6}$: We first consider $SO(32)$ heterotic string theory on $\mathbb{T}^4$ \textbf{(O.1)}, which upon strong coupling is better described by Type I string theory on $\mathbb{T}^4$ \textbf{(O.2)}. Similarly as before, upon 4 T-dualities once the 4-torus becomes small, the perturbative description is given by Type IIB on $\mathbb{T}^4/\mathbb{Z}_2\simeq\,$K3 \textbf{(O.3)} (with the 16 D5-branes on top of one of the 16 O5$^-$ planes), which in the strong coupling regime is self-S-dual to Type IIB on the same K3 \textbf{(O.4)}.\footnote{While upon S-duality the geometric part of the compactification is not affected, the D5-brane and O5$^-$-plane system is mapped to NS5-branes and ON$^-$-planes, see \cite{Hanany:1999sj}.} As for heterotic $E_8\times E_8$ on $\mathbb{T}^4$ \textbf{(E.1)}, the strong coupling regime is given by M-theory on $\mathbb{S}^1/\mathbb{Z}_2\times\mathbb{T}^4$ \textbf{(E.2)}. Upon shrinking the 4-torus, first we find ourselves on Type IIA on $\mathbb{S}^1/\mathbb{Z}_2$ times  shrinking 3-torus, which upon 3 T-dualities and an S-duality,\footnote{\label{fn. strong T} Considering that the four radii of the torus shrink like $R_i\sim \lambda^{-1}\ell_{11}$, with $\lambda\to \infty$, we have that the 10d Type IIA and Type IIB dilaton are given by
    \begin{equation}
        \Phi_{10,\rm IIA}=\frac{3}{2}\log(R_1/\ell_{11})\sim -\frac{3}{2}\log \lambda\to0 \;\xrightarrow{\rm 3\; T-dual.} \Phi_{10,\rm IIB}=\Phi_{\rm 10,IIA}+\log\frac{\ell_{11}^3}{R_1R_2R_3}\sim\frac{3}{2}\log\lambda\to\infty\,,
    \end{equation}
    and thus an additional S-duality must be taken in order to be at weak Type IIB coupling.
    } this corresponds to the weak coupling limit of Type IIB on $\mathbb{T}^3\hookrightarrow{\rm K3}\to\mathbb{S}^1/\mathbb{Z}_2$. For both HO and HE 6d theories the limit $\Phi_6\to\infty$ corresponds to an emergent Type IIB string. 
    \item $\mathbf{d=5}$: Again, we start with the HO slice, for which $SO(32)$ heterotic string theory on $\mathbb{T}^5$ \textbf{(O.1)} is S-dual in the strong coupling regime to Type I on the same $\mathbb{T}^5$ \textbf{(O.2)}. Upon five T-dualities, the system corresponds to Type IIA on $\mathbb{T}^5/\mathbb{Z}^2$ (with 16 D4-branes on top of one of the 32 O4$^-$-planes) \textbf{(O.3)}. Its strong coupling is then given by M-theory on $\mathbb{T}^5/\mathbb{Z}_2\times\mathbb{S}^1$ (with the D4-branes uplifting to M5-branes wrapping the extra circle). As shown in \cite{Dasgupta:1995zm}, the decompactification limit corresponds to a 6d theory given by Type IIB on K3. We can then take heterotic $E_8\times E_8$ on $\mathbb{T}^5$ \textbf{(E.1)}, which in the strong coupling is better described by M-theory on $\mathbb{S}^1/\mathbb{Z}_2\times\mathbb{T}^5$ \textbf{(E.2)}. Upon reduction of the $\mathbb{T}^5=\mathbb{S}^1\times\mathbb{T}^4$ down to Type IIA string theory on $\mathbb{S}^1/\mathbb{Z}_2$ times a shrinking $\mathbb{T}^4$ and performing four T-dualities, we find ourselves in the strong coupling limit of Type IIA on $\mathbb{T}^5/\mathbb{Z}_2$ (in an analogous way as in Footnote \ref{fn. strong T}), better described by M-theory on $\mathbb{S}^1\times \mathbb{T}^5/\mathbb{Z}_2$ \textbf{(E.3)}. Thus, for both descriptions the $\Phi_5\to\infty$ limit corresponds to a decompactification of the M-theory circle, which is dual to the 6d theory given by Type IIB on K3 \cite{Dasgupta:1995zm}.
\end{itemize}

The above description of the duality frames arrangement is perfectly matched by the bottom-up reconstruction of the perturbative regimes, as depicted in Figure \ref{fig. HetTn}. While of course the microscopic nature of the concrete string theory or the topology of the internal space is not directly transparent through this approach, we are nonetheless able to describe the arrangement and range of validity of the different descriptions, including the existence or absence of a weak string coupling limit, as well as the dimensionality of different cycles present in the internal dimensions. It must be noted that both reconstructions feature the same behavior along the strong coupling limit of the self-dual line at each $d$. While this is not surprising from a top-down perspective, since both $SO(32)$ and $E_8\times E_8$ heterotic theories result in the same lower-dimensional theories upon compactification (going from one slice to the other amounts to rotate along the self-dual line by changing the Wilson line moduli), this was \textit{not} an input that was used in our derivation.\\

In Appendix \ref{ap.extra} we show how, allowing for warped internal spaces without a clear higher-dimensional origin for the warping sources, additional arrangements can result. None of those feature the same $\Phi_d\to+\infty$ limits as the cases studied in this section, and in all but one case the arrangement requires exotic T-dualities exchanging KK states associated to growing internal spaces of different dimensions. In such appendix we also discuss how such T-dualities are likely inconsistent with the Emergent String Conjecture.

\subsubsection{The special cases of \texorpdfstring{$d=4$}{d=4} and \texorpdfstring{$3$}{3}} \label{sec. het d4 and d3}

\begin{figure}[ht]
\begin{center}
\centering
     \resizebox{0.97\textwidth}{!}{%
    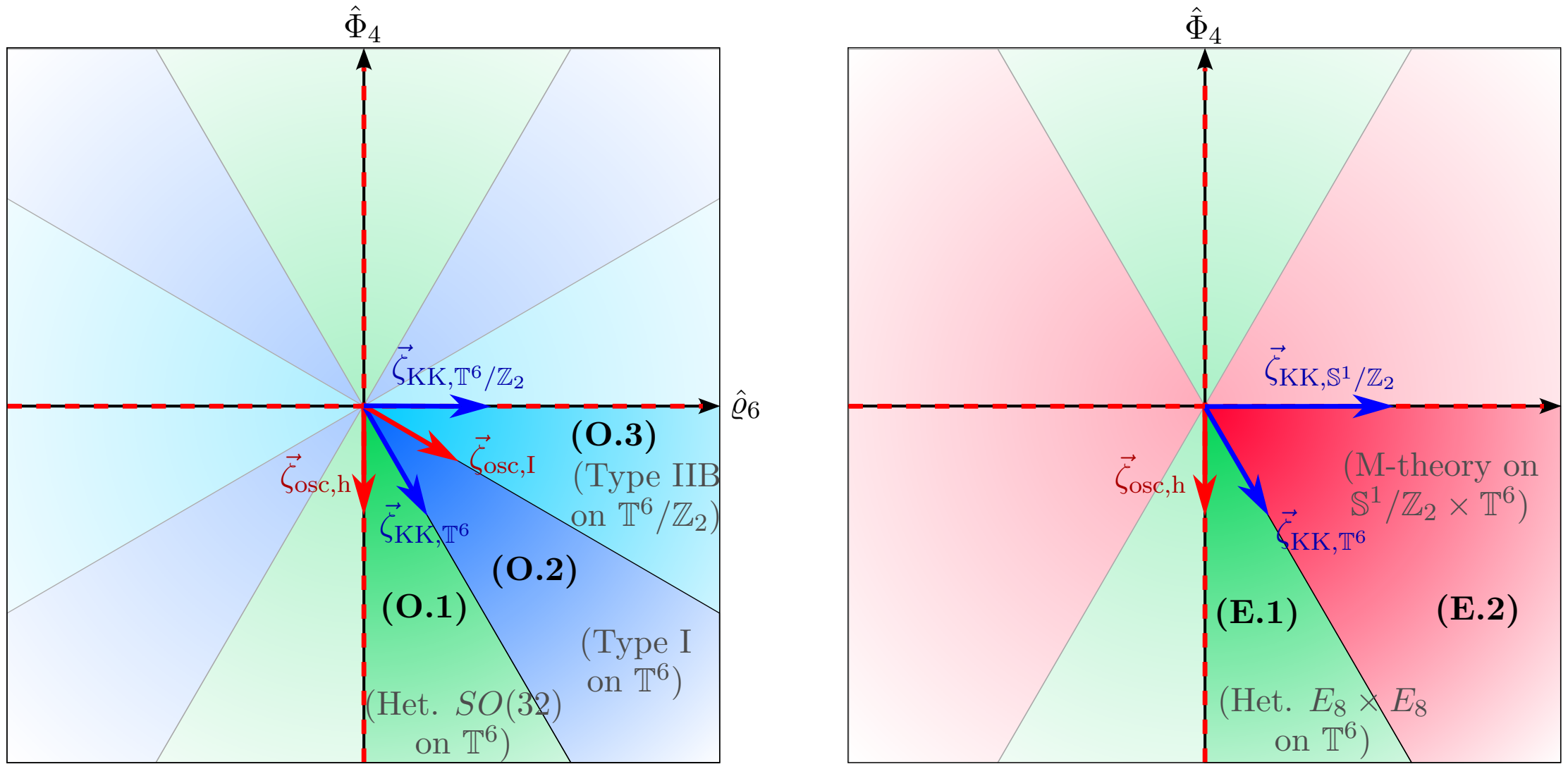
    }
\caption{Only two possible arrangements for a two-dimensional Euclidean flat slice in $d=4$ with a $\mathbb{Z}_2\times\mathbb{Z}_2$ symmetry and weakly coupled description with six compact dimensions. We depict the matching with the two heterotic on $\mathbb{T}^6$ arrangements of duality frames. We depict the self-dual lines in red, together with the scaling $\zeta$-vectors (in red for oscillator and blue for KK towers). All quantities are depicted in terms of canonically normalized moduli.
\label{fig.Het6}}
\end{center}
\end{figure}

In the previous subsection we have focused on toroidal compactifications of heterotic string theory up to $d=5$, since it is for those cases that the form \eqref{eq.moduli space het} is valid. For lower dimensions additional scalars appear and the dimensionality \eqref{eq. dim mod het} and global form of the moduli space changes, see \cite{Sen:1994fa}. In $d=4$ the external $B_2$ field can be dualized to an NSNS axion $\dd a=e^{-4\Phi_4}\star_4\dd B_2$, which together with the 4d dilaton forms a complex scalar $S=a+ie^{-2\Phi_4}$ analogous to the 10d Type IIB $\tau$. This way,
\begin{equation}
    \mathcal{M}_4=O(6,22;\Z) \, \big\backslash \,
\frac{ O(6,22;\R) }{ O(6;\R) \times O(22;\R) }\times SL(2,\mathbb{Z})\big\backslash\frac{SL(2,\mathbb{R})}{SO(2)}\,,\quad \text{with}\;\dim\mathcal{M}_4=134\,.
\end{equation}
In this case, the maximal Euclidean flat slice and the inherited duality group is given by
\begin{equation}
    F=\mathbb{R}^7=\mathbb{R}^6\times\mathbb{R}\,,\quad \mathcal{W}_{G}(F)=(\mathbb{Z}_2^6\rtimes S_6)\times\mathbb{Z}_2\,.
\end{equation}
It is important to notice that now the self-dual line has a $\mathbb{Z}_2$ symmetry acting on it, and thus the $\Phi_4\to\pm \infty$ are equivalent, with an additional fixed locus at $\Phi_4=0$. Again working in the $\mathbb{R}^2$ slice parameterized by the 4d dilaton and the volume of the square 6-torus in string units, there is a $\mathbb{Z}_2\times\mathbb{Z}_2$ symmetry. Through a reconstruction procedure analogous to the one we had before, we obtain a \textit{unique} pair of perturbative regions arrangements, as depicted in Figure \ref{fig.Het6}. Their origin from top-down compactifications is clear too, see \cite{Sen:1994fa}:\\

If we start from $SO(32)$ string theory on $\mathbb{T}^6$ \textbf{(O.1)}, the strong coupling limit is S-dual to Type I string theory on the same $\mathbb{T}^6$, \textbf{(O.2)}. Upon six T-dualities on the internal space, we find ourselves on Type IIB string theory on $\mathbb{T}^6/\mathbb{Z}_2$ \textbf{(O.3)}, where the 16 D3-branes are located on top of one of the 64 O3$^-$-planes associated to the fixed points. Now, upon strong coupling, the system is S-dual (recall that from \eqref{eq. SL2Z} $C_4$ is invariant under $SL(2,\mathbb{Z})$, and so are D3-branes and O3$^-$-planes). The upper part of the $\mathbb{R}^2$ flat slice is simply a reflection of the lower one. On the other hand, for $E_8\times E_8$ heterotic string theory on $\mathbb{T}^6$ \textbf{(E.1)}, the strong coupling regime is better described by M-theory on $\mathbb{S}^1/\mathbb{Z}_2\times\mathbb{T}^6$ \textbf{(E.2)}. In the limit in which the $\mathbb{T}^6$ becomes light, the $\mathbb{Z}_2$ symmetry maps M-theory on $\mathbb{S}^1/\mathbb{Z}_2\times\mathbb{T}^6$ to itself.\footnote{In order to see this, take M-theory on $\mathbb{T}^6$. The subgroup $SL(2,\mathbb{R})\subset E_{6(6)}$ is associated to the axio-dilaton 
\begin{equation}
 \tau_{\rm M}=a+i\mathcal{V}\,,\quad\text{with}\; a\propto\int_{\mathbb{T}^4}C_6\,,\quad\mathcal{V}=\frac{\rm vol(\mathbb{T}^6)}{\ell_{11}^6}\,,
\end{equation}
where $C_6$ is the Hodge dual to the $C_3$ field \cite{Obers:1998fb}. The analogous to the $S$ transformation in the $SL(2,\mathbb{R})\subset E_{6(6)}$ inverts the 6-torus volume. This is evident by considering the BPS towers of states associated to KK momenta and wrapped M5-branes in all but one radii:
\begin{equation}
    m_{{\rm KK},i}=\frac{1}{R_i}\,,\quad m_{{\rm M5},i}=\frac{\prod_{j\neq i} R_i}{\ell_{11}^6}=\frac{\mathcal{V}}{R_i}\equiv \frac{1}{R_i'}\,,
\end{equation}
In 4d Planck units, where $M_{\rm Pl,4}=M_{\rm Pl,11}\mathcal{V}^{\frac{1}{3}}({m_{{\rm KK},\mathbb{S}^1/\mathbb{Z}_2}}/{M_{\rm Pl,4}})^{-\frac{1}{3}}$, this translates to
\begin{equation}
    \frac{ m_{{\rm KK},i}}{M_{\rm Pl,4}}\sim\frac{\ell_{11}}{R_i}\mathcal{V}^{-\frac{1}{3}}\left(\frac{m_{{\rm KK},\mathbb{S}^1/\mathbb{Z}_2}}{M_{\rm Pl,4}}\right)^{\frac{1}{3}}\,,\quad \frac{ m_{{\rm M5},i}}{M_{\rm Pl,5}}\sim \frac{\ell_{11}}{R_i}\mathcal{V}^{2/3}\left(\frac{m_{{\rm KK},\mathbb{S}^1/\mathbb{Z}_2}}{M_{\rm Pl,4}}\right)^{\frac{1}{3}}\sim\frac{\ell_{11}}{R_i'}{\mathcal{V}'}^{\frac{1}{3}}\left(\frac{m_{{\rm KK},\mathbb{S}^1/\mathbb{Z}_2}}{M_{\rm Pl,4}}\right)^{\frac{1}{3}}\,.
\end{equation}
Keeping the KK scale of the Ho\v{r}ava-Witten interval fixed in 4d Planck units, the $S$-transformation identifies the wrapped M5-branes with the momentum modes of a dual $\mathbb{T}^6$ with $\mathcal{V}'=\mathcal{V}^{-1}$.
}\\

On the other hand, in $d=3$ the 30 gauge vector fields (7 from $g_{i\mu}$, 7 from $B_{i\mu}$ and 16 from $A^I_\mu$) are Hodge-dual to scalars. This way,\footnote{Since $\Phi_3=\Phi_4-\frac{1}{2}\log\frac{R_7}{\ell_{\rm s}}$, the 3d dilaton is a linear combination of the Cartan generators of 
\begin{equation}
    \hat{\mathcal{M}}_3=\frac{ O(8,23;\R) }{ O(8;\R) \times O(24;\R) }\supset \frac{SL(2,\mathbb{R})_S}{SO(2)}\times \frac{SL(2,\mathbb{R})_T}{SO(2)}\,,\quad\text{where}\;\left\{\begin{array}{l}
         S=a+i e^{-2\Phi_4}\\
         T=b+i (R_7/\ell_{\rm s})
    \end{array}\right.\,,
\end{equation}
where $a$ is the 4d NSNS axion and $b$ is the scalar dual to the graviphoton associated to $g_{7\mu}$. Thus it does not transform independently and must be included in the complete coset \cite{Maharana:1992my,Obers:1998fb}.
 } 
\begin{equation}
    \mathcal{M}_3=O(8,24;\Z) \, \big\backslash \,
\frac{ O(8,23;\R) }{ O(8;\R) \times O(24;\R) }\,,\quad\text{with }\dim\mathcal{M}_3=192\,.
\end{equation}
Here the maximal Euclidean flat slice and inherited duality group are given by $F=\mathbb{R}^8$ and $\mathcal{W}_G(F)=\mathbb{Z}_2^8\rtimes S_8$, and unlike the previous cases, there is no self-dual line associated with the 3d dilaton as in the higher-dimensional cases. Thus the analysis we have been performing so far, in which the (marked) moduli space and duality groups factorize into a geometry and coupling parts, does not immediately apply. As explained in \cite{Obers:2000ta}, this moduli space can be understood as that of a heterotic compactification on $\mathbb{T}^7\times\mathbb{S}^1_{\rm h}$, with an additional radius $R_{\rm h}=\ell_{\rm s} e^{-2\Phi_3}$ (note that this is the opposite as in the M-theory circle in Type IIA strong coupling, since $R_{\rm h}$ decompactifies at weak coupling). In the $\Phi_3\to\infty$ strong coupling limit, the theory is expected to be ``T-dual'' to the same compactification with $R_{\rm h}'\ell_{\rm s}^{-1}=R_{\rm h}^{-1}\ell_{\rm s}$, thus arriving to the original description. Such duality is supported by the careful study of instanton effects, see \cite{Obers:2000ta,Kachru:2016ttg}, where the dualities with Type II string theory on K3$\times \mathbb{T}^3$ along anisotropic limits are also discussed.

\subsection{Other moduli spaces with 16 supercharges}
Before closing this section, we will briefly discuss other moduli spaces associated to theories with 16 supercharges, where again $\mathcal{M}$ is given as a symmetric space. We will first consider the CHL string with $E_8$ gauge symmetry \cite{Chaudhuri:1995fk,Chaudhuri:1995bf,Aharony:2007du}, obtained by compactifying $E_8\times E_8$ string theory on $\mathbb{S}^1$ with a discrete $\mathbb{Z}_2$ monodromy that exchanges the two $E_8$ copies, resulting in a rank-9 9d theory with moduli space
\begin{equation}
    \mathcal{M}_{\rm CHL}=O(1,9;\mathbb{Z})\big\backslash\frac{O(1,9)}{O(9)}\times\mathbb{R}\,,
\end{equation}
practically analogous to the heterotic string on $\mathbb{S}^1$ discussed in Section \ref{sec.het S1}. As in that case, the maximal Euclidean flat slice is given by $F=\mathbb{R}^2$ (again parameterized by the radion in string units and the 9d dilaton), with $\mathcal{W}_G(F)=\mathbb{Z}_2$, associated to T-duality, thus defining a self-dual line parameterized by $\Phi_9$. As we explained in the previous section, only two possible perturbative descriptions are allowed for $d=9$ with an emergent string limit in $\Phi_9\to-\infty$, and these must be analogous to Figure \ref{ff.HO1} or \ref{ff.HE1}. As explained in \cite{Aharony:2007du,Etheredge:2023odp}, the duality frame arrangement is analogous to that of Figure \ref{ff.HE1}. Starting with the $E_8\times E_8$ string on the CHL circle \textbf{(E.1)}, the strong coupling limit corresponds to M-theory on a M\"obius strip $\mathbb{S}^1/\mathbb{Z}_2\hookrightarrow\mathbb{S}^1$ \textbf{(E.2)}. Finally, upon compactification along the base, the resulting description \textbf{(E.3)} corresponds to a variant of the Type I$'$ theory where one of the O8$^-$-planes with 7 D8-branes on top is replaced by an O8$^{(-1)}$-plane frozen at infinite coupling. Since the bulk is still warped due to the non-zero Romans mass, the compactification has the same coupling $\lambda=\frac{5}{\sqrt{2}}$ as in the usual heterotic case.\\

This construction can be generalized to compactification on a $\mathbb{T}^n$-torus with a $RT$ twist, where $R$ is the outer automorphism exchanging the two $E_8$ factors and $T$ is a half-period shift along the torus, with $[R,T]=1$. In this case the moduli space is given by
\begin{equation}
    \mathcal{M}_{{\rm CHL},n}=O(n,8+1;\mathbb{Z})\big\backslash\frac{O(n,8+n)}{O(n)\times O(8+n)}\times\mathbb{R}\,,
\end{equation}
with the maximal Euclidean flat slice and its inherited symmetries given by $\mathbb{R}^n\times\mathbb{R}$ and $\mathbb{Z}_2^n\rtimes S_n$ (again for $n\leq 5$), respectively. The arrangement of perturbative duality frames is analogous to heterotic $E_8\times E_8$ on $\mathbb{T}^n$, with variations of the compact geometries for the different top-down compactifications. For example, for the CHL string on $\mathbb{T}^2$, the M-theory frame is compactified on $\frac{\mathbb{S}^1/\mathbb{Z}_2\times\mathbb{S}^1_A\times\mathbb{S}^1_B}{\langle RT_A\rangle}\simeq (\mathbb{S}^1/\mathbb{Z}_2\hookrightarrow\mathbb{S}^1_A)\times \mathbb{S}^1_B$, while in the F-theory on K3 phase one of the elliptic fiber singularities is \textit{frozen} and does not translate to the gauge group, resulting in a single $E_8$ factor, see \cite{Bershadsky:1998vn}. While the microscopic realization significantly differs from the $E_8\times E_8$ theory, the interpretation as perturbative descriptions and the type of infinite distance limits are equivalent, with the arrangement and range of validity of the different duality frames the same as in the HE theory described in Section \ref{sec. half max}, and thus approachable through the symmetries of moduli space. \\

Finally, we consider asymmetric orbifolds of Type IIA and Type IIB on a circle under $(-1)^{\rm F_L}$ and a half-shift, respectively called AOA and AOB theories \cite{Aharony:2007du,Gutperle:2000bf,Hellerman:2005ja}. One can generalize these constructions on tori \cite{Persson:2017lkn,Baykara:2025lhl,Gkountoumis:2025btc} by adding additional spectator circles along which one can dualize and exchange the AOA and AOB sectors. The resulting theories have gauge group of rank $n$, with moduli space
\begin{equation}
    \mathcal{M}=\Gamma_{\rm orb}\big\backslash\frac{O(n,n;\mathbb{R})}{O(n)\times O(n)}\times\mathbb{R}\qquad\text{(again for }d=10-n>4\text{)}\,,
\end{equation}
where the duality group $\Gamma_{\rm orb}$ does not necessarily correspond to $O(n,n;\mathbb{Z})$, see \cite{Persson:2017lkn}. However, as discussed above, the set of T-dualities remains as a subgroup of $\Gamma_{\rm orb}$, trivially leaving invariant the $\mathbb{R}$ factor parameterized by $\Phi_d$. One can then repeat the same construction as in the previous theories and fill the overall volume-dilaton slice with a set of perturbative regimes symmetric with respect to the self-dual line. As explained in \cite{Aharony:2007du,Etheredge:2023odp}, the AOA and AOB arrangements are analogous to those of $E_8\times E_8$ and $SO(32)$ heterotic, respectively. For example, in $d=9$ the strong coupling regime of AOA involves M-theory on a Klein bottle and a (massive) Type IIA orientifold with two frozen O8$^{(-1)}$-planes on its ends, while for AOB the additional perturbative descriptions are given by the Dabholkar-Park background \cite{Dabholkar:1996pc} (analogous to Type I on a circle but with reduced rank) and mIIA string theory on an interval with an O8$^-$ and O8$^+$ on its ends.\\

The above generalizes to lower dimensions, with again the asymptotic regions of moduli space for theories with a self-dual line parameterized by the lower-dimensional dilaton (and a $\mathbb{Z}_2$ action leaving it invariant) is given by arrangements analogous to those of HE or HO on $\mathbb{T}^{10-d}$.

\section{Theories with little or no supersymmetry\label{sec. low}}

The maximal or half-maximal supersymmetric theories studied in the previous Sections \ref{sec. Udual} and \ref{sec. half max} are not the only theories featuring the coset of symmetric space as moduli (or better put, parameter) space. As a straightforward example, the classical moduli space of some non-supersymmetric theories is given by cosets of symmetric spaces \cite{Giveon:1994fu,Fraiman:2023cpa} (again for $d\geq 5$):\footnote{The bosonic and $E_8$ theories possess already a string-scale tachyon in their 26d and 10d spectra, which makes their vacua unstable. The $SO(16)\times SO(16)$ heterotic theory is tachyonless but, as we will discuss shortly, has a running dilaton profile $V(\Phi_{10})=\Lambda\,e^{\frac{5}{2}\Phi}>0$ in 10d.}
\begin{subequations}
    \begin{align}
        \textbf{Bosonic string theory on }\mathbb{T}^n:\quad\mathcal{M}_{\rm bos,\mathbb{T}^n}&=O(n,n;\mathbb{Z})\big\backslash\frac{O(n,n;\mathbb{R})}{O(n)\times O(n)}\times\mathbb{R}\,,\\
        O(16)\times O(16)\textbf{ het. on }\mathbb{T}^n:\quad\mathcal{M}_{O(16)^2,\mathbb{T}^n}&=O(\Upsilon_{n,n+16})\big\backslash\frac{O(n,n+16)}{O(n)\times O(n+16)}\times\mathbb{R}\,,
        \\
        E_8\textbf{ het. on }\mathbb{T}^n:\quad\mathcal{M}_{E_8,\mathbb{T}^n}&=O(\Upsilon_{n,n+8})\big\backslash\frac{O(n,n+8)}{O(n)\times O(n+8)}\times\mathbb{R}\,,
    \end{align}
\end{subequations}
where the $O(\Gamma)$ is the automorphism group of the lattice $\Gamma$, with
\begin{equation}
    \Upsilon_{n,n+16}=\mathfrak{so}_{16}\oplus\mathfrak{so}_{16}\oplus \Gamma_{n,n}\,,\qquad \Upsilon_{n,n+8}=E_8\oplus \Gamma_{n,n} \, ,
\end{equation}
while $\mathbb{R}$ is the self-dual line associated with the lower-dimensional dilaton. In all cases the duality group has $O(n,n;\mathbb{Z})$ as a subgroup, see \cite{Fraiman:2023cpa,DeFreitas:2024ztt}, and thus one can define the maximal Euclidean flat lattice $F=\mathbb{R}^n\times\mathbb{R}$, with inherited duality group $\mathbb{Z}_2^n\rtimes S_n$ that acts trivially on the self-dual line. However, in all three cases the absence of supersymmetry results in a non-vanishing vacuum energy $\Lambda$ in the string frame, which in Einstein frame results in
\begin{equation}
    V(\Phi_d)=e^{\frac{2d}{d-2}\Phi_d}\Lambda\neq 0 \, ,
\end{equation}
which diverges in the $\Phi_d\to\infty$ direction. This prevents us from considering the study of perturbative descriptions for the strongly coupled regions of the flat Euclidean slice. Since in the weakly coupled region $\Phi_d\to-\infty$ the potential goes to zero, $V(\Phi_d)\to 0$, the $\mathbb{Z}_2^n\rtimes S_n$ dualities act on the $F=\mathbb{R}^n\times\mathbb{R}$ (which has the different Wilson lines and axions set to zero) mapping the perturbative theory on $\mathbb{T}^n$ into $|\mathbb{Z}_2^n\rtimes S_n|=2^n\cdot n!$ equivalent descriptions. We refer to \cite{Fraiman:2023cpa,Fraiman:2025yrx} for an interesting discussion of the tachyonic regions, gauge enhancement locus and special points of the moduli space of non-supersymmetric heterotic string theory along such weakly coupled regions, including slices and trajectories with non-constant Wilson lines.

\subsection{4d \texorpdfstring{$\mathcal{N}=2$}{N=2} and \texorpdfstring{$1$}{1} theories}

As a more concrete example, we take Type II or heterotic string theory on the orbifold
\begin{equation}
    X_6=\mathbb{T}^6/\mathbb{Z}_3=(\mathbb{T}^2)^3/\langle\theta\rangle\,,\quad \text{with}\quad\theta:(z_1,z_2,z_3)\mapsto(e^{\frac{2\pi i}{3}}z_1,e^{\frac{2\pi i}{3}}z_2,e^{\frac{2\pi i}{3}}z_3)\,,
\end{equation}
resulting in a 4d $\mathcal{N}=2$ theory for Type II string theory and 4d $\mathcal{N}=1$ for heterotic string theory (this is, respectively 8 or 4 supercharges). While the complete moduli space depends on the particular theory, there is a common part of moduli space given by\footnote{
In Type IIA compactifications (we will stay in the untwisted sector for simplicity), $SU(3,3;\mathbb{Z})\backslash{SU(3,3)/(SU(3)\times SU(3)\times U(1)})$ is associated with the reduction of the metric and $B_2$ field, associated with nine vector multiplets. There is a single hypermultiplet associated to the 4d axio-dilaton $S=a+ie^{-2\Phi_4}$ and the 2 scalars coming from the reduction of $C_3$. As for Type IIB, since $h^{2,1}(X_6)=0$, the vector multiplet is empty, and the 40 scalars (18+2 from the NSNS sector, together with $C_0$, 9 from $C_2$, other 9 from $C_4$ and 1 from $C_6$) are organized within $n_H=10$ hypermultiplets \cite{Bianchi:1996ai}, and locally given by
\begin{equation}\label{eq. MII}
    \hat{\mathcal{M}}_{\rm IIA}^{\rm untwist}\approx \underbrace{\frac{SU(3,3)}{SU(3)\times SU(3)\times U(1)}}_{\mathcal{M}_{\rm V}}\times \underbrace{\frac{SU(2,1)}{SU(2)\times U(1)}}_{\mathcal{M}_{\rm H}}\,,\qquad \hat{\mathcal{M}}_{\rm IIB}^{\rm untwist}=\underbrace{\frac{E_{6(2)}}{SU(6)\times SU(2)}}_{\mathcal{M}_{\rm H}}\,,
\end{equation}
on which \eqref{eq. MC} embeds. As for the heterotic case, where the compactification results in a 4d $\mathcal{N}=1$ theory, taking the standard embedding $\mathfrak{e}_8\oplus\mathfrak{e}_8\to\mathfrak{e}_6\oplus\mathfrak{su}(3)\oplus \mathfrak{e}_8$ (resp. $\mathfrak{so}(32)\to \mathfrak{so}(26)\oplus\mathfrak{su}(3)\oplus\mathfrak{u}(1)$), the per internal complex direction is given by $(\mathbf{27},\mathbf{3},\mathbf{1})$ (resp. $(\mathbf{26},\mathbf{3})_{+1}\oplus(\mathbf{1},\mathbf{3})_{-2}$) \cite{Choi:2004wn,Ibanez:2012zz}, with a total of $3\times 81=243$ scalars organized in a matrix $C$ of such size. The (classical) moduli space associated with the untwisted sector is thus given by
\begin{equation}
    \hat{\mathcal{M}}_{\rm het}^{\rm untwist}=\frac{SU(3,84)}{SU(3)\times SU(84)\times U(1)}\times \frac{SU(1,1)}{U(1)}\,,
\end{equation}
common to both heterotic theories, while the K\"ahler potential \eqref{eq. kahler 4d} extends to $K=-\log [-i(S-\bar S)]-\log[-i\det(T-T^\dagger-CC^\dagger)]$.
}
\begin{equation}\label{eq. MC}
    \mathcal{M}_{\rm c}=SU(3,3;\mathbb{Z})\big\backslash\frac{SU(3,3)}{SU(3)\times SU(3)\times U(1)}\times \frac{SU(1,1)}{U(1)}\,,
\end{equation}
with complex dimension $\dim_{\rm \mathbb{C}}\mathcal{M}_{\rm c}=9+1=10$. Such space is parameterized by a $3\times 3$ complex matrix $T_{i\bar j}=\beta_{i\bar j}+iG_{i\bar j}$ (where $G_{i\bar j}$ is the hermitian matrix on $\mathbb{T}^6=(\mathbb{T}^2)^3$ and $B_2=i\beta_{i\bar j}\dd z^i\wedge\dd\bar{z}^j$),
and a 4d axio-dilaton $S=a+ie^{-2\Phi_4}$ (where $\dd a=e^{-4\Phi_4}\star_4 \dd B_2$). The $SU(3,3;\mathbb{Z})$ duality group acts only on the $T_{i\bar j}=\beta_{i\bar j}+iG_{i\bar j}$ complex scalars. We can identify a (real) maximal Euclidean flat slice and inherited duality group given by 
\begin{equation}\label{eq. flat slice 4dN2}
    F=\mathbb{R}^3\times\R\,, \quad \mathcal{W}_{SU(3,3;\mathbb{Z})}(F)\simeq \mathbb{Z}_2^3\rtimes S_3\,,
\end{equation}
respectively associated to the volume of the three hexagonal 2-tori 
$\{t_i=\Vol(\mathbb{T}^2_i)\ell_{\rm s}^{-2}\}_{i=1}^3$, as well as their double T-dualities (internal geometry does not have proper 1-cycles\footnote{\label{fn. hom T6Z2}The integer homology of the singular orbifold is given by
\begin{equation}
    H_\bullet(X_6;\mathbb{Z})=(\mathbb{Z},0,\mathbb{Z}^9,\mathbb{Z}^2\oplus\mathbb{Z}_3,\mathbb{Z}^9,0,\mathbb{Z},0,\dots)\,.
\end{equation}
Upon resolution into a smooth $\tilde{X}_6$, the 27 fixed points can be resolved through the introduction of twisted K\"ahler deformations, so that as a complex manifold, $h^{1,1}(\tilde X_6)=9+27=36$, while $h^{2,1}=0$.}) and interchange.\footnote{\label{fn.blowup}Note that in the discussion of the T-dualities we are only considering the naive singular $X_6=\mathbb{T}^6/\mathbb{Z}_3$, rather than the smooth $\tilde X_6$ where the 27 fixed points are resolved via blow-ups to $\mathbb{P}^2$, see, e.g., \cite{GrootNibbelink:2008hde}. A non-zero size of the blow-up curves modifies the duality transformations non-trivially, see \cite{Erler:1992gt}, and prevents a perturbative description in the limit where one of the $\mathbb{T}^2$ factors becomes small.} Associated to the 4d dilaton $\Phi_4$ there is an additional factor $\mathbb{R}$ in the flat slice, which is left invariant by the elements of $\mathbb{Z}_2^3\rtimes S_3$.\\

The 4d theory has a K\"ahler potential defined over this slice of moduli space, given by
\begin{equation}\label{eq. kahler 4d}
    K_{\rm c}=-\log [-i(S-\bar S)]-\log[-i\det(T-T^\dagger)]=2\Phi_4-\log\frac{\Vol(X_6)}{\ell_{\rm s}^6}+{\rm constant}\,,
\end{equation}
from where the moduli space (complex) metric can be obtained by $\mathsf{G}_{a\bar b}=\partial_a\partial_{\bar b} K_{\rm c}$. The K\"ahler potential is left invariant by permutations on the $\{t_i=\Vol(\mathbb{T}^2_i)\ell_{\rm s}^{-2}\}_{i=1}^3$, while their T-duality $t_i\to t_i^{-1}$ leaves $K_{\rm c}$ invariant up to a K\"ahler transformation:
\begin{equation}
    K_{\rm c}=2\Phi_4-\log(t_1t_2t_3)+{\rm constant.}\;\longmapsto\, K_{\rm c}'= K_{\rm c}+2\log t_i= K_{\rm c}+\log T_i+\log \bar T_i\,,
\end{equation}
with $T_i=\int_{\mathbb{T}^2_i}B_2+it_i$ is the complexified $\mathbb{T}^2_i$ volume (for simplicity we set the NSNS axions to 0 while performing the T-dualities).\\

\begin{figure}[ht]
\begin{center}
\centering
     \resizebox{\textwidth}{!}{%
    \import{figures/}{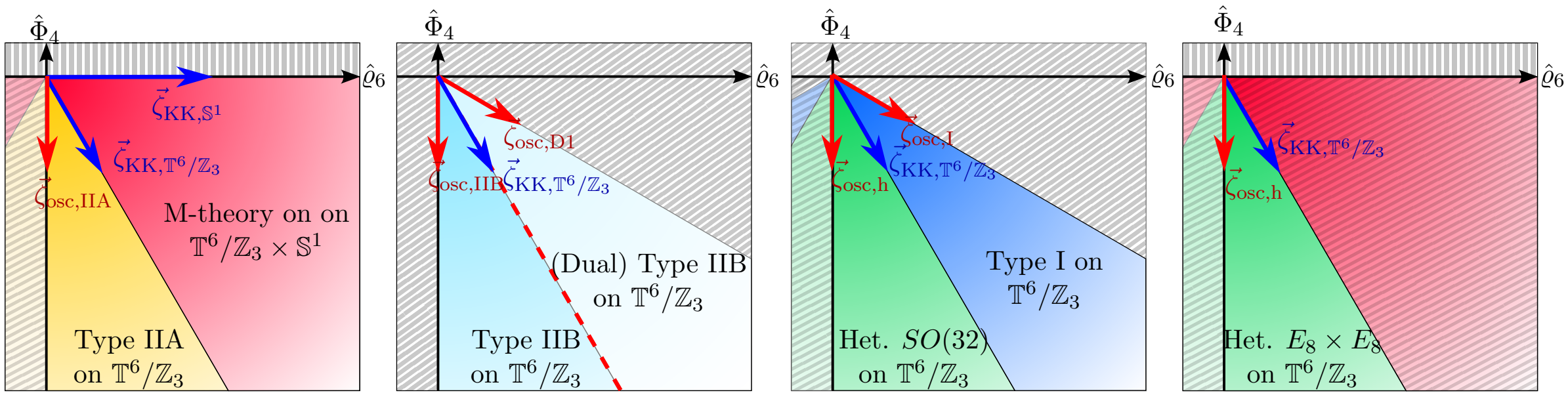}
    }
\caption{Arrangements of perturbative descriptions for Type IIA, IIB, HO and HE string theories on $X_6=\mathbb{T}^6/\mathbb{Z}_3$ along the flat Euclidean slice parameterized by the 4d dilaton $\hat{\Phi}_4$ and string units radion $\hat{\varrho}_6$. We depict in dashed lines those regions that might be obstructed, together with the scaling $\zeta$-vectors (in red for oscillator and blue for KK towers). For the Type IIB moduli space we depict in red the self-dual line under S-duality. All quantities are depicted in terms of canonically normalized moduli.
\label{fig.4dN2}}
\end{center}
\end{figure}

We can now consider filling the flat slice $\mathbb{R}^2$ parameterized by the internal 6-volume in string units and $\Phi_4$, with a $\mathbb{Z}_2$ symmetry associated to the three double T-dualities acting simultaneously on it. However, one can indeed see that the moduli space \eqref{eq. MC}, which classically allows for $\Phi_4\to\infty$, ceases to be valid upon the inclusion of quantum corrections, see Figure \ref{fig.4dN2}:
\begin{itemize}
    \item \textbf{Type IIA}: Starting with Type IIA on $X_6=\mathbb{T}^6/\mathbb{Z}_3$ \textbf{(A.1)}, the strong coupling limit corresponds to M-theory on $\mathbb{S}^1\times \mathbb{T}^6/\mathbb{Z}_3$ \textbf{(A.2)}. However, shrinking now the $\mathbb{T}^6/\mathbb{Z}_3$ does not result in an obvious perturbative description, since there are no 1-cycles on which to reduce first to Type IIA. Wrapping M2-branes on the 2-cycles does not result in additional light scalars in the $\Phi_4\to\infty$ region. On the other hand, one needs to consider Euclidean M5-branes wrapped on the whole $X_6$, which have an action scaling like
    \begin{equation}
        S_{\rm M5-inst}\sim M_{\rm Pl,11}^6\Vol(X_6)\sim\frac{\Vol(X_6)}{\ell_{\rm s}^6}\frac{\ell_{\rm s}^6}{\ell_{11}^6}\sim \frac{\Vol(X_6)}{\ell_{\rm s}^6} e^{-2\Phi_{10}}=e^{-2\Phi_4}\,,
    \end{equation}
    which goes to zero as $\Phi_4\to\infty$, and thus non-perturbative contributions $\exp(-S_{\rm M5-inst})$ to the classical K\"ahler potential \eqref{eq. kahler 4d} become of order one, making the classical \eqref{eq. MC} expression for the moduli space not reliable in this strong coupling limit.\footnote{This can also be understood from the fact that in 4d $\mathcal{N}=2$ the hypermultiplet moduli space is fibered over that of the vector multiplet, receiving quantum corrections, rather than being a direct product, and thus expression \eqref{eq. MC} needs not be reliable in all limits.} 
    \item \textbf{Type IIB}: With Type IIB on $X_6=\mathbb{T}^6/\mathbb{Z}_3$ \textbf{(B.1)} as a starting point, its strong coupling regime (as in $g_{\rm s}=e^{\Phi_{10}}\gtrsim 1$, we still have $\Phi_4\to -\infty$) is described by an S-dual Type IIB on the same $X_6=\mathbb{T}^6/\mathbb{Z}_3$ \textbf{(B.2)}.\footnote{As discussed in \cite{Baume:2019sry}, the 4d $\mathcal{N}=2$ theory inherits the S-duality from the full 10d $SL(2,\mathbb{Z})$ duality group. In our canonically normalized variables this amounts to a $\mathbb{Z}_2$ transformation
    \begin{equation}
\mathsf S:
\begin{pmatrix}\hat\varrho_6\\ \hat\Phi_4\end{pmatrix}
\longmapsto\underbrace{
\begin{pmatrix}
-\frac12&-\frac{\sqrt3}{2}\\
-\frac{\sqrt3}{2}&\frac12
\end{pmatrix}}_{M_S}
\begin{pmatrix}\hat\varrho_6\\ \hat\Phi_4\end{pmatrix}\,,\quad \text{with} \quad M_S^2={\rm Id}_2\,,\quad \det M_S=-1\,.
\end{equation}
    } Now, we could consider the small volume limit in either of the two description in order to perform six T-dualities to an analogous Type IIB description. However, following \cite{Marchesano:2019ifh,Baume:2019sry}, if we consider the Euclidean D3-instantons wrapped along 4-cycles, these become unsuppressed in the small volume limit at fixed 4d coupling:
    \begin{equation}
        S_{\rm D3-inst}\sim \mathcal{T}_{\rm D3}\Vol(X_6)^{\frac{2}{3}}\sim\ell_{\rm s}^{-4}e^{-\Phi_{10}}\Vol(X_6)^{\frac{2}{3}}\sim e^{-\Phi_4}\left(\frac{\Vol(X_6)}{\ell_{\rm s}^6}\right)^{\frac{1}{6}}\,,
    \end{equation}
    and thus makes the small volume regime before the T-dualities unreliable.
    \item \textbf{Heterotic}: The problems in completing the flat slice of moduli space remain in the 4d $\mathcal{N}=1$ theories obtained from heterotic compactifications. In the $E_8\times E_8$ case, the strong coupling regime would be given by M-theory on $\mathbb{S}^1/\mathbb{Z}_2\times \mathbb{T}^6/\mathbb{Z}_3$, which again would be obstructed in the small $\mathbb{T}^6/\mathbb{Z}_3$ volume. However, as explained in \cite{Witten:1996mz,Cvetic:2024wsj}, the large interval $\mathbb{S}^1/\mathbb{Z}_2$ limit of heterotic M-theory (at least within the standard embedding $\mathfrak{e}_8\oplus\mathfrak{e}_8\to\mathfrak{e}_6\oplus\mathfrak{su}(3)\oplus \mathfrak{e}_8$) is obstructed by the appearance of a strong coupling singularity in one of the $E_8$ gauge couplings. While it is possible that such singularity can be resolved, in general it is not clear that this results in a perturbative description, see e.g., \cite{Cvetic:2024wsj}. As for the $SO(32)$ case, its strong coupling limit is given by Type I on the same background (see \cite[Section 5.3]{Choi:2004wn} for some nuances regarding the strong coupling limit of HO orbifolds). The small volume limit of Type I is obstructed by Euclidean NS5 or D5-instantons (respectively), as
    \begin{equation}
       S_{\rm D5-inst}^{\rm (I)}=\mathcal{T}_{\rm D5,I}\Vol(X_6)=e^{-\Phi_{\rm 10,I}}\frac{\Vol(X_6)}{\ell_{\rm s,I}^6}=e^{-2\Phi_4}\sim  S_{\rm NS5-inst}^{\rm (HO)}\,,
    \end{equation}
    which goes to zero as $\Phi_4\to\infty$.
\end{itemize}
We thus conclude that, while the classical moduli space would naively allow for the completion of the flat slices \eqref{eq. flat slice 4dN2} by different perturbative regimes using the (again classical) symmetries on non-compact scalars, upon considering the effect of quantum corrections, many of the infinite distance limits along asymptotic regions can be obstructed, which in practice amounts to the possible perturbative regimes being severely constrained. Similarly, as explained in Footnote \ref{fn.blowup}, for finite sizes of the 27 $\mathbb{P}^2$ resulting from the resolution $\tilde X_6$ of the $X_6=\mathbb{T}^6/\mathbb{Z}_3$ fixed points, the small volume limit is generally not dual to a perturbative description.

The above quantum obstructions to continuing the asymptotic regions of moduli space through dualities illustrate the fact that the classical moduli space in theories with low or no supersymmetry can be corrected after quantum effects are considered, see e.g., \cite{Witten:1996qb,Mayr:1996sh,Gopakumar:1998ii,Gopakumar:1998jq,Marchesano:2019ifh,Baume:2019sry,Klaewer:2020lfg,Martucci:2022krl,Cvetic:2024wsj,Casas:2025qxz,Kaufmann:2026fli,Kaufmann:2026mha,Kaufmann:2026tsy}. For the case when the moduli $\mathcal{M}$ is a K\"ahler manifold,\footnote{
This is, $\mathcal{M}$ is a symplectic manifold endowed with a K\"ahler potential $K$ in terms of which the scalar metric $\mathsf{G}_{i\bar j}=\partial_i\partial_{\bar j}K$ and the K\"ahler form $\Omega=\frac{i}{2}\partial\bar\partial K$. While this is always the case in 4d $\mathcal{N}=1$ theories \cite{Freedman:2012zz}, it is not true for the full moduli space in 4d $\mathcal{N}=2$. In the latter case, splitting $\mathcal{M}=\mathcal{M}_{\rm V}\times\mathcal{M}_{\rm H}$ \cite{Strominger:1990pd}, the vector multiplet part $\mathcal{M}_{\rm V}$ is projective special K\"ahler, and thus admits a potential $K_{\rm V}$ \cite{Strominger:1990pd}. This is not the case for the hypermultiplet moduli space $\mathcal{M}_{\rm H}$, which is quaternionic K\"ahler and thus has no associated K\"ahler potential. When $\mathcal{M}_{\rm H}$ presents isometries an effective K\"ahler potential can be written \cite{deWit:2001brd,cortes2021symmetries}, as it is explicitly the case in some top-down constructions \cite{Alexandrov:2013yva}.}
writing the complex K\"ahler chiral scalars $t^j=a^j+i s^j$, with $a^j$ the compact axionic field and $s^j$ the non-compact saxion, the K\"ahler potential becomes approximately invariant under axionic shifts $a^j\to a^j+c$ in asymptotic regions of moduli space, with the K\"ahler potential written as
\begin{equation}
    K=-\log\left[P(s)+\mathcal{O}\big(e^{2\pi im_jt^j}\big)\right]\,,
\end{equation}
where $P(s)$ is a polynomial depending only on the saxions and $\mathcal{O}\big(e^{2\pi im_jt^j}\big)$ are non-perturbative instantonic corrections, with charges $m_j\in\mathcal{C}_I$. The perturbative regime can be thus identified with the interior of the \textit{saxionic cone}
\begin{equation}
    \Delta=\left\{s^i\in N_\mathbb{R}\,:\,m_is^i>0\;\forall\,m_i\in\mathcal{C}_I\right\}\,,
\end{equation}
with $N_\mathbb{R}$ a real vector space of the appropriate dimension. By construction, $\Delta$ is a convex set. If a flat Euclidean slice is parameterized by these saxions, then it is clear that we cannot have an arbitrarily ample fundamental domain $\mathcal{F}= \infty\big(\mathcal{W}_{G(\mathbb{Z})}(F)\backslash F\big)$ without instantonic corrections becoming important.
 Similar as in the previous case, we can find a set of symmetries of the K\"ahler cone (or a slice of it) acting on the non-compact saxions. Following a similar argument as in Section \ref{sec.sym}, along fixed loci of said symmetries, the $\zeta$-scaling vectors of the light towers (i.e., the direction of pure asymptotic limits) are left invariant, and thus point along them. One could thus then try to reconstruct or bootstrap possible duality frames within $\Delta$.

As an illustrative example, we can write the K\"ahler potential (modulo non-perturbative corrections) along the asymptotic regions of $\Delta$ as
\begin{equation}
    K\sim-\log s^0-\log \mathcal{V}(s)\,,
\end{equation} 
where $\mathcal{V}(s)$ is a polynomial on the rest of saxions. As described in \cite{Lanza:2021udy,Grieco:2026yip}, the $s^0$ (which we can call \textit{universal saxion}), controls the tension of a $\frac{1}{2}$-BPS 4d string magnetically charged with respect to the associated axion $a^0$,
\begin{equation}
    \frac{\mathcal{T}_0}{M_{\rm Pl,4}^2}\sim \frac{1}{2s^0}\sim e^{-\sqrt{2}\hat{s}^0}\to 0\quad\text{as}\; s^0\to \infty\,,
\end{equation}
where $\hat{s}^0=\frac{1}{\sqrt{2}}\log s^0$ is the canonically normalized saxion. Upon identification $s^0=e^{-2\Phi_4}$ and $a^0$ with the 4d NSNS axion, such string can be understood as a fundamental critical BPS string (note that this excludes for example the Type I string), see \cite{Grieco:2026yip}. From the perspective of the Emergent String Conjecture (ESC) \cite{Lee:2019wij,Lee:2019xtm}, the additional saxions must parameterize the internal space. We can thus interpret the orthogonal $\mathcal{V}\equiv\mathcal{V}(s)$ coordinate as controlling the volume in string units of a compact $n$-manifold $X_n$ (we do not claim $n=6$, since some directions could be stabilized in string units). Since $s^0$ and $\mathcal{V}\equiv\mathcal{V}(s)$ are orthogonal, we expect that in the associated flat coordinates the slice of the saxionic cone $\Delta$ they parameterize spans at most $\theta=\frac{\pi}{2}$ (since otherwise there would be some asymptotic directions along which one of the saxionic coordinates becomes small, with non-perturbative effects coming into effect) and is left invariant under the internal symmetries of $X_n$ (e.g., interchange of equivalent internal submanifolds).

\begin{figure}[ht]
\begin{center}
\centering
     \resizebox{\textwidth}{!}{%
    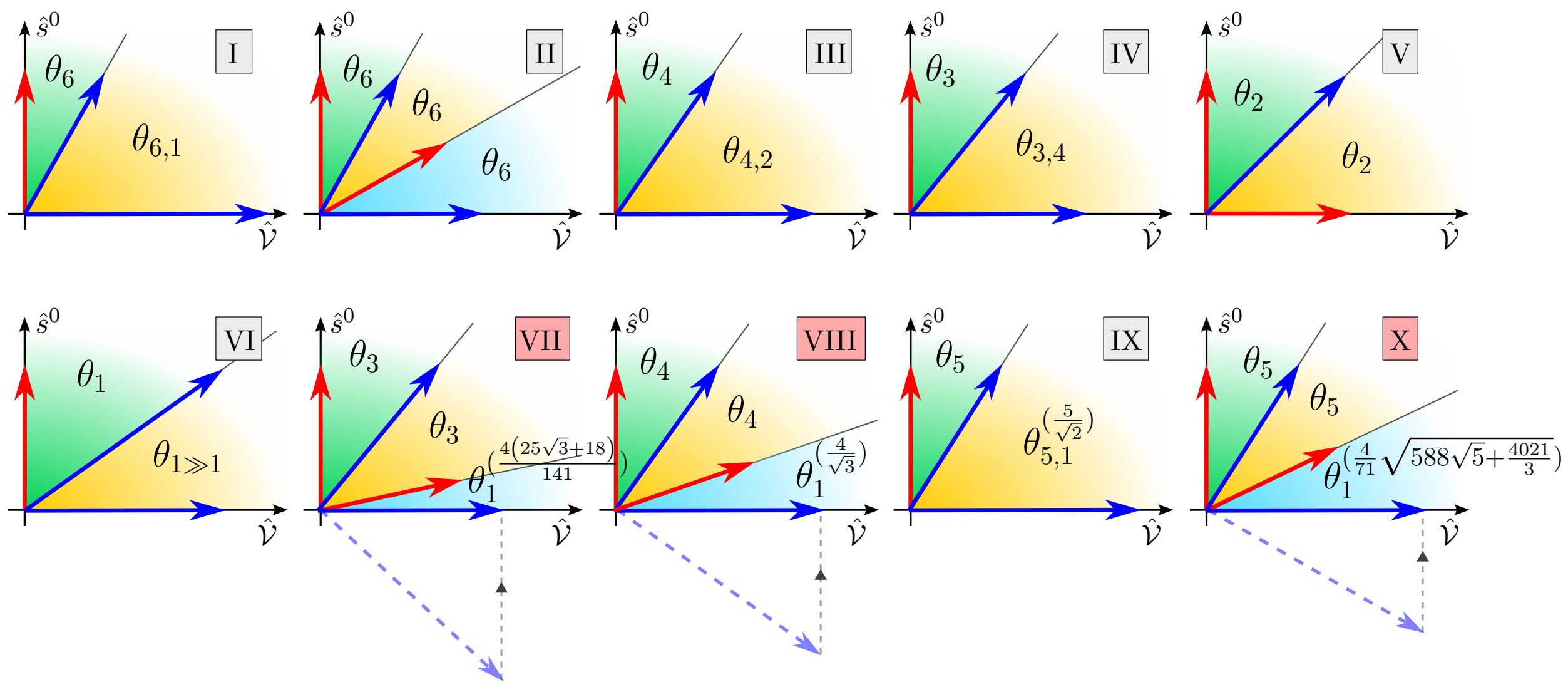
    }
\caption{Possible ten completions of a 2-dimensional K\"ahler cone spanning an angle of $\frac{\pi}{2}$, parameterized by the universal saxion $s^0$ and the remaining $\mathcal{V}(s)$ (both canonically normalized), following the possible perturbative descriptions of Section \ref{sec.Pert reg}. We use the notation from \eqref{eq. angle reg} and \eqref{eq.warping}, as well as drawing the scaling $\zeta$-vectors, whose interpretation as KK-$n$ (in red) or oscillator (in blue) towers is clear from the frame. We also depict those sliding, including the coupling $\lambda$ to the higher dimensional potential $\hat{V}(\hat\varphi)\sim V_0 \exp(-\lambda\hat{\varphi})$ causing the warped compactification. Note that for IX we do not depict the sliding, since it would occur perpendicular to the depicted slice. In red we label those exotic arrangements where an ``exotic'' T-duality exchanging KK modes associated to compact spaces of different dimensionality, see Appendix \ref{ap.extra}.
\label{fig.extra4d}}
\end{center}
\end{figure}

Following the possible perturbative descriptions of Section \ref{sec.Pert reg} and the angles that they span (see \eqref{eq. angle reg} and \eqref{eq.warping}), we depict the different 10 distinct possibilities in Figure \ref{fig.extra4d}.\footnote{As explained in Appendix \ref{ap.extra}, since the angular span for warped compactifications is always smaller than the unwarped one, it is always possible to find a coupling $\lambda$ in the higher dimensional potential $\hat{V}(\hat\varphi)\sim V_0 \exp(-\lambda\hat{\varphi})$ causing the warping such that the angular span is that of an unwarped, higher-codimension, compact space. We do not consider such cases, including only those warped backgrounds whose angular span is unique.} 
Arrangements I to VI do not feature any warped background, and as explained in \cite{Grieco:2025bjy,Grieco:2026yip}, they can all be obtained as slices of the moduli space of 4d $\mathcal{N}=1$ and 2 compactifications (to start with, I and II correspond to the compactifications already discussed and depicted in Figure \ref{fig.4dN2}).\\
More interesting to consider are the additional solutions VII to X, featuring warped decompactification limits. While for VII and X the couplings $\lambda=\frac{4\left(25 \sqrt{3}+18\right)}{141} $ and $\frac{4}{71} \sqrt{588 \sqrt{5}+\frac{4021}{3}}$ seem quite involved, and the existence of some higher dimensional source for the potential with this appropriate coupling is highly unlikely.\footnote{\label{fn.source}As explained in \cite[Appendix A]{Raucci:2026fzp}, in order for the lower-dimensional space to be Minkowski, the tension of the codimension-1 branes producing a coupling $\hat{V}(\hat\varphi)\sim V_0 \exp(-\lambda\hat{\varphi})$ must be $\mathcal{T}_{d-1}\sim M_{{\rm Pl},{d+1}}^{d}\exp(-\frac{\lambda}{2}\hat{\varphi})$.} However, the exponents $\lambda=\frac{4}{\sqrt{3}}$ and $\frac{5}{\sqrt{2}}$ are much simpler. For the latter, the possible interpretation is simple, as IX corresponds to a compactification of a 9d theory featuring a warped decompactification with $\lambda=\frac{5}{\sqrt{2}}$ on a 5-manifold. Such coupling is precisely that of the Romans mass in mIIA. This way, arrangement IX could be obtained as a heterotic compactification on a compact space that looks like $\mathbb{S}^1\times X_5$, either somehow stabilizing the $\mathbb{S}^1$ radius in string units or moving along a flat slice of moduli space where it is fixed. Similarly, arrangement VIII would require a warped background with a 5d potential $\hat{V}(\hat\Phi_5)\sim V_0 \exp(-\frac{4}{\sqrt{3}}\hat\Phi_5)$ or a 5d 3-brane with tension $\mathcal{T}_{3}\sim M_{{\rm Pl},{5}}^{4}\exp(-\frac{2}{\sqrt{3}}\hat\Phi_5)$ (see Footnote \ref{fn.source}) in terms of the 5d dilaton $\hat\Phi_5$, such as an exotic $5_4^3$-brane,\footnote{\label{fn.EXOTIC}To obtain this construction, consider a Type IIA NS5 brane, with tension $\mathcal{T}_{\rm NS5}\sim \ell_{\rm s}^{-6}e^{-2\Phi}$, wrapped along the $\mathbb{T}^2_{56}$ cycle of $\mathbb{T}^5_{56789}$. This results in a 3-brane, which upon three T-dualities along the orthogonal $\mathbb{T}^3_{789}$ to Type IIB and an S-duality, results in
\begin{equation}
    \mathcal{T}_{3}\sim\frac{\Vol(\mathbb{T}^2_{56})}{\ell_{\rm s}^6e^{2\Phi}}\xrightarrow[]{\text{T-duality along }\mathbb{T}^3_{789}}\mathcal{T}_{5^3_2}\sim\frac{\Vol(\mathbb{T}^2_{56})\Vol(\mathbb{T}^3_{789})^2}{\ell_{\rm s}^{12}e^{2\Phi}}\xrightarrow[]{\text{S-duality}}\mathcal{T}_{5^3_4}\sim\frac{\Vol(\mathbb{T}^2_{56})\Vol(\mathbb{T}^3_{789})^2}{\ell_{\rm s}^{12}e^{4\Phi}}\,,
\end{equation}
which we can write in terms of the 5d Planck mass and dilaton as:
\begin{equation}
    \mathcal{T}_{5^3_4}\sim M_{\rm Pl,5}^4e^{-\frac{4}{3}\Phi_5}\left(\tfrac{\Vol(\mathbb{T}^2_{56})}{\ell_{\rm s}^2}\right)^{-1}\sim M_{\rm Pl,5}^4e^{-\frac{2}{\sqrt{3}}\hat\Phi_5}\left(\tfrac{\Vol(\mathbb{T}^2_{56})}{\ell_{\rm s}^2}\right)^{-1}\,.
\end{equation}
Considering the modulus controlling the size in string units of the $\mathbb{T}^2_{56}$ to be orthogonal to the slice we are considering, we recover the desired tension dependence with respect to the 5d dilaton.
} see e.g., \cite{Elitzur:1997zn,deBoer:2012ma,Fernandez-Melgarejo:2018yxq,Etheredge:2025ahf}.\\
In any case, all the duality frame arrangements from Figure \ref{fig.extra4d} other than IX with warped compactifications feature what we denote as ``exotic'' T-dualities, this is, duality transformations where the string scale is left invariant but that exchange KK momenta/winding modes associated to compact spaces of different dimensionalities. As we argue in Appendix \ref{ap.extra}, such arrangements seem in tension with the Emergent String Conjecture, see also \cite{Etheredge:2024tok,Etheredge:2025ahf}.

\section{Conclusions\label{sec.conc}}

In this work, we have studied how the symmetries of moduli spaces $\mathcal{M}$ of effective theories of Quantum Gravity, together with the Emergent String Conjecture, severely constrain the possible arrangements of perturbative descriptions/duality frames in the asymptotic regions of $\mathcal{M}$. By means of these symmetries, knowing only the perturbative description of some weakly coupled limit, we are able to recover qualitative properties of the new dual theories at strong coupling. In that sense, this work can be seen as a continuation of \cite{Etheredge:2024tok}, with the additional requirement that the arrangements of light towers and infinite distance limits respect the symmetries of $\mathcal{M}$, as well as naturally allowing for warped decompactification limits and the sliding of the scaling $\zeta$-vectors (as described in \cite{Etheredge:2023odp,Raucci:2026fzp}), which had not been previously considered in classification efforts.

Our approach is best suited for symmetric spaces of the form $\mathcal{M}=\Gamma\backslash G/K$, which allows the natural identification of flat slices associated to the non-compact moduli and the inherited symmetries acting on them. In highly supersymmetric theories with 32 and 16 supercharges, we pursue our approach, recovering that the only natural possibilities are those precisely realized by top-down toroidal compactifications of M-theory or Type II/heterotic string theories. In the latter case, it is interesting to notice that, even though $SO(32)$ and $E_8\times E_8$ heterotic theories feature very different arrangements, both have equivalent perturbative descriptions for the $\Phi_d\to\pm\infty$ limits along the self-dual line. This is indeed recovered from our reconstruction algorithm, without having to impose it as a constraint. Other possible descriptions that could fit in our arrangement, see Appendix \ref{ap.extra}, would involve warped dimension-1 space with sources without a clear physical interpretation, together with exotic generalizations of T-duality which are probably inconsistent with Swampland conjectures.

For theories with lower number of supercharges, such as bosonic string theory or CY compactifications to 4d $\mathcal{N}=2$ or 1, the classical symmetries of moduli space do not survive after quantum effects and corrections are included. While this reduces the range of applicability to the well controlled regions inside the K\"ahler cone, we are still able to propose consistent duality arrangements within the slices parameterized by scalar directions interpreted as the 4d dilaton and some internal volume, see Figure \ref{fig.extra4d}. After setting aside the three solutions with ``exotic'' T-dualities, all but one of such consistent arrangements are indeed recovered by top down CY or $G_2$ constructions, see \cite{Grieco:2025bjy,Grieco:2026yip}. It would be interesting to study the feasibility of the additional one (IX in \ref{fig.extra4d}), likely involving the compactification of the warped strong coupling regime of heterotic string theory on $\mathbb{S}^1$ discussed in Section \ref{sec.het S1}. Since such geometry is not realized by the asymptotic fibrations of CY or $G_2$-manifolds, \cite{Oguiso1993ONAF,Grimm:2019bey,Liu:1998tha}, a possible fertile ground could be heterotic flux compactifications, where it is known that some internal cycles are fixed at string scale, see e.g., \cite{Melnikov:2014ywa,Evslin:2008zm,Israel:2023tjw,Israel:2026bco,NachoTBA}.\\

There are several natural directions that we can take our work in the future. First of all, we have systematically been working only on the slices $F$ of moduli space parameterized by non-compact scalars. As we know, the symmetries of the full duality group $\Gamma$ acting on $F$ correspond only to a small subset. It would be interesting to try to generalize our results considering additional compact directions, and how continuously varying their value affects the asymptotic properties of the perturbative description. In simple examples, see e.g., \cite[Section 4]{Etheredge:2022opl} for 9d maximal SUGRA, the presence of axions or compact scalars result in additional symmetries of light tower arrangement. One should be able to extend this to more involved compact scalar structures.

Similarly, while we have considered the possibility of warped decompactifications, we have only used very particular geometries, see Footnote \ref{fn.warping}, where the warped 1-dimensional submanifold appears as a direct product with the rest of the internal geometry. There are more involved warped geometries one can consider, such as the ``sausage geometry'' discussed in \cite[Section 2]{Morrison:1996pp} and Appendix \ref{app. F-theory} for the $\mathbb{P}^1$ base of the elliptically fibered K3 describing the strong coupling regime of $E_8\times E_8$ heterotic on $\mathbb{T}^2$. As discussed in Section \ref{sec. het t2} and Appendix \ref{app. Details Het T2}, the range of validity of the associated perturbative regime does not correspond with those discussed in Section \ref{sec.Pert reg}. It would be interesting to perform a more general analysis of warped compactifications outside those studied in \cite{Raucci:2026fzp}, so as to understand better the associated light towers and ranges of validity.

In disregarding as possibly spurious some arrangements, we have argued in Appendix \ref{ap.extra} against ``exotic T-dualities'' interchanging KK/winding states associated to different number of internal dimensions. Since our rationale is rather rough and qualitative, it would be very interesting to try to develop a more detailed and strong argument against this type of dualities. Similarly, it would be interesting to perform a comprehensive study of the scalar dependence in the tension of codimension-one branes causing the internal warping, see \cite[Appendix 1]{Raucci:2022jgw}, possibly in a similar spirit to \cite{Etheredge:2025ahf}.

Finally, in the 4d $\mathcal{N}=1$ and 2 analysis from Section \ref{sec. low}, we have restricted ourselves to a given K\"ahler chamber. As discussed in \cite{Lukas:2022crp,Alvarez-Garcia:2026uaq,Alvarez-Garcia:2026vwq}, the extended K\"ahler cone (this is, the union of all K\"ahler chambers related by contractions and blow-ups of curves of the Calabi-Yau that do not involve the collapse of a divisor) enjoys a rich symmetry structure in terms of Coxeter groups, so one could wonder about the extension of our analysis to such spaces. Indeed, one can perform a finite classification of the possible spans of K\"ahler chambers and extended K\"ahler cones, and then study the possible duality frames within it, see \cite{TBANaomiLucaIreneNacho}.

Ultimately, our work has aimed to go one step further in the quest of \textit{string universality} within the Swampland program: which EFTs with a consistent completion to quantum gravity are allowed, and can they always be realized as the low-energy description of string/M-theory? Similar to \cite{Grieco:2026yip}, this paper seems to point in that precise direction: All arrangements of perturbative descriptions/duality frames consistent with the Emergent String Conjecture and the symmetries of moduli space seem to be precisely those appearing from top-down constructions, at least in theories with enough supersymmetry. Further understanding the properties and possible dualities of moduli spaces in theories with low supersymmetry would be the desired next step, one that brings us closer to a better understanding of our own Universe.

\section*{Acknowledgments}

We are grateful to Rafael \'Alvarez-Garc\'ia, Veronica Collazuol, Muldrow Etheredge, Bernardo Fraiman, Mariana Grana, Lukas Kaufmann, Jeroen Monnee, Salvatore Raucci, Pablo Soler and Timo Weigand for very illuminating discussions and comments, and especially to Irene Valenzuela for collaboration in the initial stages of the project. The authors thank IFT UAM-CSIC for their hospitality during part of the project. The work of G.V. is supported by the Spanish FPU fellowship No. FPU24-02963 and by the grants from the Spanish government MCIU22-PID2021-123021NB-I00 and MCIU-25-PID2024-161500NB-I00. The work of I.R. is supported by the European Union through ERC Starting Grant SymQuaG101163591 StG-2024, and he acknowledges the additional funding of the Spanish FPI grant No. PRE2020-094163 and the ERC Starting Grant QGuide101042568 - StG 2021. 
 Funded by the European Union. Views and opinions expressed are however those of the authors only and do not necessarily reflect those of the European Union or the European Research Council. Neither the European Union nor the granting authority can be held responsible for them.

\begin{appendix}

\section{More details on the dual geometries of heterotic string theory on \texorpdfstring{$\T^2$}{T2}\label{app. Details Het T2}} 

In this appendix we will examine the heterotic compactifications on $\T^2\simeq\mathbb{S}^1_x\times\mathbb{S}^1_y$ case in more detail. In particular, we detail the computation of the different ranges of validity of the perturbative regions and the towers of states involved. 
As already discussed, the flat slice $F = \R^3$ is parametrized by the (canonically normalized, following the conventions from Section \ref{sec.Pert reg}) lower-dimensional dilaton $\hat\Phi_8$ and the two radions $\hat\varrho_x$ and $\hat\varrho_y$ (measuring the radii in string units), and it is symmetric under the action of $D_4$, which is generated by T-duality and the exchange of the two radii. These transformations relate the four heterotic duality frames associated with the compactification on $\T^2$, and the fixed locus of the full symmetry group is the self-dual line, which corresponds to the direction associated with the heterotic string oscillator tower for $\hat{\Phi}_8\to-\infty$.

\subsection{Ranges of validity of the perturbative regions}

First of all, we explicitly write the relations between the physical parameters of the original heterotic theory (either $SO(32)$ or $\EE$) and the moduli that parametrize the flat slice ($\hat{\Phi}_8$, $\hat\varrho_x$ and $\hat\varrho_y$):\footnote{
Recall that, since $\ell_{\rm s}=\ell_{d}e^{-\frac{2}{d-2}\Phi_d}=\ell_{d}e^{-\frac{1}{\sqrt{d-2}}\hat\Phi_d}$, the heterotic string length in 10d Planck units is related to the heterotic string coupling by $\ell_\h = \ell_{10} g_\h^{-1/4}$.}  
\begin{equation}
\begin{split}
g_\h =e^{\Phi_{10}} = \exp\left\{ \sfrac{3}{2} \, \hat{\Phi}_8 
+ \half \, \hat\varrho_x + \half \, \hat\varrho_y \right\} \, ,\quad \Vol(\mathbb{S}^1_x)\ell_{\rm h}^{-1} = e^{\hat\varrho_x}\, ,\quad \Vol(\mathbb{S}^1_y)\ell_{\rm h}^{-1} = e^{\hat\varrho_y}\,.
\end{split}
\end{equation}
The first perturbative region (corresponding to the weak coupling and large volume regime of the original heterotic on $\mathbb{T}^2$ description) is thus obtained (by analogy with \cite{Aharony:2007du,Etheredge:2023odp}) by imposing weak coupling and large volume: 
\begin{equation} \label{eq. pert regime 1}
g_\h \ll 1 \, , \quad
\Vol(\mathbb{S}^1_x) \gg \ell_\h \, , \quad
\Vol(\mathbb{S}^1_y) \gg \ell_\h \, ,
\end{equation}
valid for either $SO(32)$ or $\EE$ on $\mathbb{T}^2$. The remaining three heterotic regions are related to this one by T-duality transformations \eqref{eq. Tdual}.%
\footnote{Since the torus has two independent radii, we can perform two distinct \mbox{T-duality} transformations—one for each radius—and a third one given by the combination of the previous ones (regardless of the order).} 
To continue with the computation of the ranges of validity we must distinguish between the two heterotic theories, as they exhibit different duality relations.

\subsubsection*{The $SO(32)$ heterotic string}

We start by recalling the duality chain, which is summarized in the following diagram: 
\begin{center}
\begin{tikzpicture}[
  node distance=1.5cm,
  theory/.style={
    draw=black!25,
    line width=0.5pt,
    rounded corners=5pt,
    inner xsep=9pt,
    inner ysep=3pt,
    align=center
  },
  arrow/.style={
    <->, thick, shorten >=1pt, shorten <=1pt
  }
]

\node[theory, fill=hetGREEN] (het)
  {$SO(32)$ heterotic on $\T^2$};

\node[theory, fill=iBLUE, text=white, below of=het] (typeI)
  {Type I on $\T^2$ \, $\equiv$ \, Type IIB on $\T^2$ + O9$^-$ + 32 D9};

\node[theory, fill=IIA, below of=typeI] (typeIp)
  {Type $\Ip$ on $\S^1$ \, $\equiv$ \, Type IIA on $\S^1/\Z_2 \times \S^1$ + 2 O8$^-$ + 32 D8};

\node[theory, fill=iibBLUE, below of=typeIp] (fth)
  {F-theory on $\mathbb{E} \hookrightarrow \rm{K}3 \ra \P^1$ \, $\equiv$ \, Type IIB on $\T^2/\Z_2$ + 4 O7$^-$ + 32 D7};

\draw[arrow] (het)    -- node[right=5pt, font=\itshape] {S-duality} (typeI);
\draw[arrow] (typeI)  -- node[right=5pt, font=\itshape] {T-duality} (typeIp);
\draw[arrow] (typeIp) -- node[right=5pt, font=\itshape] {T-duality} (fth);

\end{tikzpicture}
\end{center}

Applying an S-duality \eqref{eq. Sdual} maps the heterotic description to Type I on $\T^2$, whose perturbative regime is analogous to the heterotic one in \eqref{eq. pert regime 1}. 
It is for the T-dual (consider, e.g., with respect to $\S^1_x$) of this theory, type $\Ip$ on $\S^1_y \equiv$ (massive) Type IIA on $\S^1_x/\Z_2 \times S^1_y$, that the perturbative conditions are slightly different. 
In this description, the string coupling increases linearly away from the brane stack, see Figure \ref{fig. geometries}. To remain in the perturbative regime, we therefore require the product of the string coupling and the length of the interval to be small compared with the string length. The resulting conditions are 
\begin{equation} 
g_{\Ip} \Vol(\mathbb{S}^1/\mathbb{Z}_2) \ll \ell_{\Ip} \, , \quad
\Vol(\mathbb{S}^1_x/\mathbb{Z}_2) \gg \ell_{\Ip} \, , \quad
\Vol(\mathbb{S}^1_y) \gg \ell_{\Ip} \, .
\end{equation}
T-duality along the remaining circle maps this description to a Type IIB orientifold compactification on $\T^2/\Z_2 \simeq \P^1$. The string coupling and the compactification volume in this duality frame are related to those in the heterotic description by (see e.g., \cite[Section 4]{Raucci:2026fzp})
\begin{equation}
\Vol(\mathbb{P}^1)=\ell_{10}^2e^{\Phi_8}=\ell_{10}^2e^{\sqrt{\frac{3}{2}}\hat\Phi_8}\,,\quad g_{\rm IIB}=e^{\Phi_{10,\rm IIB}}=(\Vol(\mathbb{T}^2)\ell_{\rm h}^{-2})^{-1}\,.
\end{equation}
The perturbative regime is thus characterized by large volume for the cycles of the torus and the base of K3: 
\begin{equation}
\Vol(\mathbb{S}^1_x) \gg \ell_\h \, , \qquad
\Vol(\mathbb{S}^1_y) \gg \ell_\h \, , \qquad
\Vol(\mathbb{S}^1_x/\mathbb{Z}_2) \gg \ell_\IIB \, , \qquad
\Vol(\mathbb{S}^1_y/\mathbb{Z}_2) \gg \ell_\IIB \, ,
\end{equation}
which correspond to weak coupling and large volume in the Type IIB orientifold description. 
In Table \ref{tab. ranges canonical} we summarize the ranges of validity in terms of the canonically normalized moduli.

\subsubsection*{The $\EE$ heterotic string}

On the other hand, the duality chain for the $\EE$ heterotic theory is summarized in the following diagram: 
%
\begin{center}
\begin{tikzpicture}[
  node distance=1.5cm,
  theory/.style={
    draw=black!25,
    line width=0.5pt,
    rounded corners=5pt,
    inner xsep=9pt,
    inner ysep=3pt,
    align=center
  },
  arrow/.style={
    <->, thick, shorten >=1pt, shorten <=1pt
  }
]

\node[theory, fill=hetGREEN] (het)
  {$\EE$ heterotic on $\T^2$};

\node[theory, fill=mthRED, below of=het] (typeI)
  {M-theory on $\S^1/\Z_2 \times \T^2$};

\node[theory, fill=IIA, below of=typeI] (typeIp)
  {Type $\Ip$ on $\S^1$ \, $\equiv$ \, Type IIA on $\S^1/\Z_2 \times \S^1$ + 2 O8$^-$ + 32 D8};

\node[theory, fill=iibBLUE, below of=typeIp] (fth)
  {F-theory on $\mathbb{E} \hookrightarrow \rm{K}3 \ra \P^1$};

\draw[arrow] (het)    -- node[right=5pt, font=\itshape] {Grow a dimension} (typeI);
\draw[arrow] (typeI)  -- node[right=5pt, font=\itshape] {Reduce on $\S^1$} (typeIp);
\draw[arrow] (typeIp) -- node[right=5pt, font=\itshape] {T-duality} (fth);

\end{tikzpicture}
\end{center}
Starting from the heterotic perturbative regime, we first determine the corresponding M-theory moduli. The length of the interval in the 11th-dimension is related to the heterotic string coupling by
\begin{equation}
\Vol(\S^1/\Z_2) = g_\h^{2/3} \ell_{11} \quad \Longrightarrow \quad
\ell_{11} = g_\h^{1/12} \ell_{10} \, .
\end{equation}
The region is thus obtained by requiring all the cycles to be large compared to the 
11-dimensional Planck length as explained in \eqref{eq. cond planck}: 
\begin{equation}
\Vol(\S^1/\Z_2) \, , \Vol(\S^1_x) \, , \Vol(\S^1_y) \gg \ell_{11} \, .
\end{equation}
After reducing on one of the circles (e.g., $\S^1_x$), the appropriate description is given by Type IIA on $\S^1/\Z_2 \times \S^1_y$ (type $\Ip$ on $\S^1_y$), with string coupling and string length given by
\begin{equation}
g_{\Ip} = \left( \Vol( \S^1_x ) \, \ell_{11}^{-1} \right)^{3/2} \, , \qquad
\ell_\Ip = g_{\Ip}^{-1/4} \ell_{10} \, .
\end{equation}
The perturbative regime is characterized by conditions analogous to those derived for the circle compactification in \cite{Aharony:2007du,Etheredge:2023odp}, namely, 
\begin{equation} \label{eq. Type Ip pert E8}
g_\Ip \ll 1 \, , \quad
g_\Ip \Vol(\S^1/\Z_2) \gg \ell_\Ip \, , \quad
\Vol(\S^1_y) \gg \ell_\Ip \, .
\end{equation}
Finally, T-duality along the remaining circle leads to the F-theory description. Since this duality frame is reached by first reducing M-theory along the circle $\S^1_x$ and then T-dualizing along the circle $\S^1_y$, we impose the hierarchy 
$\Vol(\S^1_x) \ll \Vol(\S^1_y)$. Requiring in addition that the volume of the base $\P^1$ is large, we obtain the following conditions: 
\begin{equation}
\Vol(\S^1/\Z_2) \gg \ell_{\rm IIB} \, , \quad
\Vol(\S^1_y/\Z_2) \gg \ell_{\rm IIB} \, , \quad
\Vol(\S^1_y) \gg \Vol(\S^1_x) \gg \ell_\h \, .
\end{equation}
Note that the condition $\Vol(\S^1/\Z_2) \gg \ell_{\rm IIB}$ is in fact redundant given the other requirements of the Type IIB (i.e., F-theory) frame. Since the T-duality relating type $\Ip$ to Type IIB acts along $\S^1_y$ and preserves the string length, one has $\ell_{\rm IIB} = \ell_\Ip$, while $\Vol(\S^1/\Z_2)$ is a spectator of this duality. The condition then follows directly from the first two perturbativity requirements for the type $\Ip$ regime given in \eqref{eq. Type Ip pert E8}, which together imply 
\begin{equation}
    \Vol(\S^1/\Z_2) \gg \frac{\ell_\Ip}{g_\Ip} \gg \ell_\Ip = \ell_{\rm IIB} \, .
\end{equation}
Moreover, note that for $SO(32)$ the base of the F-theory fibration is obtained as $\P^1 \simeq ( \S^1_x \times \S^1_y ) / \Z_2$, while for $\EE$ the base is related to the M-theory interval instead of $\S^1_x$: $\P^1 \simeq ( \S^1 \times \S^1_y ) / \Z_2$. This results in two different F-theory regions for each $\EE$ heterotic region, in contrast with the $SO(32)$ case. 

In Table \ref{tab. ranges canonical} we summarize the ranges of validity in terms of the canonically normalized moduli.

\begin{table}[ht]
\centering
\small
\renewcommand{\arraystretch}{1.6}
\begin{tabular}{|c|r|r|}
\hline
\multirow{2}{*}{\textbf{Duality frame}}
  & \multicolumn{2}{c|}{\textbf{Range of validity}} \\
\cline{2-3}
  & \multicolumn{1}{c|}{\textbf{SO(32)}}
  & \multicolumn{1}{c|}{$\boldsymbol{E_8 \times E_8}$} \\
\hline
\multirow{3}{*}{\shortstack[c]{Heterotic on $\mathbb{T}^2$}}
  & \multicolumn{2}{c|}{$\sqrt{6}\,\hat{\Phi}_8 + \hat\varrho_x + \hat\varrho_y \leq 0$} \\
  & \multicolumn{2}{c|}{$\hat\varrho_x \geq 0$} \\
  & \multicolumn{2}{c|}{$\hat\varrho_y \geq 0$} \\
\hline
\multirow{3}{*}{\shortstack[c]{Type I on $\mathbb{T}^2$/ \\ M-theory on $\mathbb{S}^1/\mathbb{Z}_2\times\mathbb{T}^2$}}
  & $\sqrt{6}\,\hat{\Phi}_8 + \hat\varrho_x + \hat\varrho_y \geq 0$
  & $\sqrt{6}\,\hat{\Phi}_8 + \hat\varrho_x + \hat\varrho_y \geq 0$ \\
  & $\sqrt{6}\,\hat{\Phi}_8 - 3\hat\varrho_x + \hat\varrho_y \leq 0$
  & $\sqrt{6}\,\hat{\Phi}_8 - 5\hat\varrho_x + \hat\varrho_y \leq 0$ \\
  & $\sqrt{6}\,\hat{\Phi}_8 + \hat\varrho_x - 3\hat\varrho_y \leq 0$
  & $\sqrt{6}\,\hat{\Phi}_8 + \hat\varrho_x - 5\hat\varrho_y \leq 0$ \\
\hline
\multirow{3}{*}{\shortstack[c]{Type IIA on\\$\mathbb{S}^1/\mathbb{Z}_2\times\mathbb{S}^1$}}
  & $\hat\varrho_x \geq 0$
  & $\hat\varrho_x \geq 0$ \\
  & $\sqrt{6}\,\hat{\Phi}_8 - 3\hat\varrho_x + \hat\varrho_y \geq 0$
  & $\sqrt{6}\,\hat{\Phi}_8 - 5\hat\varrho_x + \hat\varrho_y \geq 0$ \\
  & $\sqrt{6}\,\hat{\Phi}_8 + \hat\varrho_x - 3\hat\varrho_y \leq 0$
  & $\sqrt{6}\,\hat{\Phi}_8 - \hat\varrho_x - 3\hat\varrho_y \leq 0$ \\
\hline
\multirow{4}{*}{\shortstack[c]{F-theory on\\$\mathbb{E}\hookrightarrow{\rm K3}\to\mathbb{P}^1$}}
  & $\hat\varrho_x \geq 0$
  & $\hat\varrho_x \geq 0$ \\
  & $\hat\varrho_y \geq 0$
  & $\hat\varrho_x \leq \hat\varrho_y$ \\
  & $\sqrt{6}\,\hat{\Phi}_8 - 3\hat\varrho_x + \hat\varrho_y \geq 0$
  & $\sqrt{6}\,\hat{\Phi}_8 - \hat\varrho_x - 3\hat\varrho_y \geq 0$ \\
  & $\color{Gray}\sqrt{6}\,\hat{\Phi}_8 + \hat\varrho_x - 3\hat\varrho_y \geq 0$
  &  \\
\hline
\end{tabular}
\caption{Ranges of validity of the four perturbative regimes in terms of the canonically normalized moduli $\hat{\Phi}_8$, $\hat\varrho_x$, $\hat\varrho_y$, for the $SO(32)$ and $\EE$ heterotic strings. Note that the last two inequalities in the $SO(32)$ case are equivalent under $\hat\varrho_x\leftrightarrow\hat\varrho_y$.}
\label{tab. ranges canonical}
\end{table}

\subsection{Towers of states and convex hulls}
In this section we will give the explicit expressions for the different scaling $\zeta$-vectors associated to the towers of states appearing in the asymptotic limits of the heterotic on $\mathbb{T}^2$ moduli space, together with the convex hull (see \cite{Calderon-Infante:2020dhm,Etheredge:2022opl,Calderon-Infante:2023ler,Etheredge:2024tok}) generated by such vectors. 

\subsubsection*{The $SO(32)$ heterotic string}

In this section we further compute the towers of states predicted by the SDC; in particular, we compute the charge-to-mass ratio vectors defined in \eqref{eq. zeta vector}. 
From \eqref{eq. osc and KK towers}, we obtain the following vectors corresponding to the heterotic regime: 
\begin{subequations} \label{eq4:ctmH}
\begin{align}
\ctm_{\osc,\h} = \left( - \inv{\sqrt{6}} , 0, 0 \right) \, , \quad
\ctm_{\KK, \S^1_x} & = \left( - \inv{\sqrt{6}} , 1, 0 \right) \, , \quad
\ctm_{\win, \h, \, \S^1_x} = \left( - \inv{\sqrt{6}} , -1, 0 \right) \, , \\
\ctm_{\KK, \S^1_y} & = \left( - \inv{\sqrt{6}} , 0, 1 \right) \, , \quad
\ctm_{\win, \h, \, \S^1_y} = \left( - \inv{\sqrt{6}} , 0, -1 \right) \, , \quad
\end{align}
\end{subequations}
which correspond to the tower of fundamental heterotic string excitations, together with the Kaluza--Klein and winding\footnote{
We denote the winding towers by $\win,\h$, since they depend not only on the circle but also on the type of string wrapped around it, which in this case is the heterotic string. 
} towers associated with the two circles of the torus. 
The modulus of the string oscillator tower is $\inv{\sqrt{6}}$, 
while that of the KK and winding towers is $\sfrac{7}{6}$, in agreement with expectations, \cite{Etheredge:2022opl}. 
We can also obtain the KK and winding modes associated with the decompactification of the whole torus (which correspond to a linear combination of the modes of the circles): 
\begin{equation}
\ctm_{\KK, \T^2} = \left( - \inv{\sqrt{6}} , \half, \half \right) \, , \qquad
\ctm_{\win, \h, \, \T^2} = \left( - \inv{\sqrt{6}} , - \half, - \half \right) \, ,
\end{equation}
whose length is $\sfrac{2}{3}$, as expected from \eqref{eq. univ scalings}. The heterotic string is BPS, and the relevant KK and winding towers contain BPS states with protected masses along moduli space. At the same time, the oscillator scale is universally controlled by the string tension, so the above relations are expected to hold along all asymptotic limits.
In a similar way, the KK tower obtained from decompactifying the pillowcase base in the Type IIB/F-theory description is given by
\begin{equation} \label{eq. KKpillow}
\frac{m_{\KK,\P^1}}{\Mpl{8}} \sim \left(\frac{\Vol({\P^1})}{\ell_{10}^2}\right)^{-2/3} 
= e^{ - \frac{2}{3} \Phi_{8}} 
= \left( \frac{m_{\osc,\h}}{\Mpl{8}} \right)^{-2} 
\rasp
\ctm_{\KK, \P^1} = - 2 \, \ctm_{\osc, \h} \, .
\end{equation}

For the $SO(32)$ heterotic case we can obtain the oscillator towers corresponding to the Type I string. This yields four independent towers, one for each of the four Type I theories (labeled as I, $\I_x$, $\I_y$ and $\I_{xy}$ depending on the heterotic theory they come from: the original or the T-dual with respect to $\S^1_x$, $\S^1_y$, or $\T^2$ respectively): 
\begin{subequations} \label{eq4:ctmI}
\begin{align}
\ctm_{\osc,\I} & = \left( \inv{2\sqrt{6}} , \inv{4} , \inv{4} \right) \, ,  
&& \ctm_{\osc,\I_x} = \left( \inv{2\sqrt{6}} , - \inv{4} , \inv{4} \right) \, , \\
\ctm_{\osc,\I_y} & = \left( \inv{2\sqrt{6}} , \inv{4} , -\inv{4} \right) \, , 
&& \ctm_{\osc,\I_{xy}} = \left( \inv{2\sqrt{6}} , - \inv{4} , - \inv{4} \right) \, .
\end{align}
\end{subequations}
From \eqref{eq4:ctmH} and \eqref{eq4:ctmI}, the $\ctm$-vectors are manifestly symmetric under exchange of radii (it just corresponds to swapping the labels $x \leftrightarrow y$), reflecting the symmetry under $\Z_2 \subset D_4$. 

Furthermore, we can generalize the sliding KK tower of type $\Ip$ obtained in \eqref{eq.sliding} by reducing it to 8d, resulting in a vector in the \mbox{three-dimensional} moduli space. 
We will only compute the towers associated with trajectories lying within one of the four T-dual-symmetric heterotic regions of the moduli space and its corresponding dual descriptions. The analysis in the remaining three regions is completely analogous and can be obtained by the action of the $D_4$ symmetry. 
For the two type $\Ip$ on $\S^1$ $\equiv$ (massive) Type IIA on $\S^1/\Z_2 \times \S^1$ regions obtained from the original heterotic the sliding KK tower of the interval is given by: 
\begin{equation}
\ctm_{ \KK, \S^1_x/\Z_2 }
= \left( \sqrt{\frac{2}{3}} , \, \half , \,
- \frac{ 3 \sqrt{1 - e^{- 4 \hat{\varrho}_y}} }{ 4 + 2 \sqrt{1 - e^{- 4 \hat{\varrho}_y}} }
\right) \, , \quad
\ctm_{ \KK, \S^1_y/\Z_2 } 
= \left( \sqrt{\frac{2}{3}} , \,
- \frac{ 3 \sqrt{1 - e^{- 4 \hat{\varrho}_x}} }{ 4 + 2 \sqrt{1 - e^{- 4 \hat{\varrho}_x}} } , \, \half
\right) \, .
\end{equation}
see Figure \ref{fig. slice 3d T2}. From this expression, we can obtain the unwarped and highly warped limits when $\hat{\varrho}_{x,y} \rightarrow \infty,0$, respectively: 
\begin{subequations}
\begin{align}
\ctm_{ \KK, \S^1_x/\Z_2 }^{\text{unw}} & = \left( \sqrt{\frac{2}{3}} , \, \half , \, - \half
\right) \, , \qquad
\ctm_{ \KK, \S^1_x/\Z_2 }^{\text{war}}   = \left( \sqrt{\frac{2}{3}} , \, \half , \, 0
\right) \, , \\
\ctm_{ \KK, \S^1_y/\Z_2 }^{\text{unw}} & = \left( \sqrt{\frac{2}{3}} , \, - \half , \, \half
\right) \, , \qquad
\ctm_{ \KK, \S^1_y/\Z_2 }^{\text{war}}   = \left( \sqrt{\frac{2}{3}} , \, 0 , \, \half
\right) \, .
\end{align}
\end{subequations}
The norm of the unwarped vectors is $\sfrac{7}{6}$, as expected, while for the highly warped vector we find that its norm is $\frac{\sqrt{11}}{2\sqrt{3}} < \sqrt{\frac{7}{6}}$; again, this is smaller than the expected value, but still larger than the lower bound given by $\a_{\min} = \frac{1}{\sqrt{d-2}}=\frac{1}{\sqrt{6}}$.

\begin{figure}[ht]
    \begin{center}
\centering
     \resizebox{\textwidth}{!}{%
    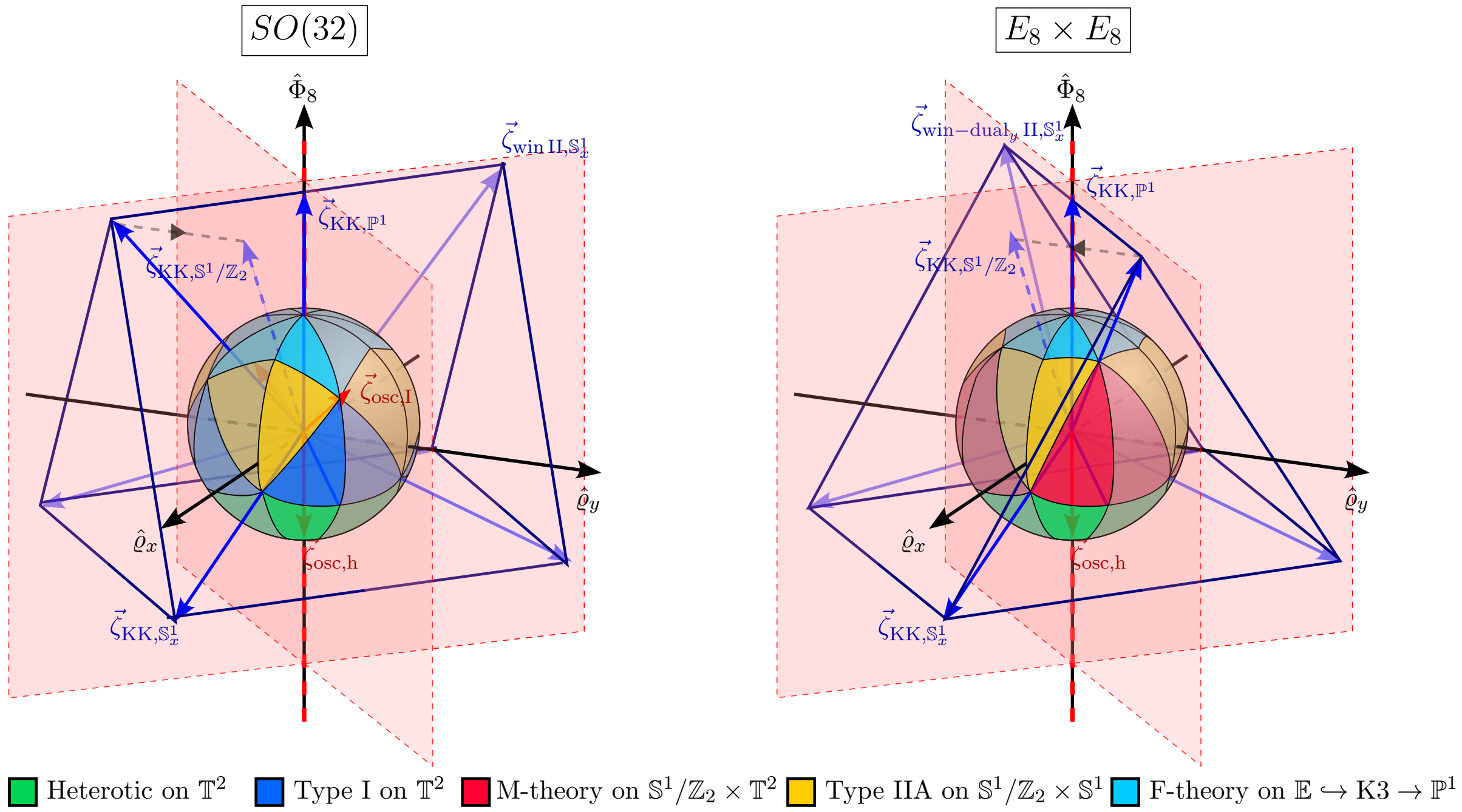
    }
    \caption{Convex hulls generated by the light towers in the interior of one of the fundamental domains corresponding to the original heterotic frame, where the internal spaces become unwarped as one decompactifies. The system is symmetric under $\Z_2^2 \times S_2 \simeq D_4$ and the convex hull is the same as that constructed from the taxonomy rules of \cite{Etheredge:2024tok}. We also depict the $\zeta$-vectors associate to the different light towers (in red for oscillator and blue for KK modes). All quantities are depicted in terms of canonically normalized moduli.
    }
    \label{fig. hulls}%
    \end{center}
\end{figure}

The previous towers were a natural generalization of those arising from the circle compactification \cite{Etheredge:2023odp}, however, the torus compactification introduces additional towers. 
Note that, since there is type $\Ip$ compactified on a circle, the corresponding winding modes are allowed—unlike in the interval case. These winding modes correspond to strings wrapping the cylinder geometry shown in the first diagram of Figure~\ref{fig. geometries}. 
Nevertheless, given that the dilaton grows as we move from the orientifold with the brane stack to the other one, the lightest winding modes are localized at the former. 
However, considering that the type $\Ip$ string is the same as the one in Type I---which is non-BPS---these winding modes are also non-BPS. 
Thus, we can wrap this Type II string (we label it as Type II since it refers to compactifications of Type I or type $\Ip$, which are obtained as orientifolds of Type IIB or Type IIA) around the circles $\S^1_x$ and $\S^1_y$, respectively, resulting in the following $\ctm$-vectors: 
\begin{equation}
\ctm_{\win,{\rm II},\, \S^1_x} = 
\left( \sqrt{\frac{2}{3}} , \, - \half , \, \half \right) \, , \qquad
\ctm_{\win,{\rm II},\, \S^1_y} = 
\left( \sqrt{\frac{2}{3}} , \, \half , \, - \half \right) \, .
\end{equation}
These towers combine into the $\T^2$ winding modes of Type I which, in turn, coincide with the KK tower of Type IIB obtained in \eqref{eq. KKpillow} as a consequence of \mbox{T-duality}. Note that $\ctm_{\win, \rm{II}, \, \S_x^1} = \ctm_{\KK, \S_y^1/\Z_2}^{\unw}$ and vice versa, reflecting the symmetry under exchange of radii. Therefore, the convex hull formed by the leading towers in the asymptotic directions of the two fundamental domains that fill the original heterotic region will have the same shape, with the KK modes of the interval and winding modes of the circle interchanged. This convex hull for the unwarped limit ($\hat\varrho_x, \hat\varrho_y \geq 0$) is depicted  in Figure \ref{fig. hulls}.

\subsubsection*{The $\EE$ heterotic string}

The towers of states of the $\EE$ case are computed in complete analogy with the $SO(32)$ case. In particular, the heterotic towers, as well as the oscillator modes of type $\Ip$ (which coincide with those of Type I given in \eqref{eq4:ctmI}), remain unchanged. Likewise, the sliding KK tower is obtained by reduction to eight dimensions. Now, the interval of the Type IIA region is the same in both cases, however, what changes is the unit conversion of the $\ctm$-vector from 10d to 8d. Then we have
\begin{equation}
\ctm_{ \KK, \S^1/\Z_2 }^x = \left( \sqrt{\frac{2}{3}} , \, \half , \, 
\frac{3}{2 + 4 \coth (\hat \varrho_y)} \right) \, , \quad
\ctm_{ \KK, \S^1/\Z_2 }^y = \left( \sqrt{\frac{2}{3}} , \, 
\frac{3}{2 + 4 \coth (\hat \varrho_x)} , \, \half \right) \, .
\end{equation}
corresponding to Type IIA on $\S^1/\Z_2 \times \S^1_x$ or $\S^1/\Z_2 \times \S^1_y$, respectively. The unwarped limits coincide, while the highly warped do not:
\begin{equation}
\ctm_{ \KK, \S^1/\Z_2 }^{\text{unw}} = \left( \sqrt{\frac{2}{3}} , \half , \half 
\right) \, , \quad
\ctm_{ \KK, \S^1/\Z_2 }^{\text{x,war}} = \left( 
\sqrt{\frac{2}{3}} , \half , 0
\right) \, , \quad
\ctm_{ \KK, \S^1/\Z_2 }^{\text{y,war}} = \left( 
\sqrt{\frac{2}{3}} , 0 , \half
\right) \, ,
\end{equation}
reflecting again the symmetry under exchange of radii. We can also compute the KK tower associated with the complete decompactification of the manifold $X_3 \equiv \S^1/\Z_2 \times \T^2$ in the M-theory regime:
\begin{equation}
\ctm_{\KK, X_3} = \left( 0 , \half, \half \right) \, , \qquad
\end{equation}
which has length $\inv{\sqrt{2}}$, as predicted by \eqref{eq. univ scalings}. 

The winding modes corresponding to the Type II string wrapped along the cylinder (now, the cylinder illustrated in the $\EE$ diagram of Figure \ref{fig. geometries}) are not present since the type $\Ip$ emergent string limit is obstructed, see \cite[Fig. 11 and footnote 13]{Etheredge:2023odp}. However, we can wrap the string of the adjacent type $\Ip$ theory\footnote{
For type $\Ip$ on $\S^1_x$ ($\S^1_y$), the adjacent type $\Ip$ theory corresponds to reducing on $\S^1_y$ ($\S^1_x$) the M-theory frame obtained from the strong coupling limit of the T-dual heterotic along $\S^1_y$ ($\S^1_x$). 
}
along the circle (despite the interpretation of these modes from the former type $\Ip$ is not trivial). Like the KK modes, these winding modes coincide, and their corresponding charge-to-mass vector is given by
\begin{equation}
\ctm_{\win,{\rm II-dual}_y ,\, \S^1_x} = \left( \sqrt{\frac{2}{3}} , - \half , - \half \right) =
\ctm_{\win,{\rm II-dual}_x ,\, \S^1_y} \, .
\end{equation}

\subsection{Dilaton-Single radius slice puzzle}

\begin{figure}[ht]
    \begin{center}
\centering
     \resizebox{\textwidth}{!}{%
    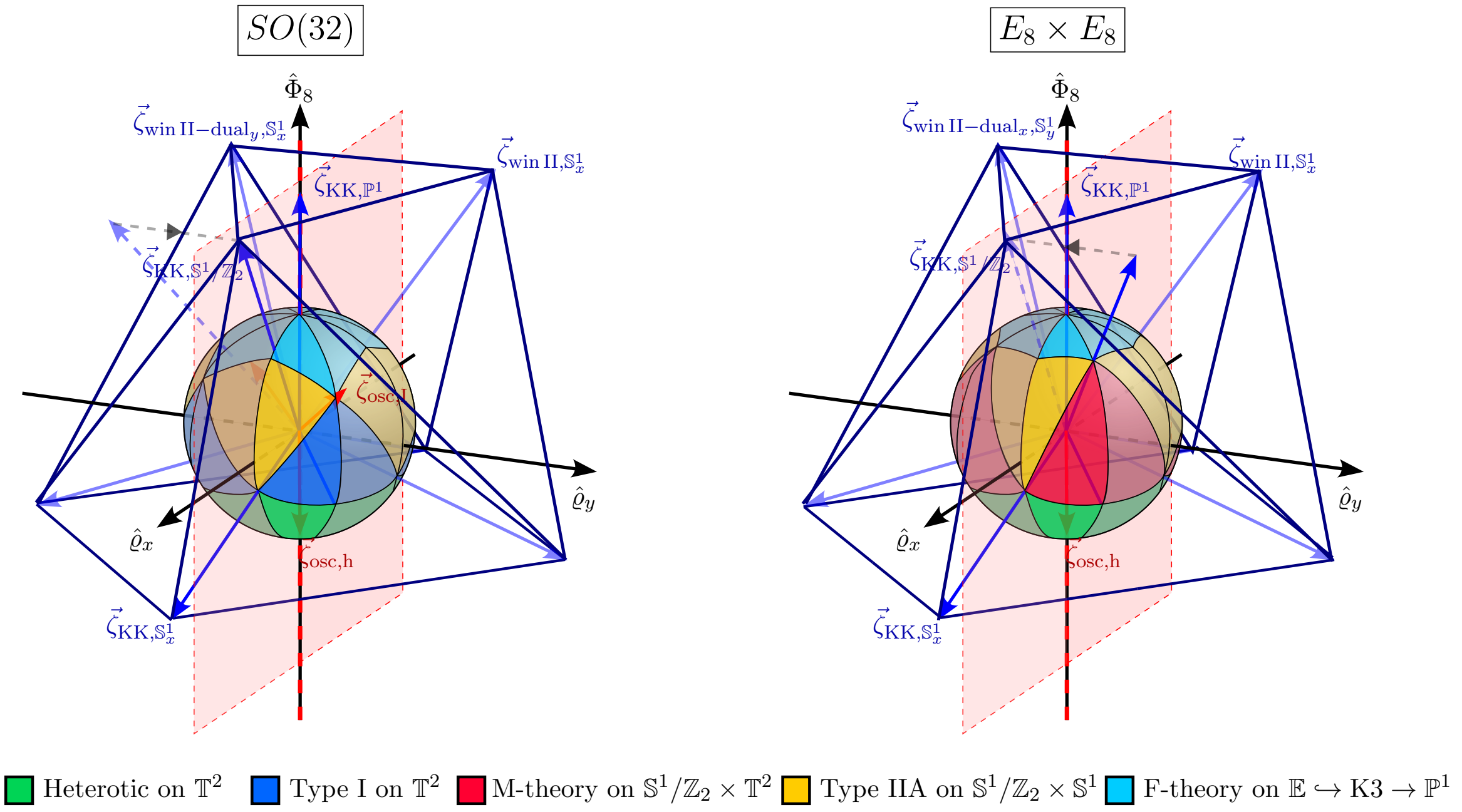
    }
    \caption{Convex hulls generated by the light towers in the fixed locus $\hat{\varrho}_y=0$, featuring warped decompactifications. We also depict the $\zeta$-vectors (in red for oscillator and blue for KK modes) associated to the BPS heterotic KK and winding modes (unlabeled), symmetric from the original $\vec\zeta_{{\rm KK},R_1}$, as well as the winding modes of the two heavy Type II strings on the initial circle. The system is symmetric under T-duality along the string-scale radius. All quantities are depicted in terms of canonically normalized moduli.
    }
    \label{fig. hulls warped}%
    \end{center}
\end{figure}

We can also approach the plane fixed by T-duality on ${\S^1_y}$, which is parametrized by $\Vol ({\S^1_y}) = \ell_\h$ (i.e., $\hat\varrho_y=0$). This corresponds to the highly warped regime, where the relevant light tower is the highly warped KK tower associated with the interval of the type $\Ip$ on $\S^1_y$ description. Since this plane lies at the intersection of two distinct type $\Ip$ duality frames, both descriptions are simultaneously valid, and we therefore observe the two strings wrapping the corresponding circle. As a result, two distinct winding towers become light, in contrast to the previous case where only a single winding tower was present: 
\begin{equation}
    \ctm_{\win,\rm II,\, \S^1_x} = 
\left( \sqrt{\frac{2}{3}} , \, - \half , \, \half \right) \, , \qquad \ctm_{\win,\rm{II-dual}_y,\, \S^1_x} = 
\left( \sqrt{\frac{2}{3}} , \, - \half , \, -\half \right) \, .
\end{equation}
The resulting leading towers and their convex hull are shown in Figure~\ref{fig. hulls warped}. \\

We would expect the projection of the two moduli spaces ($SO(32)$ and $\EE$) onto the plane fixed by T-duality ($R_{\S^1_y} \sim \ell_\h$) in Figure \ref{fig. hulls warped} to coincide with the moduli space shown in Figure \ref{fig.slice1t2}, as they correspond to the same slice. However, they do not, a fact that constitutes a puzzle. The explanation owes to the fact that our treatment of the moduli space has been carried out for homogeneous (de)compactifications, whereas in this case the $\P^1$ base of the F-theory $\E \hookrightarrow \rm{K3} \ra \P^1$ fibration becomes highly elongated as one approaches this slice---developing a ``sausage geometry'', as explained in \cite[Section 2]{Morrison:1996pp}---so that the full arrangement depicted in Figure \ref{fig. hulls warped} is no longer valid in this regime. Nevertheless, the treatment based on the reconstruction rules and the symmetries of moduli space remains applicable, since it relies only on the duality structure. It is precisely through these arguments that we obtained Figure \ref{fig.slice1t2}, which should therefore be regarded as the correct description of that corner of moduli space where the full treatment breaks down. We thus expect that in the full three-dimensional picture the boundary between the Type IIA and F-theory regions is not straight but rather ``curve'' as it approaches the plane parallel to the locus $\Vol({\S^1_y}) = \ell_\h$ as one approaches it, in such a way that its intersection with that plane reproduces exactly the moduli space of Figure \ref{fig.slice1t2}. Correctly understanding the boundary along this specific plane would require good control of the warped ``sausage geometry'' in the F-theory frame. Since we do not currently have this technology, we leave this open for a future project.

\subsection{F-theory on K3 and the Weierstrass model\label{app. F-theory}
} 

We now will briefly discuss the details of the F-theory picture describing the strong coupling regime of heterotic string theory on $\mathbb{T}^2$, see Section \ref{sec. het t2}. Following \cite{Sen:1996vd} (see also \cite{Weigand:2018rez} for a more general picture), our elliptically fibered $\mathbb{E}\hookrightarrow{\rm K3}\to \mathbb{P}^1$ is described by a Weierstrass model
\begin{equation}
    y^2=x^3+f(z)x+g(z)\,,\quad\text{with}\quad z\in\mathbb{P}^1\,, 
\end{equation}
and $x$ and $y$ local Weierstrassfiber coordinates, such that $f(z)\in H^0(\mathbb{P}^1,\mathcal{O}(8))$ and $g(z)\in H^0(\mathbb{P}^1,\mathcal{O}(12))$. This defines an elliptic fibration over the $\mathbb{P}^1$ base, whose modular parameter, identified with the axio-dilaton $\tau=C_0+ie^{-\Phi}$, is controlled by the Klein $j$-invariant
\begin{equation}
    j(\tau(z))=\frac{4(24 f(z))^3}{27 g(z)^2+4f(z)^3}=\frac{4(24 f(z))^3}{\Delta}\,,
\end{equation}
where the 24 zeroes of the discriminant $\Delta=27 g^2+4f^3$ control the degeneration of the fiber. This F-theory configuration describes the strongly coupled regime of Type IIB string theory on the $\mathbb{T}^2/\mathbb{Z}_2$ orientifold. As explained in \cite{Sen:1997gv}, we can recover the weak coupling limit by parameterizing
\begin{equation}\label{eq.SEN}
    f(z)=-3h(z)^2+\epsilon\,\eta(z)\,,\quad g(z)=2h(z)^3-\epsilon\,h(z)\eta(z)+\epsilon^2\chi(z)\,,
\end{equation}
with $h(z) \in H^0(\mathbb{P}^1, \mathcal{O}(4))$, $\eta(z) \in H^0(\mathbb{P}^1, \mathcal{O}(8))$, $\chi(z) \in H^0(\mathbb{P}^1, \mathcal{O}(12))$
and $\epsilon>0$ a scaling parameter. Expanding in powers of $\epsilon$, we have
\begin{subequations}
    \begin{align}
        \Delta&=9\epsilon^2 h(z)^2\big[12h(z)\chi(z)-\eta(z)^2\big]+\mathcal{O}(\epsilon^3)\\
        j(\tau(z))&=\frac{8[12h(z)]^4}{\epsilon^2[\eta(z)^2-12 h(z)\chi(z)]}+\mathcal{O}(\epsilon^{-1})
    \end{align}
\end{subequations}
Thus, at any generic point of the base where $h(z)\neq 0$ and $12h(z)\chi(z)-\eta(z)^2\neq 0$, $j(\tau(z))\to\infty$ as $\epsilon\to 0$. In such case we can approximate $j(\tau)\sim e^{-2\pi i\tau}$\, and as such 
\begin{equation}
    \Phi\sim -\log\left[\log|j(\tau(z))|\right]\sim -\log\left[\log(\epsilon^{-1})\right]\to-\infty\quad\text{as}\;\epsilon\to0\,.
\end{equation}
Thus, any Weierstrass model allowing for the \eqref{eq.SEN} parametrization has a weak coupling (and thus an emergent string) limit.

\subsection*{The $SO(32)$ case}
 We first consider the $SO(32)$ configuration. Within the \eqref{eq.SEN} parameterization, take $h(z)$ to have four distinct roots $\{z_1,z_2,z_3,z_4\}\in\mathbb{P}^1$, which will correspond to the location of the O7$^-$ planes. Now, we take
 \begin{equation}
     h(z)=(z-z_1)\tilde{h}(z)\,,\quad \eta(z)=(z-z_1)^8\tilde{\eta}_0\,,\quad \chi(z)=0\,,
 \end{equation}
 with $\tilde{h}(z_1)\neq 0$. Then ${\rm ord}_{z_1}(f,g,\Delta)=(2,3,18)$, precisely corresponding to a $I_{12}^*$ singularity carrying gauge group $SO(32)$, see \cite[Table 4.1]{Weigand:2018rez}. Regarding the O7$^-$ plane singularities, we have that in the strict Sen limit $\epsilon=0$, ${\rm ord}_{z_i\neq z_1}(f,g,\Delta)=(2,3,2)$, which would correspond to a non-Kodaira $I_{-4}^*$ singularity, where the $\mathbb{Z}_2$ involution of the fixed point in the Type IIB geometry acts on the fiber, see Figure \ref{fig. geometries}. At finite $\epsilon$ (i.e., in the strong coupling regime), quantum effects split the bare O7$^-$ plane into two $I_1$ fibers, each associated with monodromy $\begin{psmallmatrix}
     4&9\\-1&-2
 \end{psmallmatrix}$ and $\begin{psmallmatrix}
     2&1\\-1&0
 \end{psmallmatrix}$ and separated at finite $\epsilon>0$.

\subsection*{The $E_8\times E_8$ case}
Following Kodaira-N\'eron classification \cite{Kodaira+2015+1142+1156,Neron}, the $E_8$ gauge symmetry is obtained by a $II^*$ singular fiber, with ${\rm ord}(f,g,\Delta)=(\geq 4,5,10)$. Having then two $II^*$ singularities leaves respectively 0, 2 and 4 zeroes for $f$, $g$ and $\Delta$. This way,
\begin{equation}
    f(z)=f_0(z-z_1)^4(z-z_2)^4\,,\quad g(z)=\tilde g_2(z)(z-z_1)^5(z-z_2)^5\,, \quad \Delta(z)=\tilde\Delta_4(z)(z-z_1)^{10}(z-z_2)^{10}
\end{equation}
with $\tilde g_2(z)$ a degree-2 polynomial with $\tilde g_2(z_i)\neq 0$ and $\tilde\Delta_4(z)=27\tilde g_2(z)^2+4f_0^3(z-z_1)^2(z-z_2)^2$. Then it is clear that for the additional four zeros $\{z_j\}$ of $\Delta$, $g(z_j)\neq 0$, so that ${\rm ord}_{z_j\neq z_1,z_2}(f,g,\Delta)=(0,0,1)$, corresponding to $I_1$ singularities. Now, it is easy to see that this setting does not admit a Sen limit.

In order to realize a parametrization \eqref{eq.SEN} in the $\epsilon\to 0$ limit, we need $h(z)\sim (z-z_1)^2(z-z_2)^2$, so that $g(z)\sim (z-z_1)^6(z-z_2)^6$ and ${\rm ord}_{z_i=z_1,z_2}(f,g,\Delta)=(\geq 4,\geq 6,\geq 12)$, which does not admit a minimal Kodaira singularity. This produces a non-minimal degeneration, so the ordinary smooth elliptic-K3 description is insufficient. We thus conclude that Sen limit is not present in this setting.

To understand the location of the $II^*$ singularities, we consider the projection $\mathbb{T}^2\xrightarrow[]{\mathbb{Z}_2}\mathbb{T}^2/\mathbb{Z}_2\simeq\mathbb{P}^1$, branched over the four fixed points $\{z_1,z_2,z_3,z_4\}\subset\mathbb{P}^1$, defining $h(z)=\prod_{i=1}^4(z-z_i)$. The $\mathbb{Z}_2$ symmetry forces the discriminant $\Delta(z)$ to vanish at these points with order $\geq 2$, which prevents the $I_1$ singularities from sitting here. On the other hand, coalescing $z_1=z_2$ and $z_3=z_4$, we have $f(z)\sim h(z)^2$, recovering the precise vanishing 4 order for $f$. This is depicted in Figure \ref{fig. geometries}. From the F-theory/M-theory duality, the base of the $\mathbb{S}^1\hookrightarrow\mathbb{P}^1\to\mathbb{S}^1/\mathbb{Z}_2$ base can be understood as the Ho\v{r}ava-Witten interval, with the $II^*$ singularities at the coalesced fixed points the $E_8$ 9-planes. See \cite[Section 2]{Morrison:1996pp} for a description of the ``sausage geometry'' of $\P^1$, where the base elongates anisotropically, precisely corresponding with the decompactification of the $\mathbb{S}^1\hookrightarrow\mathbb{P}^1\to\mathbb{S}^1/\mathbb{Z}_2$ base.

\section{Additional possible perturbative arrangements and ``exotic'' T-dualities.\label{ap.extra}}

\begin{figure}[ht]
\begin{center}
\centering
     \resizebox{\textwidth}{!}{%
    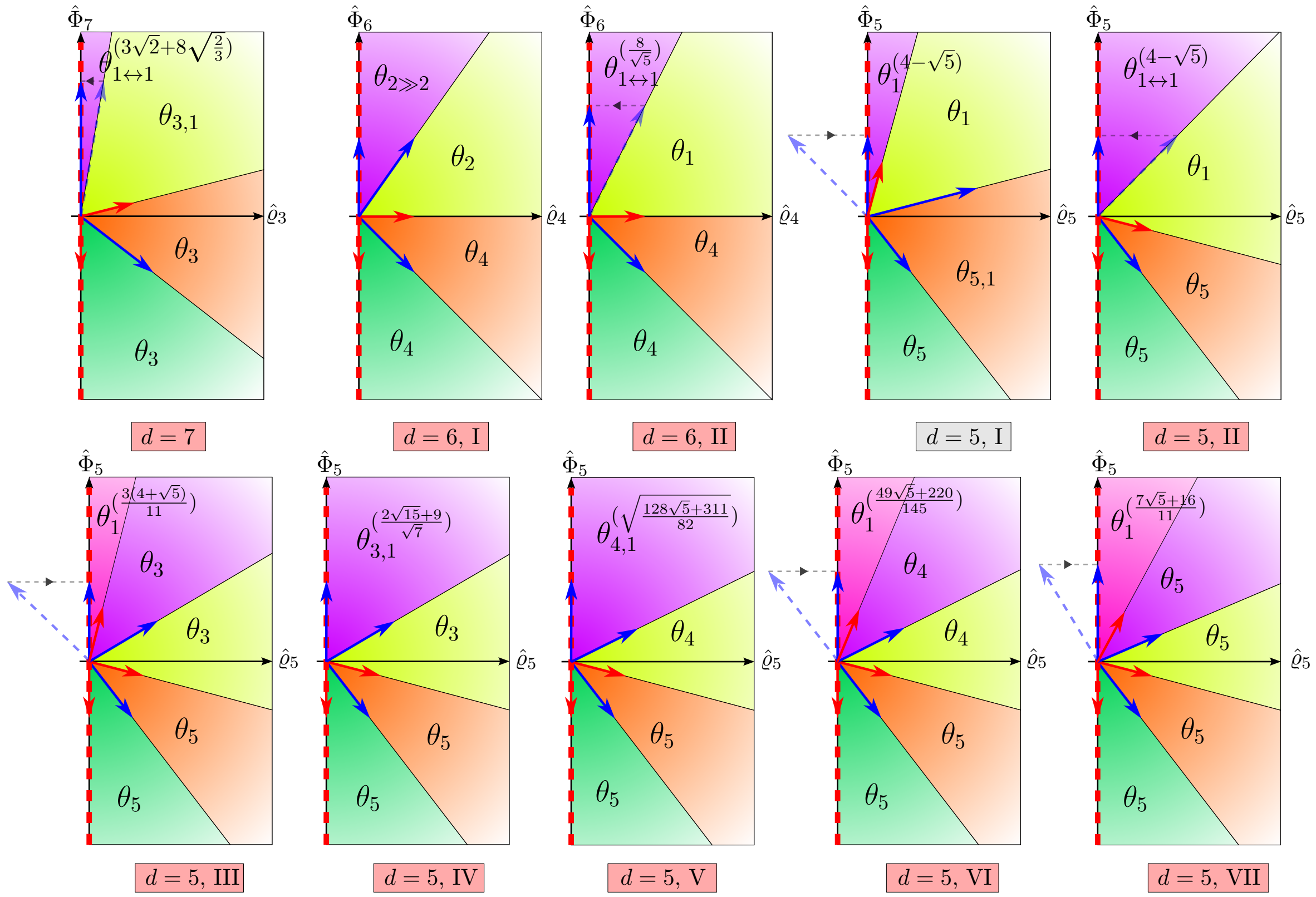
    }
\caption{
Additional perturbative regime configurations for $d=7$ to 5, for 2-dimensional flat Euclidean slices of moduli space with an emergent string limit on $X_{10-d}$ in the $\hat{\Phi}_d\to-\infty$ limit, and a $\mathbb{Z}_2$ symmetry leaving invariant the self-dual line parameterized by $\hat{\Phi}_d$ (depicted in dashed red). For reasons of space, we only depict the fundamental domain. We use the notation from \eqref{eq. angle reg} and \eqref{eq.warping}, as well as drawing the scaling $\zeta$-vectors, whose interpretation as KK-$n$ (in blue) or oscillator (in red) modes is clear from the frame. We also depict those sliding, including the coupling $\lambda$ to the higher dimensional potential $\hat{V}(\hat\varphi)\sim V_0 \exp(-\lambda\hat{\varphi})$ causing the warped compactification. Note that for IV and V in $d=5$ we do not depict the sliding, since it would occur perpendicular to the depicted slice. In red we label those exotic arrangements where an ``exotic'' T-duality exchanging KK modes associated to compact spaces of different dimensionality. All quantities are depicted in terms of canonically normalized moduli.
\label{fig.extraW}}
\end{center}
\end{figure}

In Section \ref{sec. half max} we were able to recover the set of duality frames/perturbative descriptions within the overall volume-$\Phi_d$ flat slices of moduli spaces of heterotic string theory on $\mathbb{T}^{10-d}$. For $d=9$ a frame involving a warped (de)compactification was needed, and indeed its interpretation as Type I$'$/mIIA compactification is clear due to the identification of the coupling $\lambda=\frac{5}{\sqrt{2}}$ with that of the 10d Romans mass in Einstein frame. For lower dimensionalities, however, we saw that no warped limits were needed in order to fill the half-plane. Since they depend continuously on the $\lambda$ coupling, see \eqref{eq.warping}, we have \footnote{Note that for the limiting case $\lambda=\frac{2}{\sqrt{d-1}}$ there is no ``fringe'' value for $\theta_{n,1}^{(\lambda)}$. Since there the warping occurs in the $(d+n)$-dimensional theory, the limiting case is $\lambda=\frac{2}{\sqrt{d+n-1}}$, for which $\theta_{n,1}^{(\frac{2}{\sqrt{d+n-1}})}=\arccos\sqrt{\frac{n}{d+n-2}}=\theta_n$.}
\begin{equation}    \big(\theta_1^{(\lambda)},\theta_{1\leftrightarrow1}^{(\lambda)},\theta_{n,1}^{(\lambda)}\big)\to\left\{\begin{array}{ll}
\big(\theta_1=\arccos\frac{1}{\sqrt{d-1}},0,\theta_{n,1}=\arccos\sqrt{\frac{n}{(d+n-2)(d-1)}}\big)  & \text{for }\lambda\to\infty \\
\big(0,\theta_1=\arccos\frac{1}{\sqrt{d-1}},\arccos{\sqrt{\frac{d n (d+n-2)+n}{(d-1) (d+n-2) (d+2 n-1)}}}\big) &\text{for }\lambda\to\frac{2}{\sqrt{d-1}} 
    \end{array}\right.\,,
\end{equation}
any angle can be accommodated through a combination of unwarped frames and warped ones for some appropriate $\lambda$. We thus will refrain from considering perturbative regimes associated to warped compactifications with the same angular span than those of unwarped compactifications with different internal dimensionalities. We depict in Figure \ref{fig.extraW} the additional solutions for $d=7$ to 5 (the single arrangement in $d=9$ is that already discussed in Section \ref{sec.het S1}, and there are not other arrangements in $d=8$ even if warping is allowed), i.e., the additional possible arrangements of perturbative regimes on 2-dimensional flat Euclidean slices of moduli space with an emergent string limit on $X_{10-d}$ in the $\hat{\Phi}_d\to-\infty$ limit, and a $\mathbb{Z}_2$ symmetry leaving invariant the self-dual line parameterized by $\hat{\Phi}_d$. All of these feature warped compactifications, which we identify by the coupling $\lambda$ to the higher dimensional potential $\hat{V}(\hat\varphi)\sim V_0 \exp(-\lambda\hat{\varphi})$. Similar to what we did for arrangements VIII and IX in Figure \ref{fig.extra4d} for $d=4$, where we respectively identified the sources of the higher-dimensional warping with an exotic $5^3_4$-brane in 5d and D8-branes in 10d, respectively, we can do a similar approach here for the exponents of Figure \ref{fig.extraW}. We find that we are only able to find a possible candidate for arrangement II in $d=6$, where the $\hat{V}(\hat\varphi)\sim V_0 \exp(-\frac{8}{\sqrt{5}}\hat{\varphi})$ 7d warping potential can be sourced by an exotic $5^3_4$-brane in $d=7$.\footnote{To see this, we proceed as in Footnote \ref{fn.EXOTIC}, and consider a Type IIA NS5 and a compact $\mathbb{T}^3_{789}$ down to $d=7$, with the 6 worldvolume directions of the brane unwrapped. Upon performing 3 T-dualities and a Type IIB S-duality, we find
\begin{equation}
\mathcal{T}_{\rm NS5}\sim\ell_{\rm s}^{-6}e^{-2\Phi}{}\xrightarrow[]{\text{T-duality along }\mathbb{T}^3_{789}}\mathcal{T}_{5^3_2}\sim\frac{\Vol(\mathbb{T}^3_{789})^2}{\ell_{\rm s}^{12}e^{2\Phi}}\xrightarrow[]{\text{S-duality}}\mathcal{T}_{5^3_4}\sim\frac{\Vol(\mathbb{T}^3_{789})^2}{\ell_{\rm s}^{12}e^{4\Phi}}\,,
\end{equation}
which we can write in terms of the 7d Planck mass and dilaton as
\begin{equation}
    \mathcal{T}_{5^3_4}\sim M_{\rm Pl,7}^6e^{-\frac{8}{5}\Phi_7}\sim M_{\rm Pl,7}^6e^{-\frac{4}{\sqrt5}\hat\Phi_7}\,,
\end{equation}
precisely the tension needed to generate a  $\hat{V}(\hat\varphi)\sim V_0 \exp(-\frac{8}{\sqrt{5}}\hat{\varphi})$ potential, see Footnote \ref{fn.source}.
} None of the other $\lambda$ from Figure \ref{fig.extraW} can be associated with (wrapped) D-/NS5-/exotic branes, though we do not completely exclude the possibility that some other construction could realize the quite involved exponents.

\subsection*{On ``exotic'' T-dualities}
We finish this section by noticing that all arrangements in Figure \ref{fig.extraW} (other than arrangement I for $d=5$, see also arrangements VII, VIII and X in Figure \ref{fig.extra4d}) feature dualities between descriptions where the light modes of a KK-$p$ and KK-$q$ (with $p\neq q$) towers are interchanged under some type of ``exotic'' T-duality, that leaves the string scale invariant. Such possibilities were already discussed in \cite{Etheredge:2024tok,Etheredge:2025ahf}, where it was noted that while such arrangements were in principle consistent with the taxonomy rules from \cite{Etheredge:2024tok}, they do not appear in any known corner of the string landscape, and present problematic behaviors under further compactification. Indeed, one could argue that such duality could be in tension with the Emergent String Conjecture \cite{Lee:2019wij,Lee:2019xtm}.

For this qualitative argument, we consider the emergent string limit common to both descriptions, where the scale of the light tower is $m_{\rm osc}\sim \ell_{\rm s}^{-1}$. Now, consider the KK tower on each description, associated to a growing compact $X_p$ and $Y_q$, with scale
\begin{equation}\label{eq. several KK}
    \frac{m_{{\rm KK},p}}{M_{{\rm Pl,}d}}\sim\frac{\ell_{\rm s}}{\Vol(X_p)^{-\frac{1}{p}}}\frac{m_{\rm osc}}{M_{{\rm Pl,}d}}\,,\qquad \frac{m_{{\rm KK},q}}{M_{{\rm Pl,}d}}\sim\frac{\ell_{\rm s}}{\Vol(Y_q)^{-\frac{1}{q}}}\frac{m_{\rm osc}}{M_{{\rm Pl,}d}}\;.
\end{equation}
Now, if we assume that from the $X_p$ perspective the $m_{{\rm KK},q}$ light states come from wrapped modes of $k$-brane with tension $\mathcal{T}_k$ over a $k$-cycle $C_k\subseteq X_p$, this is, $m_{{\rm KK},q}\sim \mathcal{T}_{k}\Vol(X_p)^{\frac{k}{p}}$, so that
\begin{equation}
    \mathcal{T}_k\sim\ell_{\rm s}^{-(k+1)}\frac{\ell_{\rm s}^{k+1}}{\Vol(Y_q)^{\frac{1}{q}}\Vol(X_p)^{\frac{k}{p}}}\,.
\end{equation}
Now, since the $k$-brane lives in the $(d+p)$-theory, its tension should not depend on the volume of $X_p$. Since both the volume in string units of $X_p$ and $Y_q$ are controlled by (opposite directions of) a modulus perpendicular to the dilaton controlling the string scale, we conclude that
\begin{equation}
    \mathcal{T}_k\sim \ell_{\rm s}^{-(k+1)}\,,\qquad \frac{\Vol(X_p)}{\ell_{\rm s}^p}\sim \left[\frac{\Vol(Y_q)}{\ell_{\rm s}^q}\right]^{-\frac{p}{q k}}\,.
\end{equation}
We can start with the second relation. As it measures volumes in string units, its dependence is orthogonal to the $d$-dimensional dilaton. Using \eqref{eq. can norm}, and considering that $X_p$ and $Y_q$ grow in opposite directions, we obtain
\begin{equation}
 \sqrt{p}|\hat\rho|=\frac{p}{\sqrt{q}k}|\hat\rho|\Longrightarrow\;   k^2q=p\,.
\end{equation}
Since $p,q\leq 10-d\leq7$, we have two possibilities to the above integral equation: Either $k=1$ and $p=q\in\{1,\dots,10-d\}$ or $(p,q,k)=(4,1,2)$. The first case is equivalent to the usual T-duality on $\mathbb{T}^p$ or quotients thereof, while the second requires more attention, as it would imply the existence of a 2-brane with $\mathcal{T}_2\sim m_{\rm osc}^3$. This is inconsistent with the expectations from \cite{Alvarez-Garcia:2021pxo}, which expect the mass scale of 2-branes to be parametrically higher than the KK scale, which along emergent string limit decays with the same dependence as the string scale, see \eqref{eq. several KK}. We thus expect that the $(p,q,k)=(4,1,2)$ solution is not present, and all ``T-dualities'' exchanging KK momenta in the small/large volume regime are between internal toroidal spaces of the same dimensionality.

The above discussion is a simple qualitative argument on why these ``exotic'' T-dualities should not occur in a consistent moduli space. We have used simple classical scaling relations that ignore quantum effects or instabilities that can affect the mass dependences, and assume that the topology of the internal space remains constant and does not experience a change, such as a conifold transition, at finite distance. Since the study of these possible caveats would require a more detailed analysis, see \cite{Alvarez-Garcia:2021pxo}, a complete argument excluding such type of dualities is out of the scope of this paper, and is left for future work.

\end{appendix}

\bibliographystyle{JHEP}
\bibliography{ref}
\end{document}